# Determining Molecular Complexity using Assembly Theory and Spectroscopy

Michael Jirasek†[1], Abhishek Sharma†[1], Jessica R. Bame†[1], S. Hessam M. Mehr[1], Nicola Bell[1], Stuart M. Marshall,[1] Cole Mathis[1], Alasdair Macleod,[1] Geoffrey J. T. Cooper,[1] Marcel Swart[2,3], Rosa Mollfulleda[2], Leroy Cronin*[1] †Equal contribution; *Corresponding author* **Lee.Cronin@glasgow.ac.uk*

***[1] School of Chemistry, The University of Glasgow, University Avenue, Glasgow G12 8QQ, UK.***

***[2] University of Girona, Campus Montilivi (Ciencies), c/M.A. Capmany 69, 17003 Girona Spain***

***[3]ICREA, Pg. Lluis Companys 23, 08010 Barcelona, Spain***

**Abstract:** Determining the complexity of molecules has important applications from molecular design to understanding the history of the process that led to the formation of the molecule. Currently, it is not possible to experimentally determine, without full structure elucidation, how complex a molecule is. Assembly Theory has been developed to quantify the complexity of a molecule by finding the shortest path to construct the molecule from building blocks, revealing its molecular assembly index (MA). In this study, we present an approach to rapidly and exhaustively calculate the MA of molecules from the spectroscopic measurements. We demonstrate that molecular complexity (MA) can be experimentally estimated using three independent techniques: nuclear magnetic resonance (NMR), tandem mass spectrometry (MS/MS), and infrared spectroscopy (IR), and these give consistent results with good correlations with the theoretically determined values from assembly theory. By identifying and analysing the number of absorbances in IR spectra, carbon resonances in NMR, or molecular fragments in tandem MS, the molecular assembly of an unknown molecule can be reliably estimated from experimental data. This represents the first experimentally quantifiable approach to defining molecular assembly, a reliable metric for complexity, as an intrinsic property of molecules and can also be performed on complex mixtures. This paves the way to use spectroscopic and spectrometric techniques to unambiguously detect alien life in the solar system, and beyond on exoplanets.

## Significance

We provided new experimental measures for inferring molecular complexity based on Nuclear Magnetic Resonance, Infrared Spectroscopy and tandem Mass Spectrometry. Here the molecular complexity is estimated without the requirement of a structural elucidation. This enables us to judge whether a sample of unknown origin, is expected to be coming from a living system, i.e. a biosignature.

## Introduction

The exploration of chemical space reveals the striking fact that most molecules greater than the molecular weight of 300 Da, which are not simple oligomers or composed of heavy atoms, are all connected to the existence of life on Earth.(1) It has been shown that complex molecules such as natural products(2) are too complex to form by chance in any detectable abundance and, therefore can only be made by the complex biochemical pathways found in biological cells. Currently, the exploration of complex chemical space is done *in-silico*(3, 4) and this focuses on chemical structure,(5) topological features,(6) application-specific physicochemical descriptors and graph theory and tends to explore medicinal chemical space for drug discovery and development.(7) In this regard pharmaceutical products can also be considered to be biosignatures, or more specifically technosignatures, since many are complex and would not have been made without humans using technology.(8–10) In addition to targeted selectivity, synthetic accessibility is important to explore the complexity of the molecule.(11) There are many competing notions of molecular complexity(12), which have led to different algorithmic methodologies being developed using metrics based on molecular weight, counting chiral centres or primarily focusing on substructure properties etc.(13–15) However, with the recent development in algorithmic chemical exploration,(16) a proxy for complexity is required that is fast to estimate molecular complexity directly from the acquired experimental data, instead of performing complete structure elucidation. Additionally, for biosignature detection,(17) it is important that the complexity metric can be estimated directly from the experimental data without any assumptions about the local environment or chemistry due to the minimalistic information available for an unknown sample.

Recently, we developed a novel approach to quantify and explore the complexity of molecules using Assembly Theory (AT).(18) Assembly Theory estimates the complexity of a molecule by quantifying the minimum constraints required to construct an object from the building blocks. The assembly pathway gives the shortest path to create an object in the absence of physical constraints and reusing the substructures formed along the pathway. The complexity of an object is therefore defined by the number of steps along the assembly pathway and is called the Assembly Index,(19) which for molecules is called Molecular Assembly (MA). To date, all other approaches to experimentally address molecular complexity require the formula and connectivity of the molecule to be known.(20) The Molecular Assembly (MA)(21, 22) for a molecule is computed by representing the molecule as a graph and performing an algorithmic search to find the shortest pathway to construct the graph by reusing previously made structures along the pathway, see **Fig. 1A**. Thus, various constraints in the molecular graph are found along the pathway to quantify the complexity of the molecule.

In previous work, we used tandem Mass Spectrometry (MS/MS) for the experimental measurement of MA and were able to rank molecules in order of their complexity by placing them on a scale where molecules beyond MA of 15 were shown to be biosignatures for life-detection. Experimentally, over a range of high MA molecules, it was demonstrated that there exists a correlation between $MS^2$ peaks and computed MA values(23). Herein, we developed experimental measurement strategies to infer molecular complexity using MA by using IR and NMR spectroscopies and expanded our understanding of inferring MA from mass spectrometry using a new algorithm. Using both simulated and experimental data, we demonstrate that MA can be experimentally inferred over a wide range of complex molecules as well as mixtures. Additionally, we demonstrate that by combining multiple spectroscopic techniques into one measure, the MA prediction can be improved further.

Infrared Spectroscopy (IR) is routinely used to confirm the presence of specific bond types in molecules by observing their characteristic vibrational energies in higher energy ranges (1500–3600 cm$^{-1}$). Vibrational motion corresponding to those absorption bands is typically of a local nature, for example, a stretch vibration of one bond. Contrary, lower energy (fingerprint) region 400–1500 cm$^{-1}$ typically possesses a plethora of absorption bands, without direct easy interpretation toward structure elucidation.(25) These modes include various collective motions, bending vibrations, and coupled modes. Since the number of different substructures increases with molecular complexity, we hypothesise that the number of unique absorption bands in the fingerprint region can be used to infer the complexity of organic molecules. Moreover, IR has previously been used to fingerprint complex molecular ensembles in their native natural environment.(26)

Nuclear Magnetic Resonance (NMR) spectroscopy provides resonance frequencies of magnetically inequivalent atoms nuclei in the structure. The exact chemical shift of each nucleus of the same element depends on the effective magnetic field experienced by it, strongly influenced by its chemical microenvironments (affected by e.g. bond correlation *via* scalar coupling, or through space (de)shielding effects).(27) NMR has been used in the past to analyse chemical space for fragment screening in drug discovery(28–30) and characterising structural complexity of compound classes(31). Advantageously, NMR has minimal solvent limitation allowing to keep the sample in its native state/solvent if desired. NMR is uniquely equipped to address the structural diversity as symmetric (magnetically equivalent) units in an isotropic environment (e.g., in a homogenous solution) possess the same chemical shift (thus not creating duplicated resonances). Further, from the perspective of molecular assembly, the effect of symmetry and bond rotation on NMR spectra was hypothesised to provide near equivalent resonances for duplicated fragments, even if not magnetically equivalent as a result of very similar chemical microenvironment experiences by the fragment. This represents the fact that assembled fragments may be utilised in multiple symmetric positions in a structure without having to 'rebuild' them each time. Therefore, we hypothesise that the number of observed NMR resonances will reflect a degree of structural complexity.

Both IR and NMR will agnostically indicate the complexity of a molecule defined by MA since Assembly Theory states that the MA utilizes unique irreducible motifs to construct the molecule that are indicated by the observed spectral features. This suggests that spectroscopy techniques that can quantify the properties of unique environments, and molecular substructures should in principle produce a good correlation with MA. Thus, we hypothesise that with more unique bond types and atomic environments for a given molecule, the larger the number of peaks that should be found in IR and NMR spectra for that molecule, see **Fig. 1**.

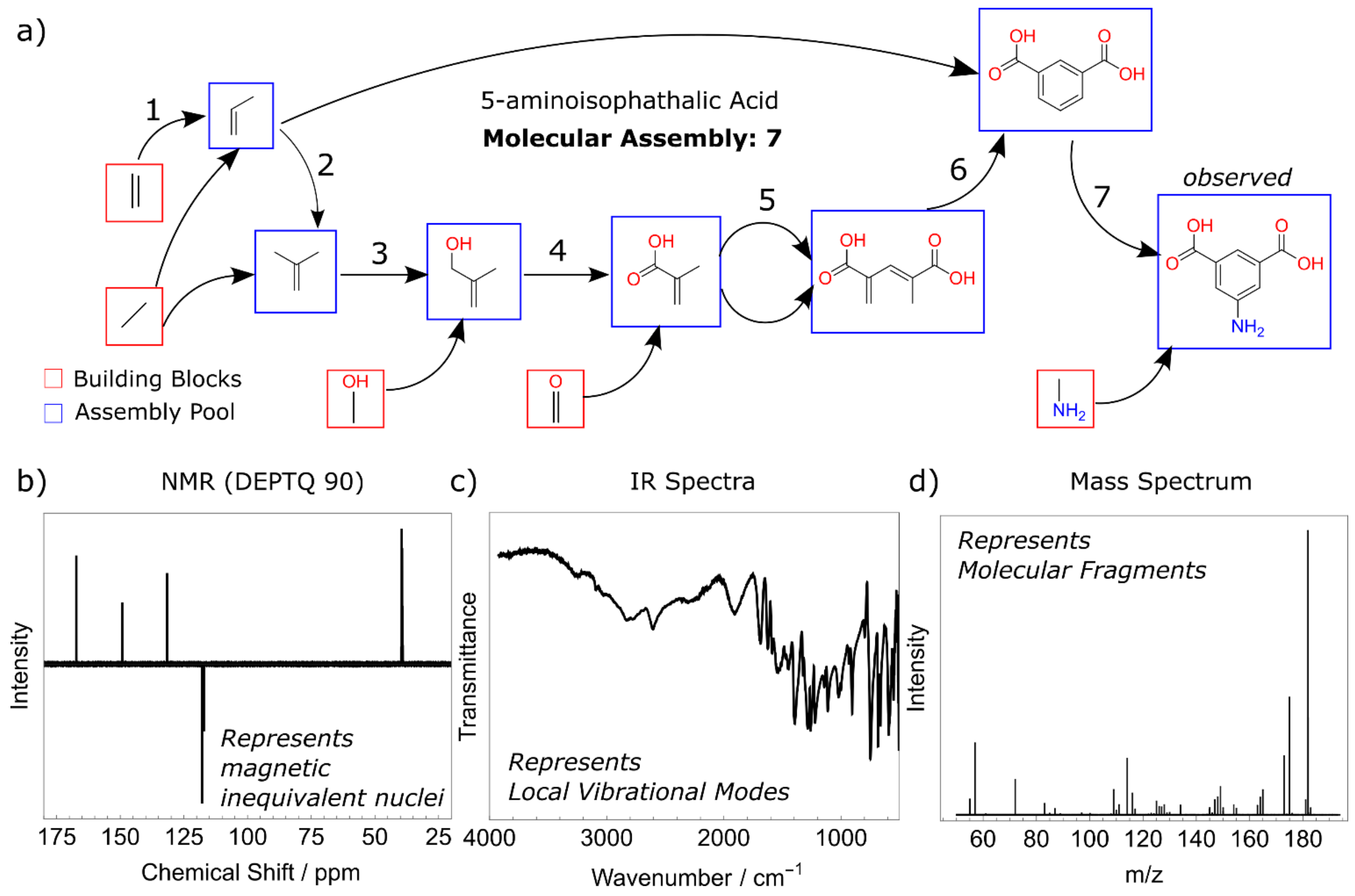

**Fig 1. Molecular Assembly of 5-aminoisophathalic acid.** (A) Molecular Assembly pathway of 5-aminoisophathalic with a total of 7 steps. The various chemical bonds are considered as fundamental building blocks (shown in red) and the substructures (shown in blue) along the pathways constitute the assembly pool. (B-D) Experimental NMR, IR, and $MS^2$ spectra of 5-aminoisophathalic acid highlight different features of the molecule from which the molecular constraints and the MA can be inferred.

### Calculating Assembly Index from a molecular graph

The Assembly index, and associated minimal assembly pathways, are calculated using an algorithm written in the Go programming language. In prior work(23), the assembly index was calculated using a serial algorithm written in C++ and yielded the "split-branch" assembly index, an approximation that provides a reasonably tight upper bound for the assembly index. The Go algorithm used in this work is a faster algorithm that incorporates concurrency and can provide the exact assembly index if it can be calculated in a reasonable time. The process can also be terminated early to provide the lowest assembly index found so far, which has been found to be a good approximation for the assembly index in most cases.

The assembly index is calculated by iterating over subgraphs within a molecular graph and finding duplicates of that subgraph within the remainder of the molecule. For each of the matching subgraphs found an assembly pathway can be represented by a duplicate structure and a remnant structure (see **Fig. 2**).

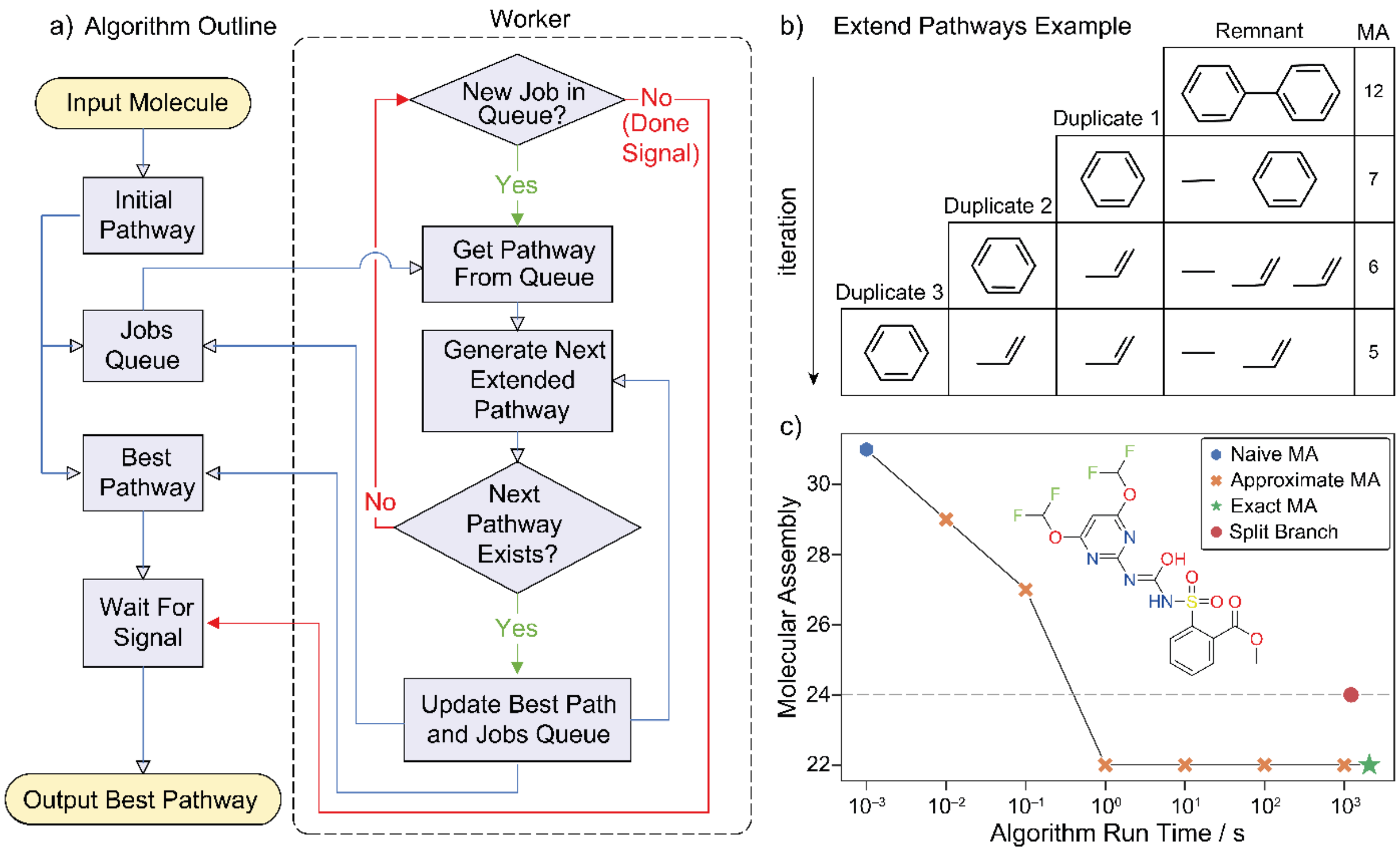

**Fig 2.** (a) The general structure of the Go assembly algorithm, with a pool of workers extending pathways. Some features are omitted for brevity, such as branch and bound methods to improve efficiency. (b) A sequence of assembly pathways as processed by the Go algorithm. The top pathway is the starting pathway for the molecule shown, and each subsequent pathway is extended from the pathway above. Pathways are generally extended in multiple ways, and only one such sequence of extensions is shown here. (c) An example of MA values found over time for Primisulfuron-methyl, run to completion, and approximated by stopping early at various stages prior. The new algorithm found pathways at the correct MA of 22 by 10 s, significantly before completion at ~2064 s. The red circle shows split branch algorithm performance on the same molecule. The naïve MA (blue hexagon) is calculated trivially for pathways in which one bond is added at a time (placed illustratively at $10^{-3}$ s, as 0 s cannot be represented on the logarithmic scale).

The remnant structure comprises the original structure with one duplicate removed, and the other "broken off", which ensures that all structures on an assembly pathway that are duplicated will be first constructed. The process can then be repeated recursively with the remnant structure as an input, which may result in more pathways containing two duplicate structures and a smaller remnant. Thus, each pathway is represented by a sequence of duplicated structures and a remnant structure. In this regard, it is important to point out that molecular assembly uses bonds as building blocks and not atoms. In order to determine the assembly index, we consider that a molecular graph with $N$ bonds could be constructed in $N-1$ steps by adding one bond at a time (the naive MA, or $MA_{naive}$ ). Each duplicate structure of size $N_{dup}$ allows us to add that structure in one step, reducing the number of steps compared to $MA_{naive}$ by $N_{dup}-1$. Thus, the MA for a particular pathway is $MA_{naive} - \sum_{dup}(N_{dup}-1)$ . For more details, see **SI Section 1**.

**Inferring Molecular Assembly using Infrared Spectroscopy**

As a first step, we computationally explore the potential for inferring the molecular assembly from IR absorption. A set of 10,000 molecules was chosen uniformly from the dataset published with previous study (23) of approx. $10^6$ molecules with MA. The new algorithm vastly speeded up the calculation and we were able to sample chemical space by calculating the MA (previously called Pathway Assembly, calculated using a split-branch algorithm). This was done so that we calculated MA for *ca.* 650 molecules at each MA unit between 2 and 23 MA for each molecule with the new implementation.(32) We calculated the IR spectra of the molecules using an extended semiempirical Tight Binding model implemented in xTB software including geometry optimization and calculating frequency resonances (see **SI Section 2.3**).(33, 34) For each spectrum, we estimated the total number of peaks in the fingerprint region (400–1500 $cm^{-1}$) assuming a resolution of 2 $cm^{-1}$. The number of peaks correlated significantly (Pearson correlation coefficient of 0.86) with the calculated MA, yielding a simple prediction function that was phenomenologically derived via linear regression:

$\mathrm{MA} = 0.21 \times n_{peaks} - 0.15$ (**Eq. 1**). This observation corroborated our hypothesis that number of absorption peaks in the IR fingerprint region reflects molecular complexity, see **Fig. 3**.

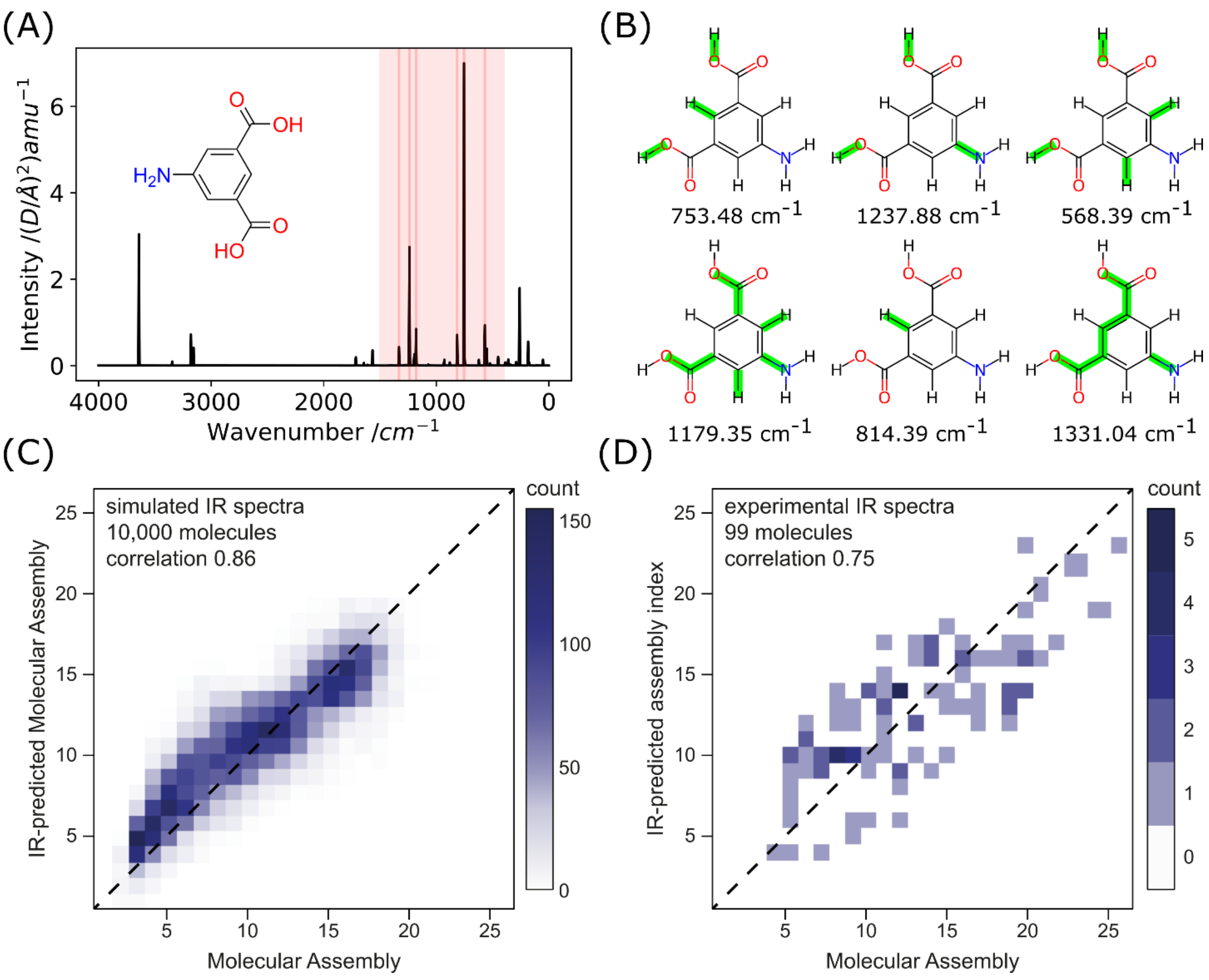

**Fig. 3. Inferring Molecular Assembly from infrared spectroscopy. (A)** XTB-calculated IR spectrum of 5-aminoisophthalic acid with highlighted fingerprint region (400–1500 cm$^{-1}$). **(B)** Example of the six most intense vibrational bands in the fingerprint region demonstrating its collective-motion nature. **(C)** Molecular Assembly *vs.* IR-inferred molecular assembly estimated from the number of IR peaks in the fingerprint region (400–1500 cm$^{-1}$) based on XTB calculation on 10,000 molecules (see **Eq. 1**). Correlation between the predicted and expected molecular Assembly is 0.86. **(D)** Molecular Assembly *vs.* IR-inferred molecular assembly estimated from the number of IR peaks in the fingerprint region (400–1500 cm$^{-1}$) based on the experimental measurement on 99 molecules (see **Eq. 2**). Correlation between the predicted and expected molecular assembly is 0.75.

Further, we expanded the study with experimental validation, using a set of 99 compounds MA over the range 4–26. The experiments were performed using diamond-attenuated total reflectance IR spectroscopy with a resolution of 2 cm$^{-1}$. The obtained spectra were processed at 50% sensitivity and up to 80% transmittance threshold for selecting peaks using OMNIC software as the coarse filter against low-intensity noise in real spectra. The total number of IR peaks in the fingerprint region (400–1500 cm$^{-1}$) correlated well with the compounds MA with a 0.75 correlation coefficient. This provided a handle for inferring the molecular assembly from an experimental IR using a simple linear function: MA $= 0.45 \times n_{peaks} - 2.3$ (**Eq. 2**). For more details see **SI Section 3**.

**Inferring molecular assembly from NMR spectra**

Most organic molecules (by definition) are composed of mainly carbon and hydrogen atoms, we hypothesised that $^{13}$C NMR is a practical technique to infer the molecular assembly of organic molecules. This was because the computation of molecular assembly is based upon bonds as building blocks and NMR will be uniquely able to explore the connectivity within complex organic molecules by exploring and quantifying the types of carbon atom present such as $CH_3$, $CH_2$, CH, and C, along with their relative connectivities. For the experimental measurement, a spectral width within which typically observed $^{13}$C nuclei resonances are found is relatively broad (~200 ppm), and it is reasonable to assume that inequivalent nuclei of sufficiently different microenvironments would rarely possess the same resonance frequency within a resolution of 0.5 ppm. Further, we expect that magnetically non-equivalent, yet structurally very similar sub-units with the exact environment in nuclei vicinity will be found within the resolution width (see **SI Section 2.2**). Observing such overlap will reflect the unit's similarity and the peaks will not be overcounted as the corresponding substructures likely share the assembly space (the space of motifs that are used to construct the target) and do not contribute to the molecular assembly (e.g. repeating units of the polymer chain such as –$CH_2$–).

Further information that can be experimentally extracted from the $^{13}$C-NMR spectrum is the classification of the carbon nuclei by the number of attached hydrogens. Based on the assembly theory, we hypothesise that the presence of carbons with no attached hydrogens (for clarity will be referred to as *quaternary*, note that this name will not be used exclusively for carbons with four different substituents, but for any carbon without connected hydrogen) reports most significantly on the molecular complexity as such centres are highly connected to four different atoms, but also can be connected to a range of different heteroatoms. Thus, these centres are hard to produce, and require many constraints to construct them. Analogously, we hypothesise that the more hydrogens are attached to the carbon, the less localized information it stores and hence, contributes less to the molecular assembly. From the experimental point of view, the classification of carbon nuclei by the number of attached hydrogens can be experimentally achieved using standard DEPTQ-90 and -135 routines, which provide information about the number of hydrogens attached to the carbon *via* the hydrogen-carbon coupling.(35, 36)

**Theoretical investigation**

To test our hypothesis, we examined a set of predicted $^{13}$C NMR spectra of 10,000 molecules (the same set as in the case of theoretical IR investigation). We have used the established predicting tool NMRShiftDB employing the Hierarchical Organization of Spherical Environments (HOSE) method.(37) An example of NMR prediction for two molecules (5-aminoisophathalic acid ($\mathrm{MA} = 7$) and Quinine ($\mathrm{MA} = 16$) with various carbon atoms labelled is shown in **Fig. 4(A&B)**. We classified the carbons by the number of hydrogens attached to them and summed the number of predicted peaks of a certain type assuming a bin width of 0.5 ppm. We performed multivariate linear regression (weighing out differently different types of carbons) and provided a model with a good correlation of 0.87, see **Fig. 4C.** The formula for inferring the molecular assembly from the number of found peaks associated with individual carbon types was phenomenologically derived via linear regression to be: $\mathrm{MA} = 1.3 \times \mathrm{C} + 0.8 \times \mathrm{CH} + 0.6 \times \mathrm{CH_2} + 0.3 \times \mathrm{CH_3} + 2.1$ (**Eq. 3**), where C (quaternary), CH (tertiary), $CH_2$ (secondary) and $CH_3$ (primary) are the number of binned (by 0.5 ppm) $^{13}$C resonances of carbons with none, one, two or three hydrogens attached, respectively. This observation on a large dataset significantly corroborates our prediction that quaternary carbons possess the highest degree of constraints and have the highest potential to report on molecular complexity.

**Experimental validation**

For experimental validation, we have assessed 101 compounds covering a range of molecular assembly of 3–26. We have acquired $^{13}$C NMR spectra and experimentally assigned the carbon type (C, CH, $CH_2$ and $CH_3$) *via* DEPTQ-90 and DEPTQ-135. The correct assignment was further cross validated with $^1$H-$^{13}$C HSQC since occasionally post processing of DEPTQ spectra can result in the inversion of the peaks phase. As the number of peaks is a simple and reliable measure directly comparable to the experimental observable property, we could test the trained model (**Eq. 3**) directly on independently chosen experimental molecules.

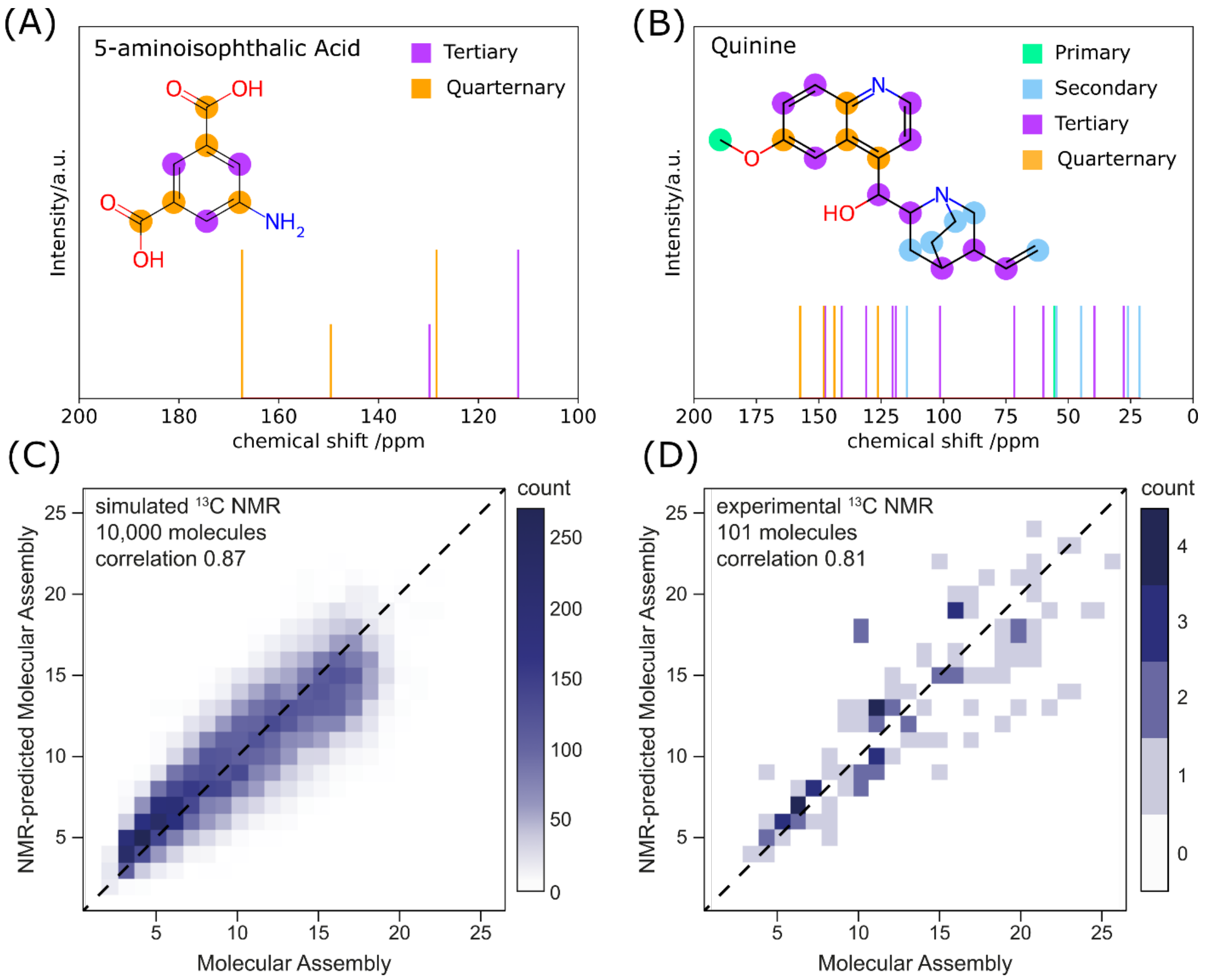

**Fig. 4. Inferring molecular complexity from $^{13}$C NMR spectra**. **(A)** and **(B)** shows the predicted $^{13}$C NMR spectrum of 5-aminoisophathalic acid and quinine, with highlighted different types of carbons. **(C)** Molecular assembly vs. NMR-inferred molecular assembly estimated from the number of different types of carbons (see **Eq. 3**) based on NMRshiftDB calculation on 10,000 molecules. The correlation between the predicted and expected molecular assembly is 0.87. **(D)** Molecular assembly *vs.* NMR-inferred molecular assembly estimated from the number of different types of carbons experimentally on 101 molecules, using the same model as in the theoretical set. The correlation between the predicted and expected molecular assembly is 0.81.

Testing the trained model provided a good correlation of 0.81, see **Fig. 4D**. Allowing the change in the multivariate regression on the experimental set could provide an even better correlation of 0.86 (**see SI Section 4**), however, we have considered using the model train on a large dataset as the more robust model, less biased by the sampling of the chemical space.

## Determining Molecular Assembly using tandem mass spectrometry

In our previous study, we demonstrated that tandem mass spectrometry can be used to estimate the molecular assembly of molecules directly from the peak count of $MS^2$ spectra.(23) As a generalized extension, herein, inspired by the construction of the assembly pathway of a molecule, we develop a new and robust algorithm to estimate molecular assembly utilising multiple fragmentations of the molecule. The key idea is to construct a hierarchical tree structure by matching the fragment masses in the $MS^n$ data with nodes representing the molecular fragments. This is analogous to the assembly contingent pathways(19) representing the steps to construct the molecule from the molecular bonds as the fundamental building blocks. It is important to note that the tree structure does not necessarily represent the shortest path within the assembly space, however, is crucial for inferring molecular assembly accurately from the mass spectrum. Within the tree structure of the fragmentation spectra, we compute the molecular assembly of all the fragments from the mass and combine them to estimate the molecular assembly of the parent molecule. In the next sections, we explore the relationship between molecular weight and molecular assembly over a large dataset and utilize a recursive algorithm to compute the molecular assembly of the molecule.

## Relation of Molecular Assembly and Molecular Weight

Since both molecular mass and molecular assembly increase with an increasing number of construction steps along an assembly pathway, one intuitively anticipates that heavier molecules are likely to have a higher MA. On a large dataset of 16.7 million molecules, sampled from the PubChem database,(37) we compute MA using the assembly algorithm with a short cut-off of 10 seconds. In the absence of any further information such as $MS^n$ spectra, the correlation between the MA and MW ($\mathrm{MA} = 0.047 \times \mathrm{MW} - 0.4$) can be considered as a first-order approximation, suitable for inferring MA of heterogeneous organic molecules (**Fig. 5a)**. This proxy for the MA inference provides reliable prediction for heterogenous, non-symmetric organic molecules, but not general solution for molecules with repeating units or heavy elements. The empirical distribution of MA per MW follows an approximate skew-normal distribution, which can be described with fitted parameters as: location parameter $loc = 0.0539 \times \text{MW} - 0.406$, skewness $\alpha = -0.0083 \times \text{MW} + 0.1$ and scale $s = 0.0074 \times \text{MW} + 0.511$ (See **SI Section 7.1** for more details). The upper limits of the MA within the sample could be bound as empirically found 99% percentile fit ($\text{MA}_{max} = 0.055 \times \text{MW} + 0.9$), interpretable as a naïve MA of a molecule constructed from lighter elements with relatively similar atomic weights such as C, N, and O etc. On the other hand, a lower limit could be defined, in the case of a polymer constructed from single-type monomeric units, to be approximately proportional to the logarithm of the MW as ($\text{MA}_{\text{min}} \propto \log_2(\text{MW})$).(19) Yet, given the heterogeneity of the chemical space, and the presence of heavy atoms significantly lowering the expected MA, the lower limit is not strictly logarithmic. As clear from the sample, the low MA of molecules within each distribution (*i.e.*, with the same molecular weight) can be attributed to the presence of heavy atoms, or repeating units, *i.e.* structural re-use (**Fig. 5b**).

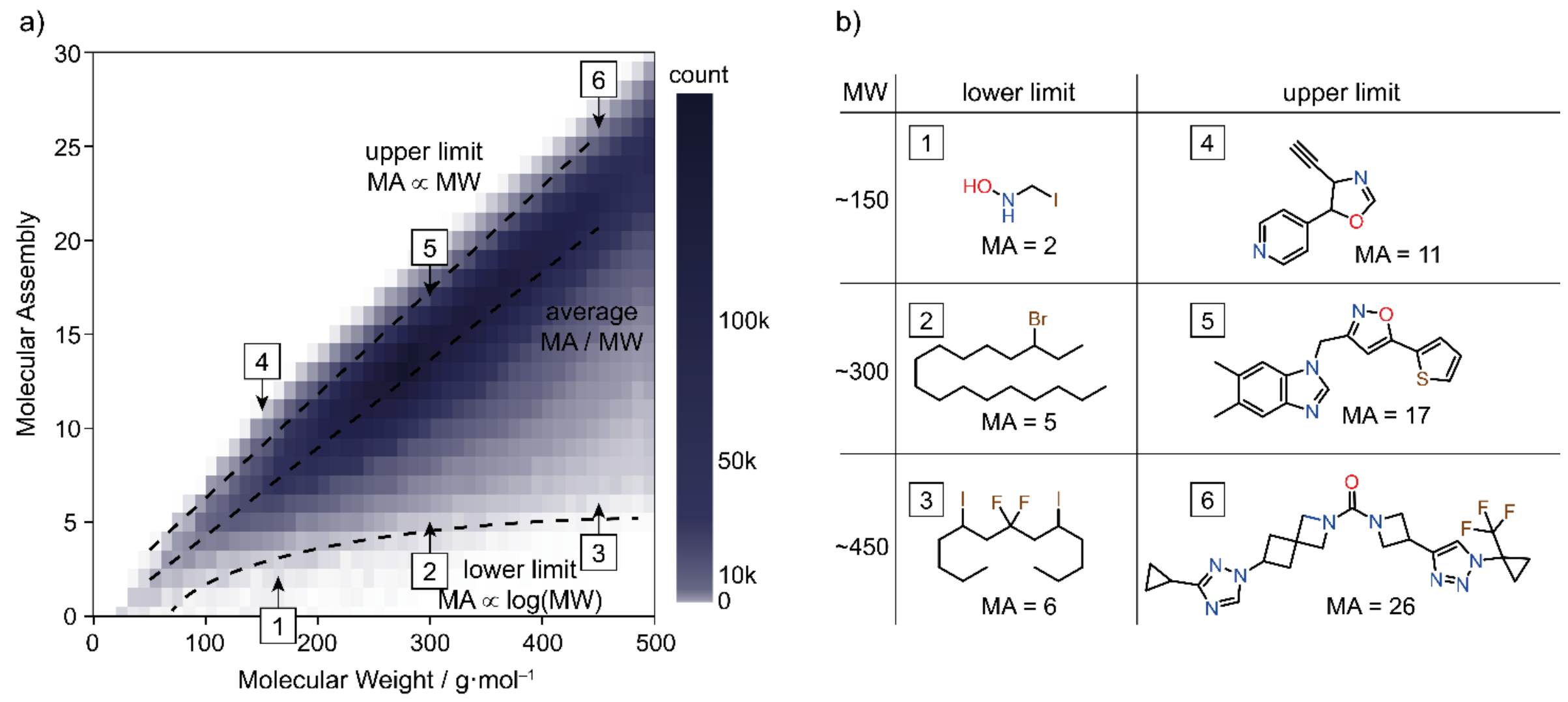

**Fig. 5** a) Distribution of MA against molecular mass, based on 16.5 M molecules sampled from PubChem database. The upper limit is linear and the lower limit is approximately logarithmic in nature. The theoretical MA values were calculated with a 10 s cut-off. b) Sample illustrating features of molecules on the MA/MW ranges. The characteristic features of lower MA molecules at a given molecular weight include the presence of periodic units, heavy elements or both. The high MA molecules usually comprise of higher heterogeneity with atoms of similar atomic weights

## Inferring Molecular Assembly from Tandem Mass Spectrometry

Multiple-level tandem mass spectrometry provides structured information about molecular fragments, which can be mapped to contingent pathways in the assembly space.(19) Considering this, we developed a new recursive algorithm which combines molecular fragments based on their masses to create a tree and compute the MA of the molecule. As an input, we provide a mass spectrum, with multiple fragmentation events ($MS^n$, where $n$ indicates the number of consecutive ion fragmentations). The MA of a given ion is calculated inferring all possible pathways to construct it from its daughter ions, applying the same process to each daughter ion. The chain of recursive search terminates whenever a daughter's mass matches the monoisotopic mass of an element (MA = 0) or when there are no daughters present for a given ion, in which case the molecular mass approximation is used (**Fig. 6a**). The range of possible MA of a fragment is predicted as a sample from normal distribution with location parameter $loc = 0.074 \times \mathrm{MW} - 1.4$ and scale $s = 0.0074 \times \mathrm{MW} + 0.511$. The scale parameter of the distribution was fitted on the large dataset from PubChem database (see previous section), and the *loc* parameter was parameterised to describe well small molecular fragments (See **SI Section 7.2** for further discussion). To test the recursive fragmentation algorithm, we experimentally assessed 101 molecules with various MA in the range from 4 to 24. This set included 30 molecules, selected to have nearly the same MW (300 ± 5 g/mol) and various MA in the range from 5 to 17. Similar MW molecules were chosen to demonstrate the capability of the algorithm to distinguish high and low MA molecules using $MS^n$ with similar characteristic MW values. The fragmentation events were carried out up to $MS^5$, where possible. The test sample of molecules with similar MW is a particularly difficult task for inferring MA by MS, as the MW prediction would fail to distinguish the difference in complexity among the sample. Examples of determining the molecule MA from this set, based on matching fragments or finding heavy elements are highlighted in **Fig. 6b**. The correlation coefficient between the predicted and expected values on the total sample was found 0.73 (**Fig. 6c**).

## Combining Analytical Techniques for Molecular Assembly Inference

Molecular constraints are probed by different physical interactions depending on the spectroscopic techniques, which independently have been shown to correlate with MA. In general, due to different limitations in the considered techniques (NMR, IR, and MS), individual spectral features of a molecule of unknown origin can be biased. For example, MS/MS fragmentation is biased by the strengths of the different chemical bonds, which molecular assembly calculation does not consider. Similarly, $^{13}C$ NMR spectroscopy might not fully reflect the MA should the constraints be realised through heteroatoms; further diastereotopic carbons can be overcounted although considered equivalent. IR fingerprint region can contain overtones of the functional groups, causing the peak overcount. All herein listed examples responsible for variance in the correlation with the MA have principally different physical interactions. We, therefore, hypothesised that a combination of the analytical techniques can increase confidence in the MA inference, see **Fig. 7**.

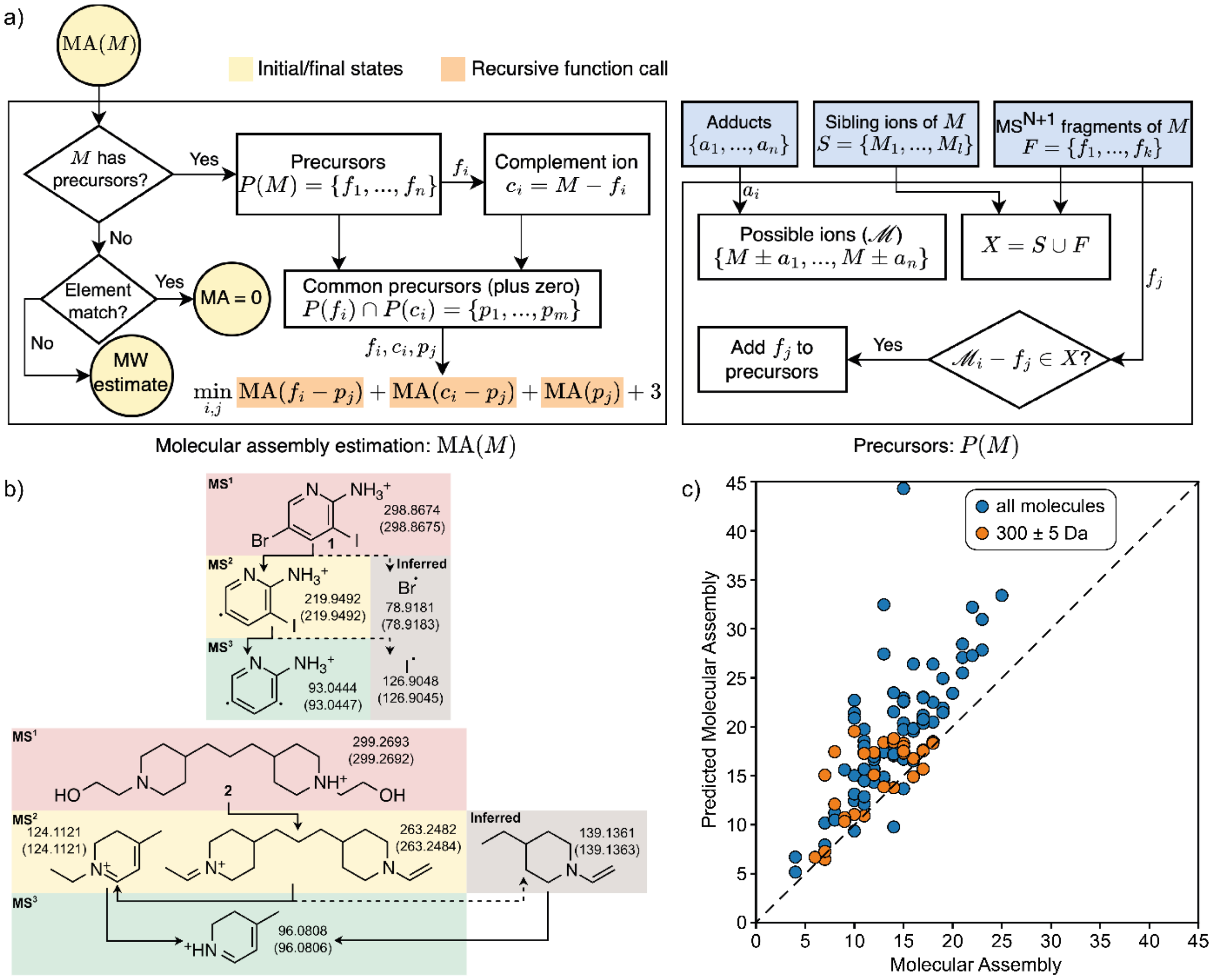

**Fig. 6** a) The recursive algorithm for estimating molecular assembly from tandem mass spectrometry data. b) Application of the recursive MA algorithm to resolve MA differences between molecules with similar molecular mass. Example of reduced MA based on the presence of heavy elements (bromine and iodine); Example of reduced MA based on the presence of repeated structural features. c) Molecular assembly *vs*. recursive-MS$^n$-inferred molecular assembly. Blue points represent the dataset of 71 molecules sampled across the MA values. The orange points are 30 molecules, specifically chosen to cover a large range of MA, but within a very narrow molecular weight (300 ± 5 g/mol). The correlation coefficient is 0.73.

On the set of 10,000 calculated NMR and IR spectra, we have examined our hypothesis that combined information can provide a more reliable MA prediction. We have used the same models (**Eq. 1** and **Eq. 3**) for the individual spectroscopic techniques and allowed them to optimise for their relative weighting. The combined model provided a higher correlation of 0.91 using the weighted average of 0.55×NMR and 0.45×IR inferred MA, see **Fig. 7A**. Further, we have validated this approach on the available intersection of the experimental NMR and IR data, comprising 54 molecules. The combined model provided a higher correlation of 0.89 using a combination of 0.7×NMR and 0.3×IR inferred MA, see **Fig. 7B**. Lastly, we have explored the combination of all three techniques to infer MA. Where the experimental dataset was available, we used a recursive algorithm applied to $MS^n$ fragmentation data to infer MA as accurately as possible. In cases where the data were not available, we approximated inference by the linear correlation to the exact mass of the molecule (**Fig. 7C**). Although the average value might not always provide a better estimate than certain individual components, it provides a more robust prediction, should no information about the sample be available. Such a difficult case is for symmetric polyaromatic heterocycle, where building blocks are reused, yet the strength of the multiple aromatic bonds does not provide (under the condition of collision-induced fragmentation used in this study) useful fragments to report on it. Spectroscopic techniques, on the contrary, will provide correct prediction as the symmetry will be reflected in simplified spectra. This observation highlights the utility of inferring the MA of unknown species using multiple experimental techniques and acquiring an average of their MA predictions.

An additional challenge for inferring the complexity of an unknown sample is to consider a mixture of compounds. To address this issue with spectroscopy, we demonstrated utilizing $^{13}C$ DOSY spectroscopy to experimentally deconvolute the mixture of chemical resonances to their individual components. We investigated $^{13}C$ DOSY spectroscopy on mixtures, separating individual compounds *via* their diffusion coefficient.(39) Together with the experimental assignment of the carbon types, we could predict the molecular assembly of each component in the mixture, using the same logic as for the individual compounds.

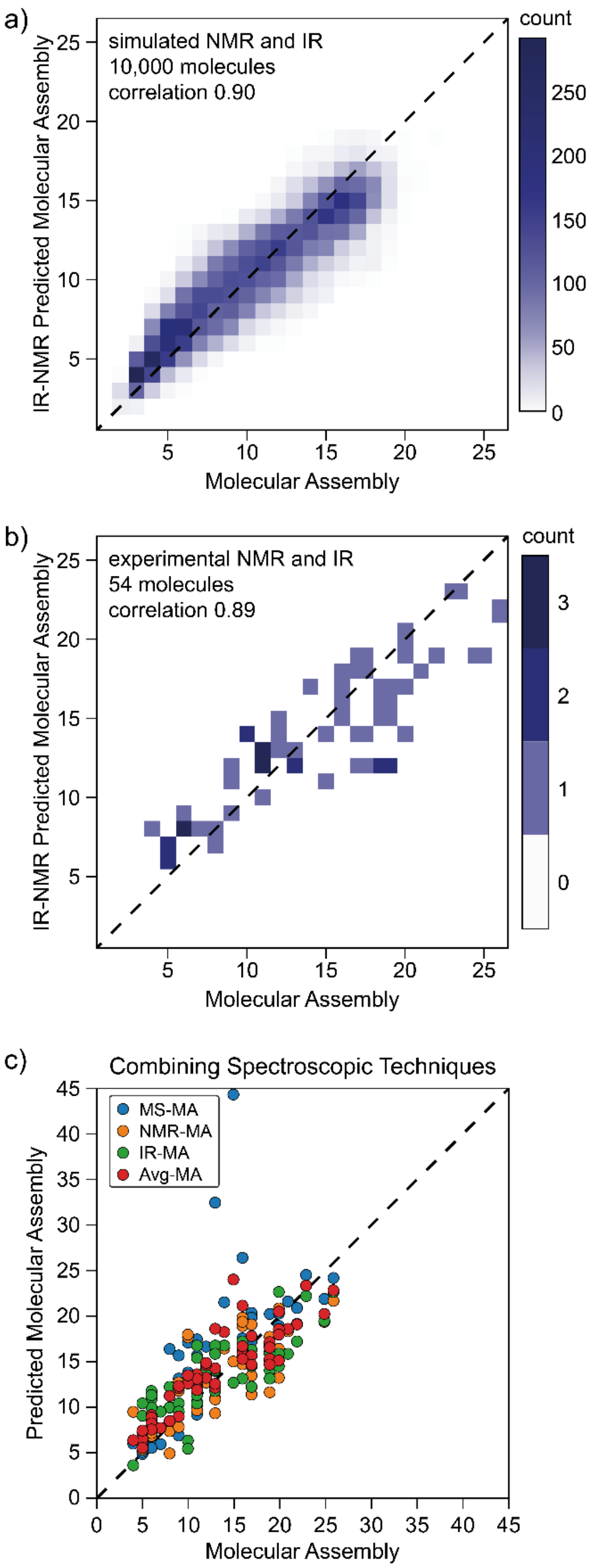

**Fig. 7 (A)** Molecular assembly vs. combined IR and NMR-inferred molecular assembly (using weights of 0.55 and 0.45 from NMR and IR, respectively) based on 10,000 calculated spectra showing an increased correlation of 0.90. **(B)** Molecular assembly vs. combined IR and NMR-inferred molecular assembly (using weights of 0.7 and 0.3 from NMR and IR, respectively) based on 54 experimental spectra showing an increased (relative to the individual components) correlation of 0.89. **(C)** Molecular assembly *vs*. individual and combined IR, NMR and MS-inferred molecular assembly based on the 54 molecules. Note that due to experimental limitations, multi-level MS fragmentation data was available for only 10 compounds, for the rest the MS part of the MA inference was performed based solely on the exact ion mass approximation. The correlation coefficient for the combined techniques is 0.88.

An example of such workflow is assessing a mixture of 5-aminoisophathalic acid and quinine, which yielded a prediction of the molecular assembly to be 8 and 19, in reasonably good agreement with the expected real value of 7 and 16, respectively. For more details, see **SI Section 6.1**. The work here shows that the general concept of measuring molecular complexity as a function of the number of different parts in a molecule, using spectroscopic measurements, gives a very strong correlation with the theoretical assembly complexity. This is important since it means we can use experimental measurements on environmental samples to read out the amount of selection and evolution that the samples have been subjected to making this approach suitable for the search for new life on and beyond Earth.

## Conclusion

We have demonstrated on a set of 10,000 simulated and approximately 100 experimental IR and NMR spectra that it is possible to predict the MA of compounds without structural elucidation. On a set of approximately 100 molecules, we have demonstrated a novel algorithm interpreting multi-level tandem mass spectrometry to infer MA. This experimental dataset included 30 molecules with nearly the same molecular mass, testing the algorithm to differentiate MA based solely on the $MS^n$. All the above mentioned methods are particularly useful for molecules of unknown origin and cases when a fast metric for probing complexity is required. In the case of IR, the constraints and molecular complexity are reflected by the number of peaks in the fingerprint region and their simple summation can be used to predict molecular complexity. In NMR, we have shown that the weighted sum of the number of carbon resonances, sorted by the number of hydrogens attached to them, provides a good prediction of MA. We found that the fewer hydrogens attached to the carbon, the higher the weight it possesses for the MA prediction. This finding corroborates our interpretation based on Assembly Theory that the quaternary carbons effectively encode the most information whereas the primary carbons, which have more hydrogen atoms, are least encoded and hence contribute less to the molecular assembly. Furthermore, we provided a new algorithm examining multi-level tandem mass spectrometry to infer MA as a function of matching fragments or identifying the presence of heavy elements. Finally, we have demonstrated the possibility of addressing the complexity of the components in mixtures using $^{13}$C DOSY, deconvoluting the $^{13}$C NMR signals to their individual compounds. These findings are of particular significance for the development of missions looking for life in our solar system.(40) NASA has already managed to put several mass spectrometers on Mars,(41) and several mass specs have been in the solar system including on the Cassini probe which visited Saturn and Enceladus.(42) Dragonfly is set to visit Titan, launching in 2026 and arriving in 2034, which is important since it will be a mobile mass spectrometer that flies around Titan.(43) Of critical importance will be the ability to resolve high molecular weight compounds (300-600 Da) with the possibility of generating in situ fragmentation. In this instance, the use of Assembly Theory when analysing the data will allow us to put some limits on the complexity of the molecules found on Titan. Further afield, as exo-planet spectroscopy becomes more advanced, it will be possible to look for infra-red signatures associated with exoplanets. Whilst these applications seem far away, it is only now with assembly theory making firm and experimental predictions about the complexity of molecules that can result from an evolutionary or informational process, that we can seriously contemplate truly agnostic biosignature searches now we have validated the measurement of molecular complexity across three different experimental domains.

**Experimental Section**

**Infrared Experimental Setup:** IR spectra were acquired on a Thermo Scientific Nicolet iS5 with Specac Golden Gate Reflection Diamond ATR System. All data were processed with Thermo Scientific OMNIC 8.3.103. All samples were measured in the native state at room temperature (solid state unless liquid at room temperature).

**NMR Experiment Setup:** NMR data were acquired on a Bruker Ascend Aeon 600 MHz NMR spectrometer with a DCH cryoprobe ($^{13}$C + $^{1}$H channels) at 300 K unless otherwise stated in which case a Room temperature BBFO probe head ($^{1}$H + $^{19}$F-$^{183}$W channels) was used. $^{1}$H NMR spectra were acquired using 16 scans, spectral width 20 ppm and relaxation delay 2 s. Spectra on the $^{13}$C channel were acquired with a spectral width of 200 ppm. The $^{13}C$ NMR spectra were acquired using 16 scans and a relaxation delay of 0.8 s. The DEPTQ routines were carried out using 16 scans and a relaxation delay of 1 second. The $^{13}C$ DOSY spectra were acquired using 256 scans, a relaxation delay of 8 seconds and a 500-1000 μs gradient pulse. All spectra were processed using Bruker Topspin 3.6 and Mestrenova 14.1.1. The spectra were phase and baseline corrected and calibrated relative to the residual solvent peak. Residual solvent peaks were not included in resonance counts. Unless otherwise stated samples were prepared in DMSO-$d_6$ at the concentration stated in the ESI.

**MS Experiment Setup:** Tandem mass spectrometry experiments were carried out up to the $MS^5$ level on a Thermo Scientific Orbitrap Fusion Lumos Tribrid system *via* direct injection of samples dissolved in acetonitrile (details in SI). Following the acquisition, raw vendor outputs were converted to mzML using ProteoWizard (no filters applied) for consumption by the Recursive MA algorithm.

**Supplementary Data**

This describes the algorithm for calculating molecular assembly for molecules (molecular assembly – MA), the details of the theoretical calculations for NMR, Infra-Red (xTB and DFT simulations), sample preparations for the experimental data collection, experimental IR and NMR data, the regression analysis, mixture analysis. The Molecular Assembly calculator called *AssemblyGo* was written in GO programming language (https://github.com/croningp/assembly_go). The codes used for processing data and further details can be found on Github https://github.com/croningp/molecular_complexity and https://github.com/croningp/RecursiveMA.

**Acknowledgements**

The authors gratefully acknowledge financial support from the John Templeton Foundation (Grant 60025), EPSRC (Grant Nos EP/L023652/1, EP/R01308X/1, EP/J015156/1, EP/P00153X/1), the Breakthrough Prize Foundation and NASA (Agnostic Biosignatures award #80NSSC18K1140), MINECO (project CTQ2017-87392-P) and ERC (project 670467 SMART-POM). We would like to thank Sara. I. Walker and Estelle M. Janin (ASU) for comments on the manuscript.

## Author Contributions

MJ generated calculated NMR, IR spectra, analysed theoretical MA data and interpreted all data; AS generated calculated NMR, IR spectra, analysed theoretical MA data and interpreted all data; SHMM devised and implemented the recursive MA algorithm and validated; JB collected the NMR and IR experimental data and did preliminary fitting with NB; CM helped with the assembly algorithm development and wrote software interpreting the mass spectrometry data; AM acquired the mass spectrometry data; GJTC provided some samples for the blinded tests; SMM developed the assembly go program; MS and RM generated IR data from DFT analysis. LC developed assembly theory, conceived the idea, raised the funding, and supervised the research. LC wrote the paper with input from all the authors.

Supplementary Information for:

**Determining Molecular Complexity using Assembly Theory and Spectroscopy**

Michael Jirasek†[1], Abhishek Sharma†[1], Jessica R. Bame†[1], S. Hessam M. Mehr[1], Nicola Bell[1], Stuart M. Marshall,[1] Cole Mathis[1], Alasdair Macleod,[1] Geoffrey J. T. Cooper[1], Marcel Swart[2,3], Rosa Mollfulleda[2], Leroy Cronin*[1]

**Lee.Cronin@glasgow.ac.uk*

*[1] School of Chemistry, The University of Glasgow, University Avenue, Glasgow G12 8QQ, UK.*

*[2] University of Girona, Campus Montilivi (Ciencies), c/M.A. Capmany 69, 17003 Girona Spain*

*[3]ICREA, Pg. Lluis Companys 23, 08010 Barcelona, Spain*

# Contents

# 1 Calculating Assembly Index From Molecular Graph

## 1.1 Algorithm Description

The assembly index, and associated minimal assembly pathways, are calculated using an algorithm written in the Go programming language. In prior work,(1) the assembly index was calculated using a serial algorithm written in C++, and yielded the "split-branch" assembly index, an approximation that provides a reasonably tight upper bound for the assembly index. The Go algorithm used in this work is a faster algorithm that incorporates concurrency, and can provide the exact assembly index if it can be calculated in a reasonable time. The process can also be terminated early to provide the lowest assembly index found so far, which has been found to be a good approximation for the assembly index in most cases.

The assembly index is calculated by iterating over subgraphs within a molecular graph, and finding duplicates of that subgraph within the remainder of the molecule. For each of the matching subgraphs found an assembly pathway can be represented by a duplicate structure and a remnant structure. The remnant structure comprises the original structure with one duplicate removed, and the other "broken off", which ensures that all structures on an assembly pathway that are duplicated will be first constructed (**Fig. S1**).

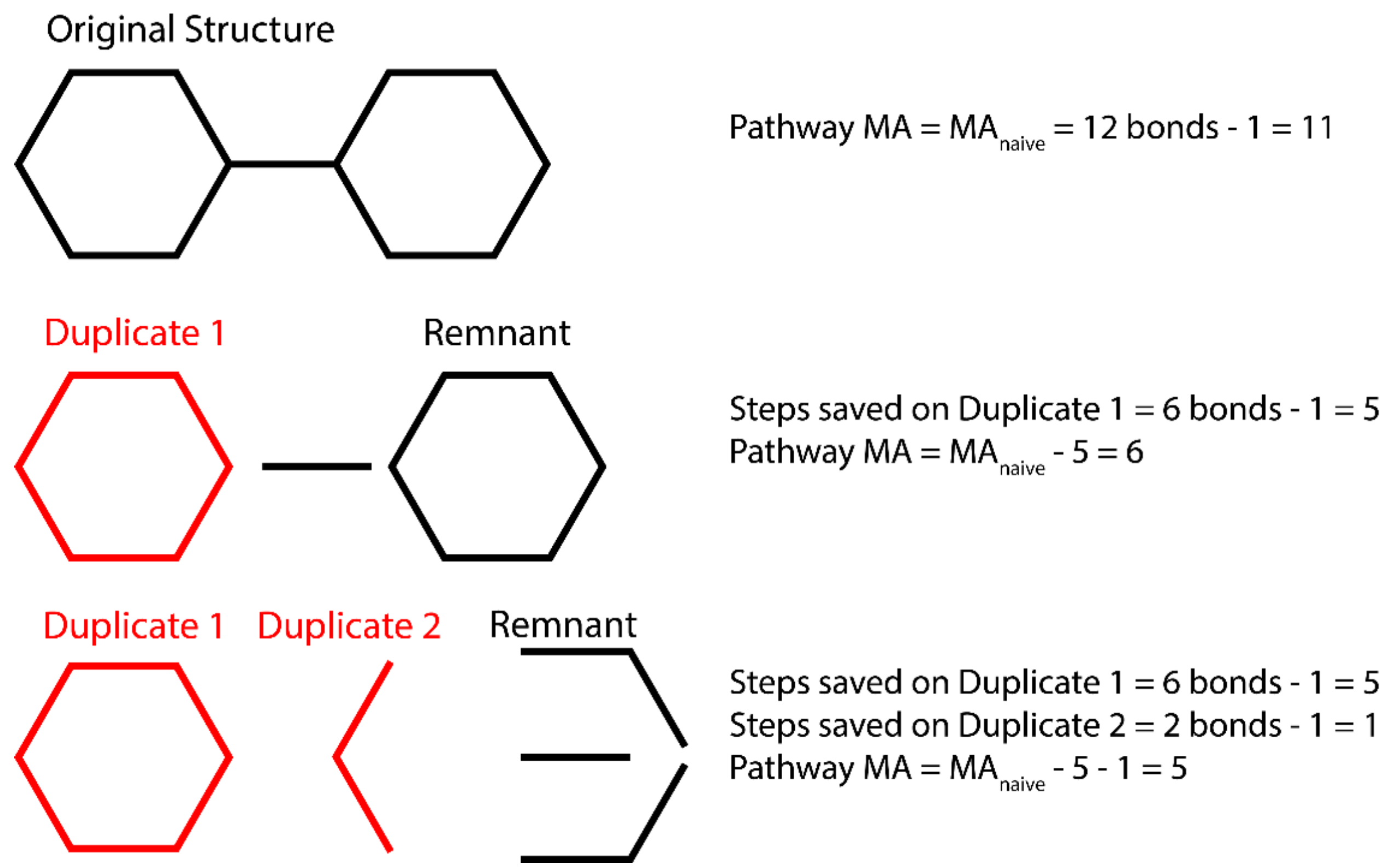

**Fig. S1**. Illustration of the process used by the assembly algorithm to extend an assembly pathway. At each step (top to bottom) a duplicate is found (red) and stored separately. The remnant is the remaining part of the structure, but with the matching duplicates separated. The process can then be repeated recursively on the remnant.

The process can then be repeated recursively with the remnant structure as an input, which may result in more pathways containing two duplicate structures and a smaller remnant. Thus each pathway is represented by a sequence of duplicated structures and a remnant structure. In order to determine the assembly index, we consider that a molecular graph with $N$ bonds could be constructed in $N-1$ steps by adding one bond at a time (the naive $MA$, or $MA_{naive}$). Each duplicate structure of size $N_{dup}$ allows us to add that structure in one step, reducing the number of steps compared to $MA_{naive}$ by $N_{dup}-1$. Thus the MA for a particular pathway is $MA_{naive} - \sum_{dup}(N_dup - 1)$ .

Concurrency is implemented through a worker pool, with each worker iterating over the subgraphs of a particular pathway and placing generated extended pathways into a jobs queue to be picked up and extended by other workers. In order to prevent unbounded resource use, the jobs queue size is limited, and if full a worker will process generated pathways in a depth-first fashion until there is space in the queue, before resuming the breadth-first search. The algorithm has some branch and bound methods to reduce the search space (it will not extend pathways that cannot have lower MA than the lowest found so far), and can be terminated early to output the best pathway found so far. The approximation through stopping early has been found to output values at or close to the actual assembly index fairly quickly (**Fig. S2**).

The subgraph iteration process is based on,(2) and the subgraph matching functions are based on processes used in Nauty.(3) The overall algorithm concept is similar to the exact MA algorithm we published previously,(4) but with substantial improvements in terms of performance.

Molecular assembly can be expected to correlate with molecular weight. This is because there are upper and lower bounds for molecular assembly indices that scale with the number of bonds in the molecule. The trivial upper bound for the assembly index relates to pathways where one bond is joined at a time without any leverage of duplication (so the assembly index is equal to the number of bonds minus one). A basic lower bound can be determined by considering that the quickest way to increase the size of a structure using an assembly pathway is to take the largest structure created so far and combine it with itself, essentially doubling the size at each step. For example, a structure of 8 bonds cannot be made in less than 3 joining operations, and in general, a structure of $N$ bonds has a lower bound on the assembly index of $\log_2(N)$. Both these bounds increase with the number of bonds, and since the number of atoms and hence the molecular weight tends to increase with the number of bonds, we can expect the assembly index to increase with the molecular weight.

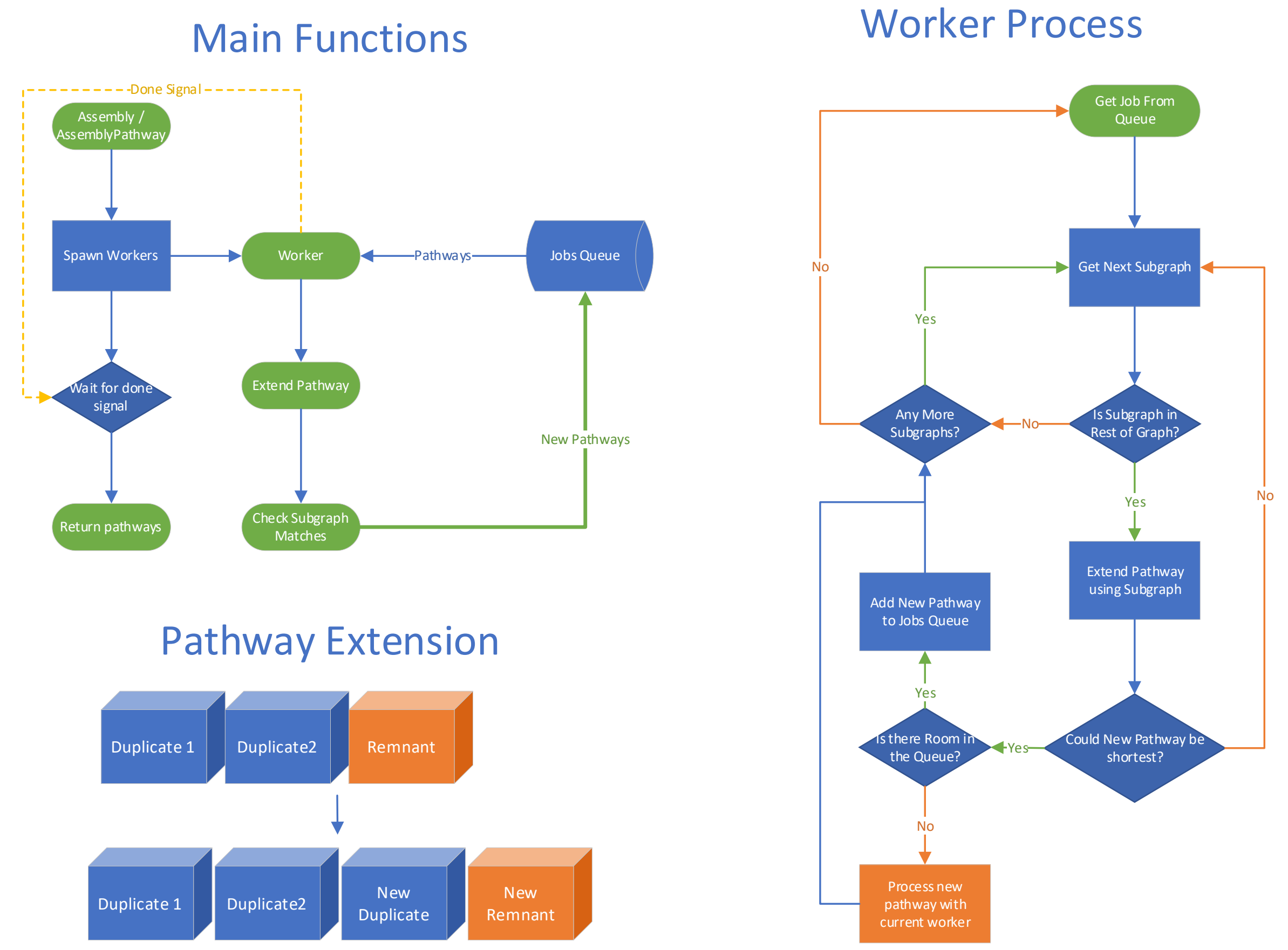

**Fig. S2**: Outline of algorithm process (top left). Illustration of extension of a single pathway (bottom left). Outline of process for single worker (right).

# 2 Theoretical calculations for IR and NMR

## 2.1 Database and sampling

The calculation and theoretical basis for Molecular Assembly (MA) calculation are described in detail in our previous work.(1, 5) Most of the analysis was performed using Python 3 and Mathematica 12. The Molecular Assembly calculator called *AssemblyGo* was written in GO programming language (https://github.com/croningp/assembly_go). The codes used for processing data and further details can be found at https://github.com/croningp/molecular_complexity.

To study the relationship between the MA and physically measurable properties, we used a previously published database of compounds for which the MA was calculated (~2.2M compounds). In order to address the molecular complexity of organic molecules, and given that we try to address the molecular complexity through carbon-sensitive $^{13}C$ NMR, hence, a relatively high abundance of carbon is essential. We filtered the compounds to have at least 50% and not more than 85% of the heavy atoms as carbons and must contain at least 4 carbon atoms. Such a filtered database contained *ca.* 0.77 million compounds. The range of the previously calculated Molecular Assembly (called Pathway

Assembly, using the split-branch algorithm) was found in between 3–25. The distribution of MA in the database is not uniform with the highest counts between 8-12 see **Fig. S3**.

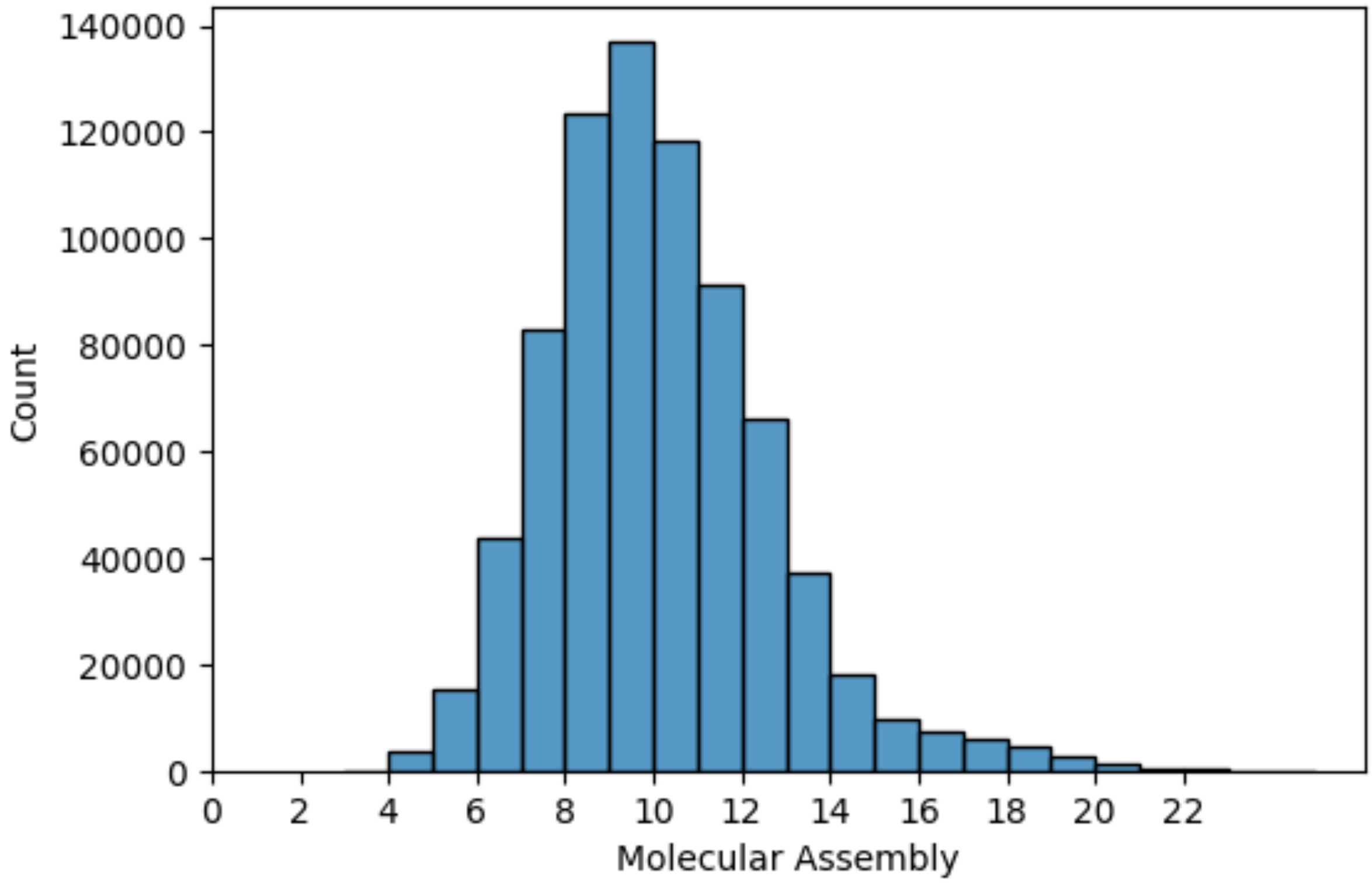

**Fig. S3.** Distribution of molecules over the Molecular Assembly from the previous dataset. The Fig shows the histogram of MA (previously called Pathway Assembly) distribution in the available dataset, containing ~0.77 million compounds with 50–85% carbons as heavy atoms, originating from the previously published dataset.(1)

The MA distribution of the compounds filtered from the previously published(1) database reflects synthetic availability (as the compounds originate from the published Reaxys database) and our previous capacity to reliably calculate MA using the split-branch algorithm (compounds for which it was assumed it would be impossible (at the time) to calculate MA reliably were rejected from the database). To assess the characteristic relationship between the MA with the spectroscopic techniques, we sampled 10,000 compounds uniformly across the MA range, up to 629 compounds per MA. The sampling was performed on the subset of which both theoretical NMR and IR data could be calculated using the simulation tools discussed in the later sections. For those compounds, the MA was recalculated using the newly developed assembly algorithm AssemblyGo which provides more accurate estimates at faster timescales, generally leading to estimating the assembly index to be lower by 1 or 2 relative to the original value (previously calculated Pathway Assembly). The distribution of the molecules over the MA range with newly calculated MA is shown in **Fig. S4**.

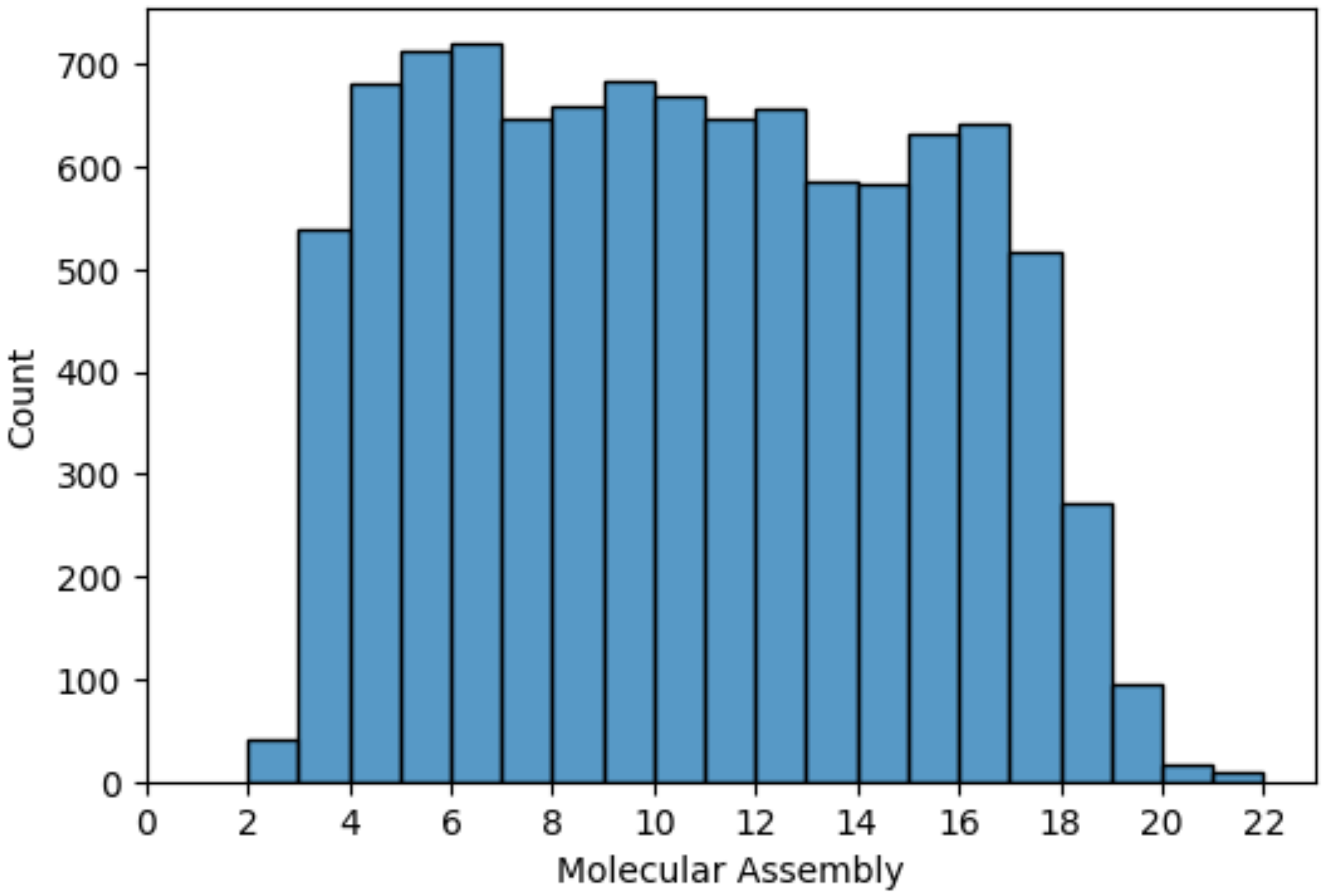

**Fig. S4.** Histogram of distribution in the sample of 10,000 molecules to cover uniformly the range of MA (recalculated values using a more accurate algorithm).

A representative subset of the molecules in the dataset over the range of MA values is shown in **Fig. S5** on the following page.

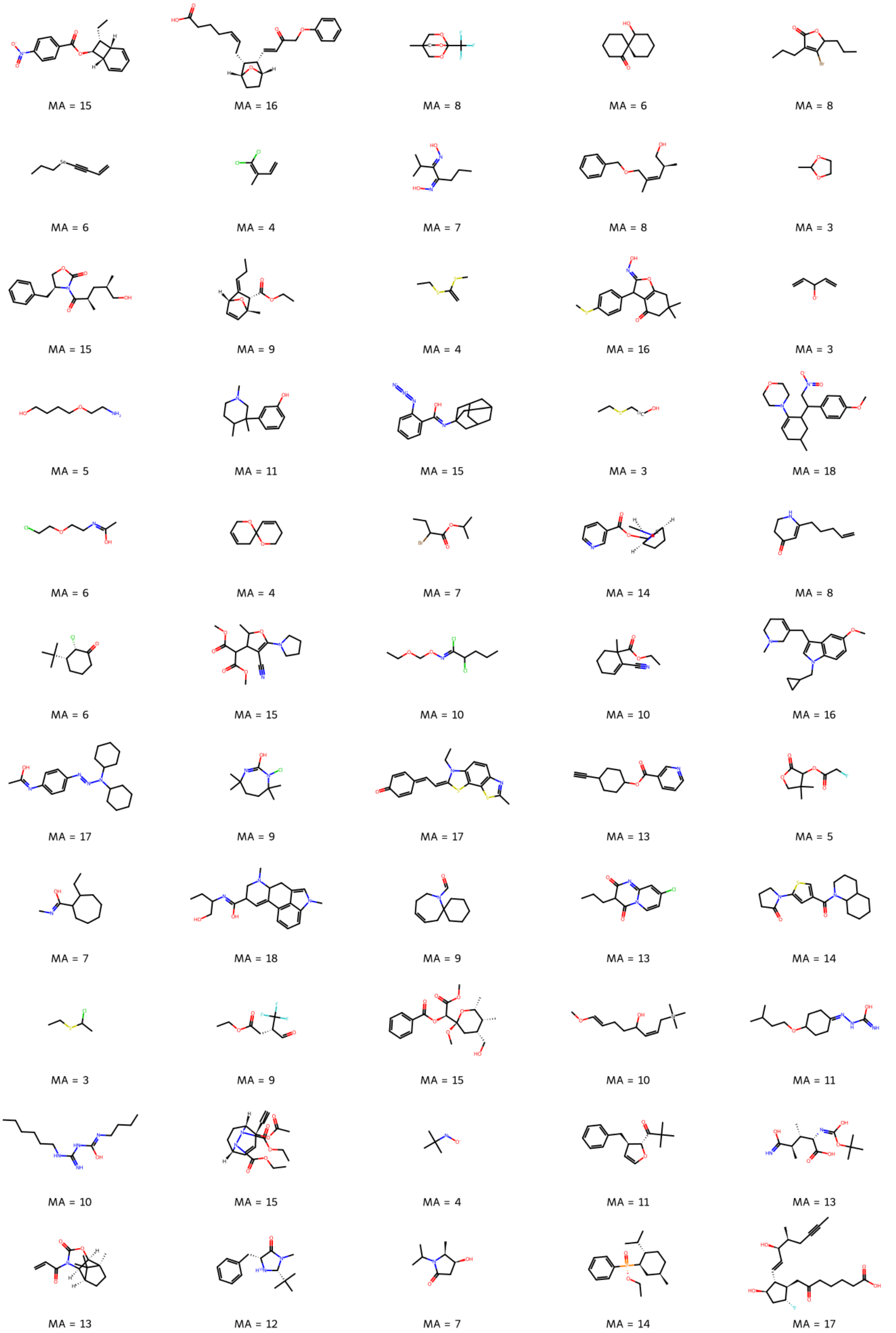

**Fig. S5**. Example of 55 compounds sampled from the database of 10,000 compounds used in the theoretical study.

## 2.2 NMR Prediction

The $^{13}C$ NMR spectra were predicted using *nmrshiftdb2* tool.(6) The corresponding chemical shifts were grouped by the type of carbon (primary ($CH_3$), secondary ($CH_2$), tertiary (CH) and quarternary (C)). Such sorting was performed by Python script using the rdkit(7) tool to estimate the number of hydrogen atoms attached. Further, the number of chemical shifts was binned, applying the minimum 0.5 ppm chemical shift difference (i.e. resonances within the 0.5 ppm were considered as a single peak for analysis).

The importance of an actual $^{13}C$ NMR measure/prediction instead of the sole counting of the number of chemically non-equivalent carbons in structure can be demonstrated on a large dataset of ~1.1 million compounds (allowing all compounds with more than 4 carbons and no constraints on the C content). This set was analysed by both NMR prediction, as well as by counting the number of nonequivalent carbons (for simplicity, the carbons were not classified by the type) (**Fig. S6**). Also, note that considered assembly index values are based on the old database that used the previous algorithm which is relatively less accurate. The potential outliers deviating from the linear trends highlight the utility of the actual NMR measure (as an oriented oligomer possesses plenty of non-equivalent carbons, yet of very similar chemical shift).

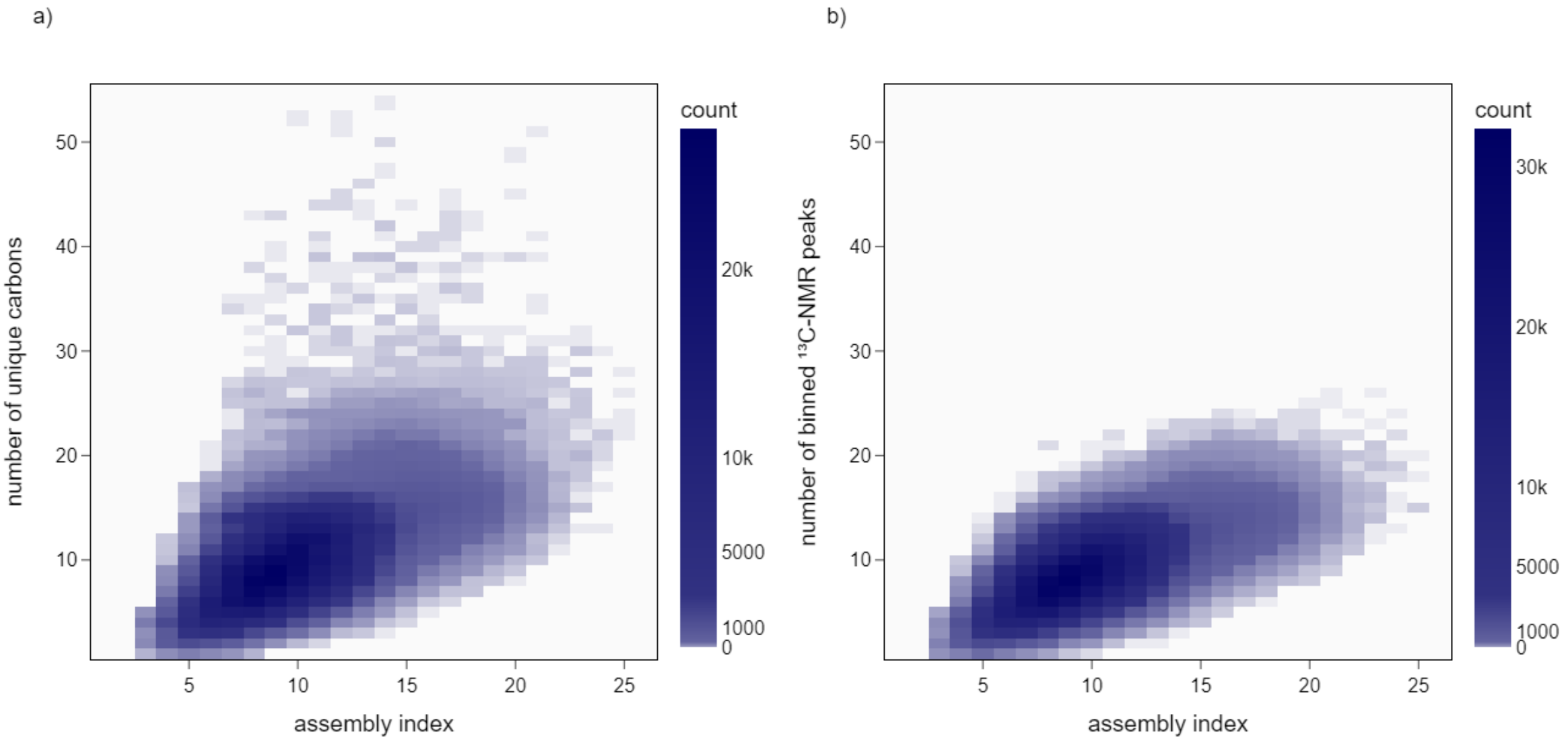

**Fig. S6**. **Analysis of *ca.* 1.1 million compounds.** a) Molecular Assembly (assembly index) *vs.* the number of unique carbons. b) MA *vs.* the number of predicted 0.5 ppm binned $^{13}C$ NMR resonances. Note that the z-axis (histogram count) is scaled logarithmically with base 10 to emphasise even the very uncommon cases.

On the sample of 10,000 molecules, whose MA was recalculated using the new algorithm, we used multivariate fit of the number of 0.5 ppm binned resonances of C, CH, $CH_2$ and $CH_3$ carbon resonances. For the fit, the *statsmodels.api.OLS* module in Python was used (**Fig. S7**).(8)

```
                    Results: Ordinary least squares
====================================================================
Model:              OLS              Adj. R-squared:     0.753
Dependent Variable: y                AIC:                44452.0003
Date:               2023-02-07 15:52 BIC:                44488.0520
No. Observations:   10000            Log-Likelihood:     -22221.
Df Model:           4                F-statistic:        7607.
Df Residuals:       9995             Prob (F-statistic): 0.00
R-squared:          0.753            Scale:              4.9869
-----------------------------------------------------------------------
           Coef.    Std.Err.      t       P>|t|     [0.025    0.975]
-----------------------------------------------------------------------
x1         1.3172     0.0135   97.2754    0.0000    1.2907    1.3438
x2         0.7996     0.0113   70.5589    0.0000    0.7774    0.8218
x3         0.6451     0.0118   54.6230    0.0000    0.6220    0.6683
x4         0.2620     0.0254   10.3091    0.0000    0.2122    0.3119
const      2.1549     0.0612   35.2323    0.0000    2.0350    2.2748
--------------------------------------------------------------------
Omnibus:              269.066       Durbin-Watson:          1.546
Prob(Omnibus):        0.000         Jarque-Bera (JB):       292.973
Skew:                 0.399         Prob(JB):               0.000
Kurtosis:             3.260         Condition No.:          16
====================================================================
```

**Fig. S7**. Print output from the multivariate fit of MA = $x_1$×C + $x_2$×CH + $x_3$×$CH_2$ + $x_4$×$CH_3$ + *const.*; using *statsmodels.api.OLS* in python.(8)

The best prediction of MA based on the NMR data is given by:

$$\mathrm{MA} = 1.32 \times \mathrm{C} + 0.80 \times \mathrm{CH} + 0.65 \times \mathrm{CH_2} + 0.26 \times \mathrm{CH_3} + 2.15 \quad \textbf{(1)}$$

where C, CH, $CH_2$ and $CH_3$ are the number of calculated unique (binned with 0.5 ppm resolution) $^{13}$C resonances corresponding to carbons with 0, 1, 2 and 3 attached hydrogens, respectively. The distribution of MA *vs*. NMR-predicted MA is visualised as a histogram is shown in **Fig. S8**.

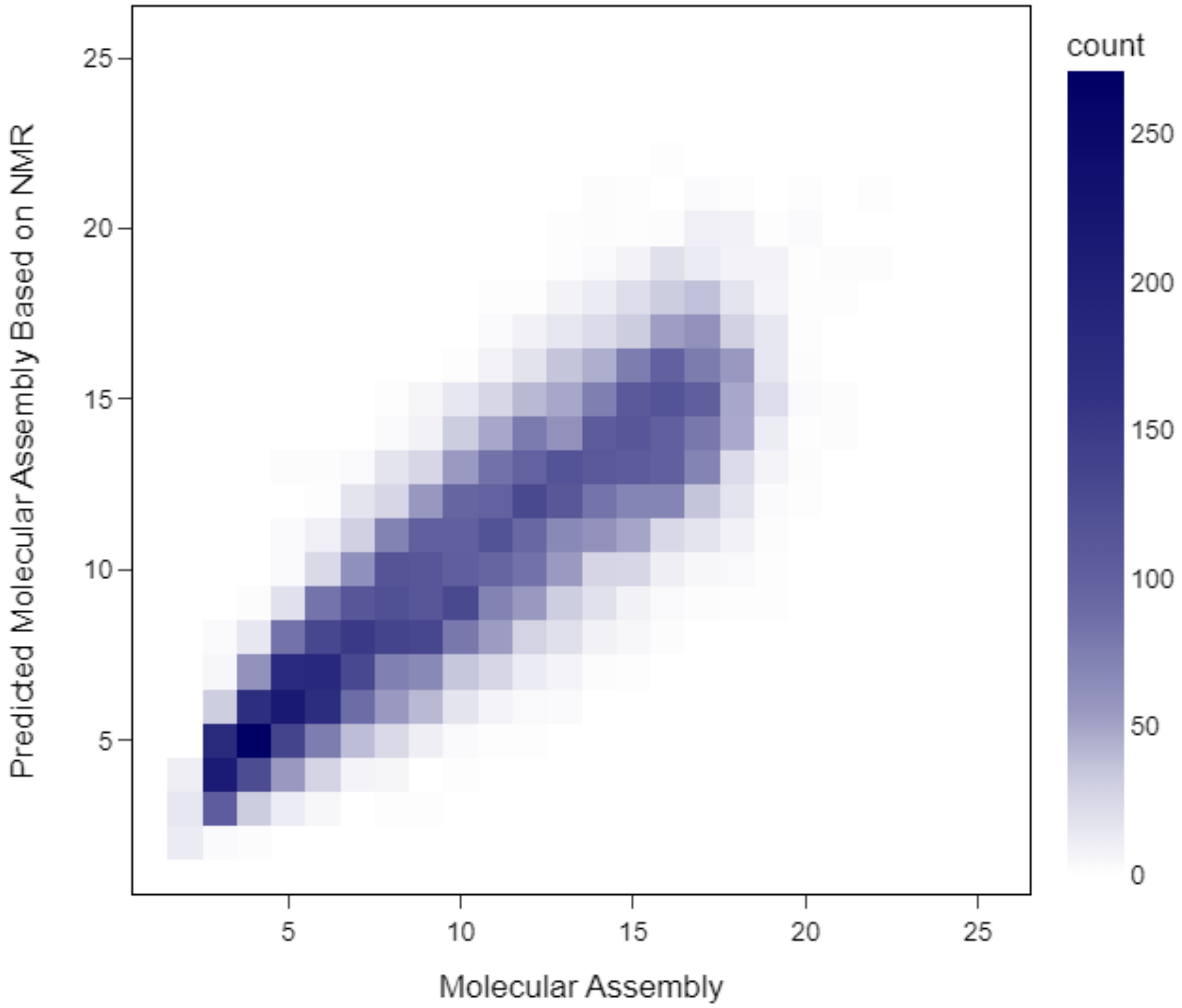

**Fig. S8**. Histogram of predicted MA (based on the **Eq. 1**) *vs.* MA on 10,000 compounds sample.

## 2.3 Infrared Spectroscopy – xTB simulations

To predict the IR spectra of the sampled molecules, we have used the xTB-service tool(9, 10) for faster prediction over a large dataset. The provided Python interface(11) was used with the default setting, using 100 seconds as timeout for the geometry optimisation using the GFNFF forcefield. The default gaussian broadening was not applied to the observed intensity. Using the default threshold checks, calculated spectra assumed for the interpretation were based on the molecule with no large imaginary frequency (set as maximum $i \cdot 10$ cm$^{-1}$, although many structures possess small imaginary frequencies). The peaks in the range of 400–1500 cm$^{-1}$ were counted with a threshold of 0.0005 (D/Å)$^2$ · amu$^{-1}$ to not consider signals with 0 oscillatory strength and binned together peaks within 2 cm$^{-1}$. The coefficients for the simple linear function of the number of IR peaks were fit using the *statsmodels.api.OLS* module in python (**Fig. S9**).(8)

```
                         Results: Ordinary least squares
====================================================================
Model:               OLS                Adj. R-squared:     0.739
Dependent Variable:  y                  AIC:                45000.1544
Date:                2023-02-07 15:54 BIC:                45014.5751
No. Observations:    10000              Log-Likelihood:     -22498.
Df Model:            1                  F-statistic:        2.826e+04
Df Residuals:        9998               Prob (F-statistic): 0.00
R-squared:           0.739              Scale:              5.2695
----------------------------------------------------------------------
             Coef.    Std.Err.       t        P>|t|     [0.025    0.975]
----------------------------------------------------------------------
x1           0.2076     0.0012   168.0982   0.0000    0.2052    0.2100
const       -0.1454     0.0654    -2.2246   0.0261   -0.2735   -0.0173
--------------------------------------------------------------------
Omnibus:              38.032        Durbin-Watson:           1.439
Prob(Omnibus):        0.000         Jarque-Bera (JB):        47.476
Skew:                 0.066         Prob(JB):                0.000
Kurtosis:             3.311         Condition No.:           151
====================================================================
```

**Fig. S9**. Print output from the fit of MA = $x_1 \times n_{\text{peaks}}$ + *const.*; using *statsmodels.api.OLS* in python.(8)

The best prediction of MA based on the xTB-based IR predicted data is thus:

$$\text{MA} = 0.21 \times n_{\text{IR_peaks}} - 0.15 \qquad \textbf{(2)}$$

where $n_{\text{IR_peaks}}$ is the number of IR peaks in the region of 400–1500 cm$^{-1}$ with intensity above 0.0005 (D/Å)$^2$ · amu$^{-1}$. The distribution of MA *vs*. IR-predicted MA is visualised as a histogram in **Fig. S10**.

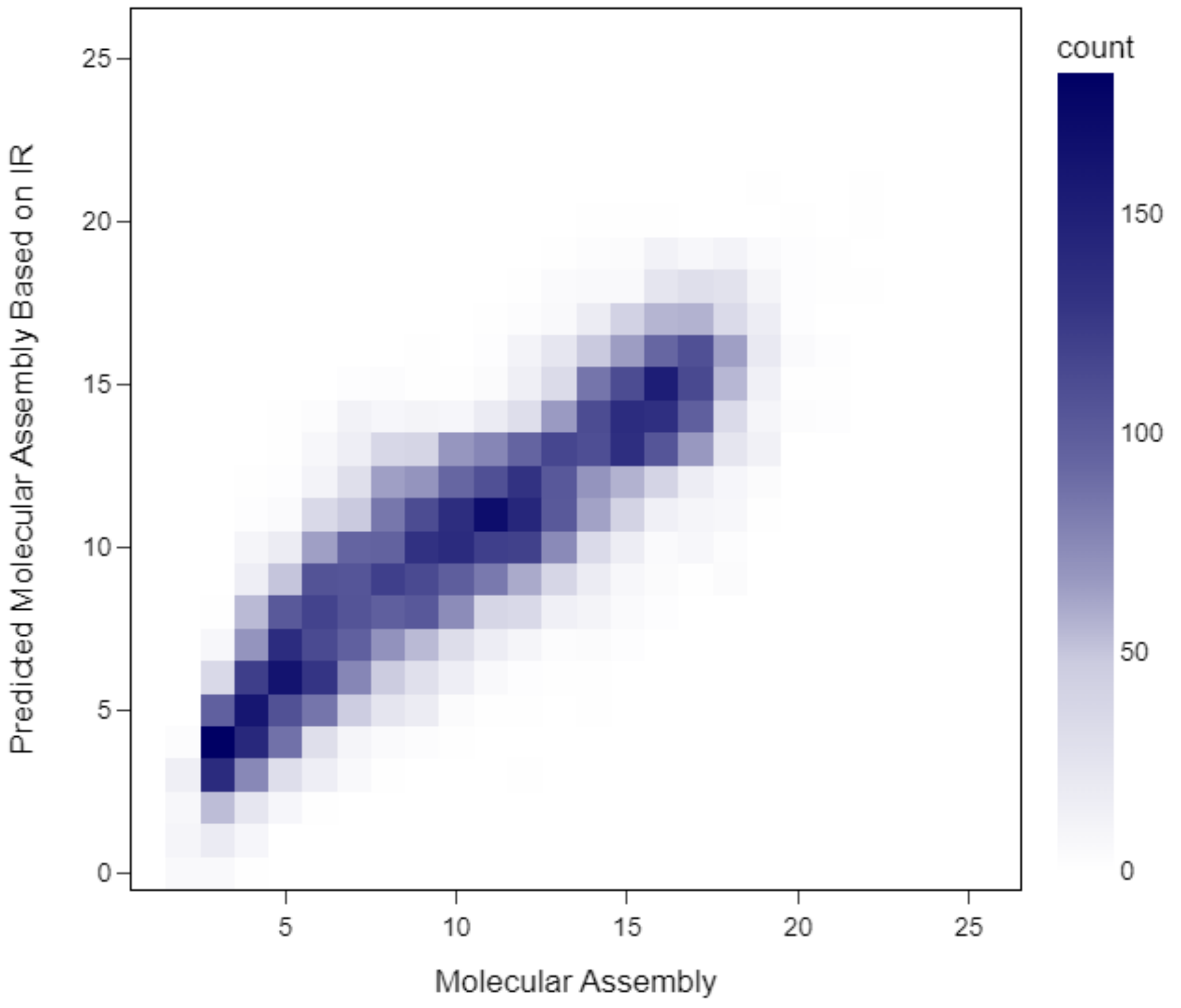

**Fig. S10**. Histogram of predicted MA (based on **Eq. 2**) *vs.* MA on 10,000 compounds dataset.

Our general hypothesis is that modes in the IR fingerprint region could be associated largely with collective motions, involving bonds from the various subgraphs of the whole structure. Therefore, from the number of the total modes in the fingerprint region, the overall molecular complexity could be inferred. To illustrate that on a simple and a complex molecule, vibrational modes in the fingerprint region (400–1500 $cm^{-1}$) above the set intensity threshold of 0.0005 $(D/Å)^2 \cdot amu^{-1}$ for chemical structures of 5-aminoisopthalic acid (**Fig. S11**.) and quinine (

**Fig. S12–Fig. S15**) are visualised. On the molecular structure, bonds involved in the vibrational modes are highlighted.

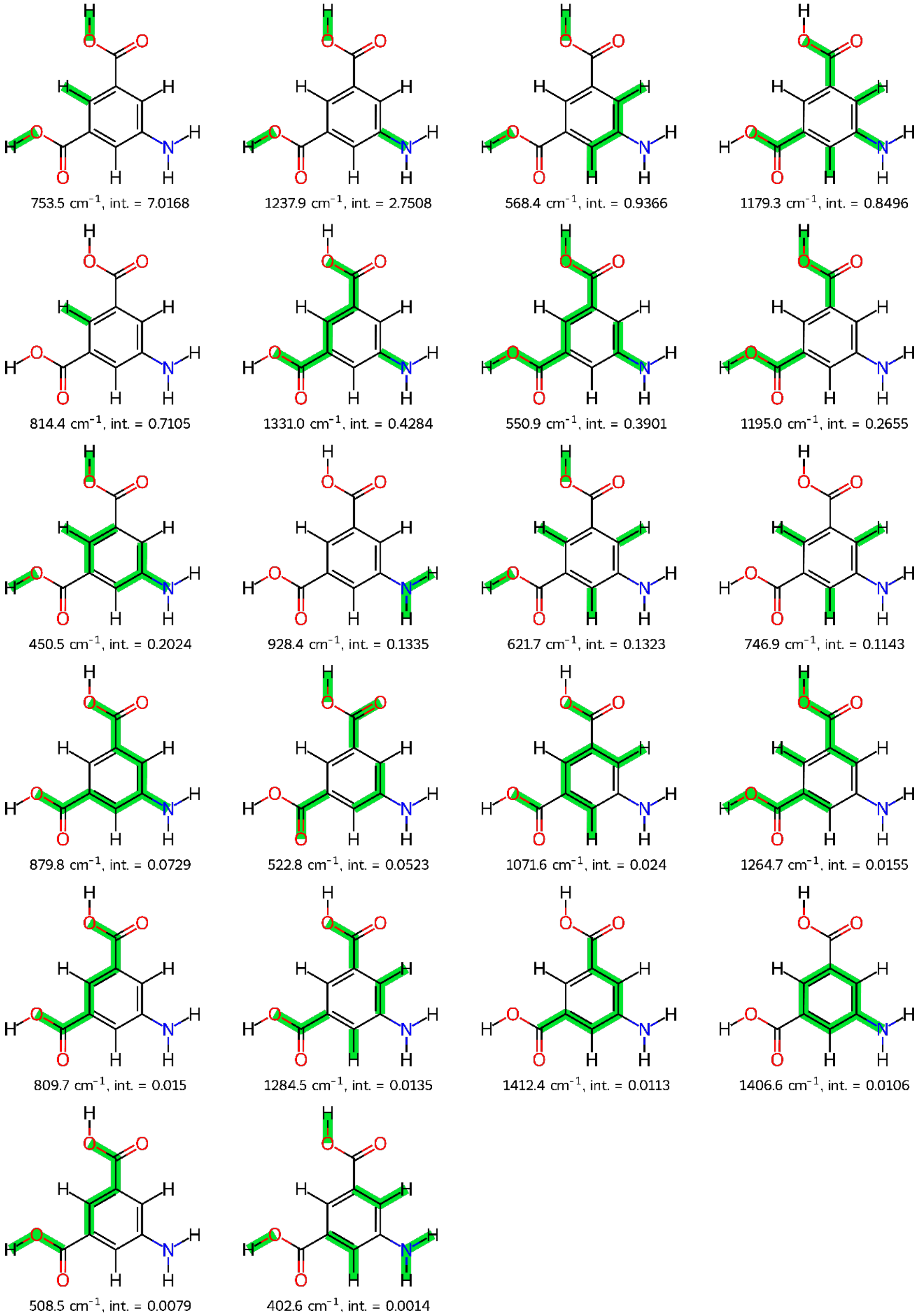

**Fig. S11**. Example of all vibrational bands of 5-aminoisopthalic acid in the fingerprint region demonstrating its collective-motion nature. Vibrational modes are ordered by intensity as calculated by xTB software.

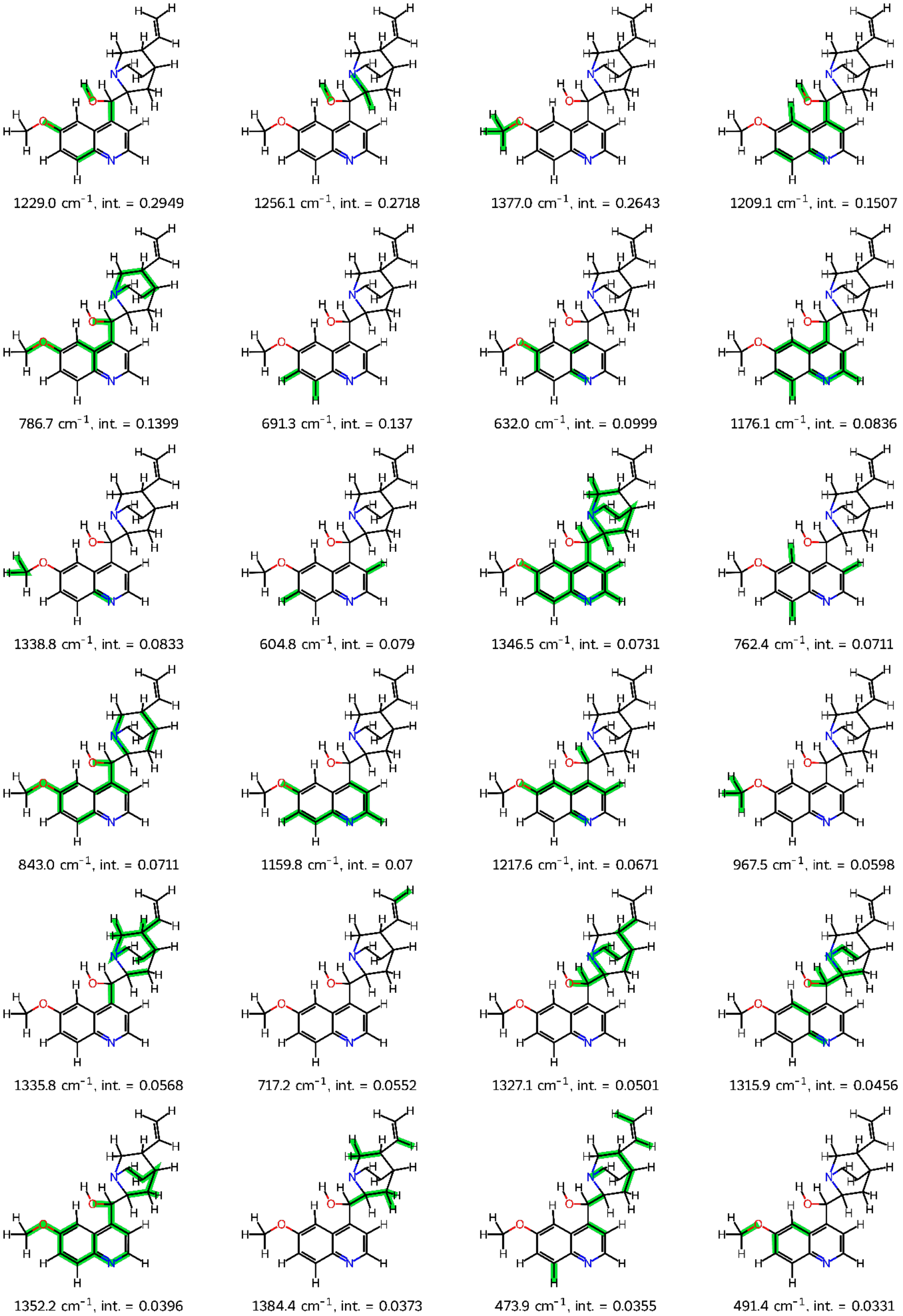

**Fig. S12**. Example of all vibrational bands of quinine in the fingerprint region demonstrating its collective-motion nature. Vibrational modes are ordered by intensity as calculated by xTB. (part 1)

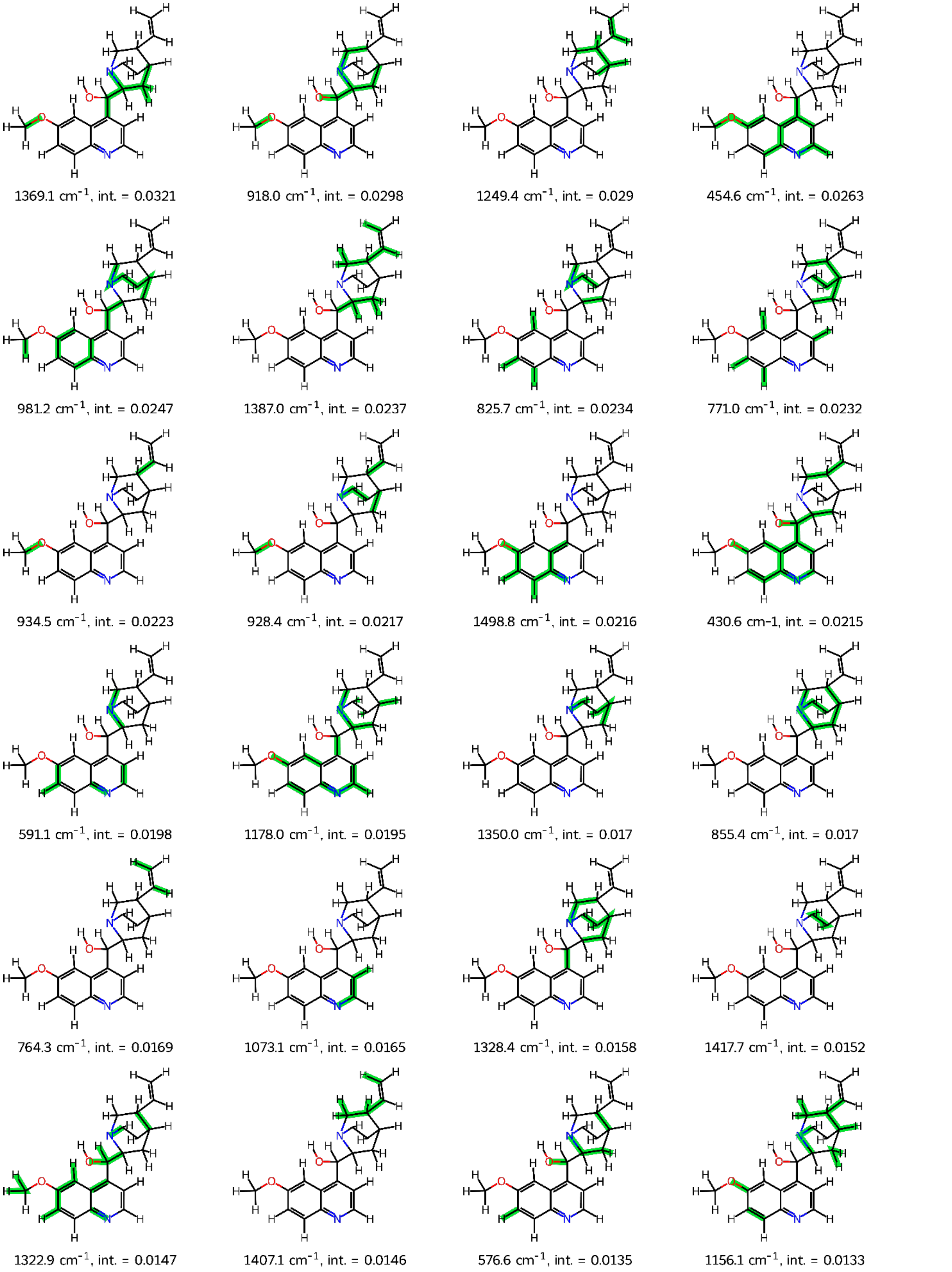

**Fig. S13**. Example of all vibrational bands of quinine in the fingerprint region demonstrating its collective-motion nature. Vibrational modes are ordered by intensity as calculated by xTB. (part 2)

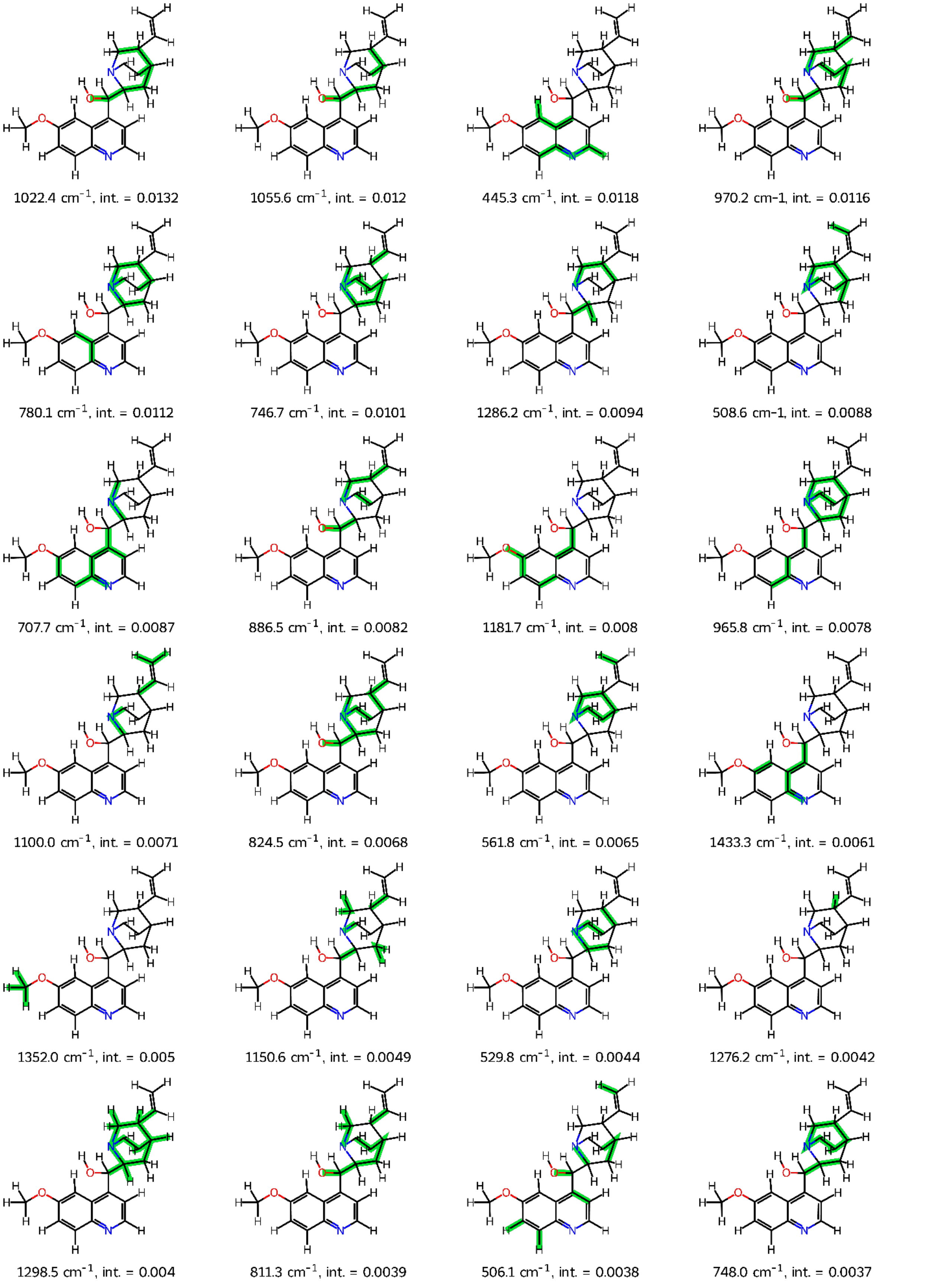

**Fig. S14**. Example of all vibrational bands of quinine in the fingerprint region demonstrating its collective-motion nature. Vibrational modes are ordered by intensity as calculated by xTB. (part 3)

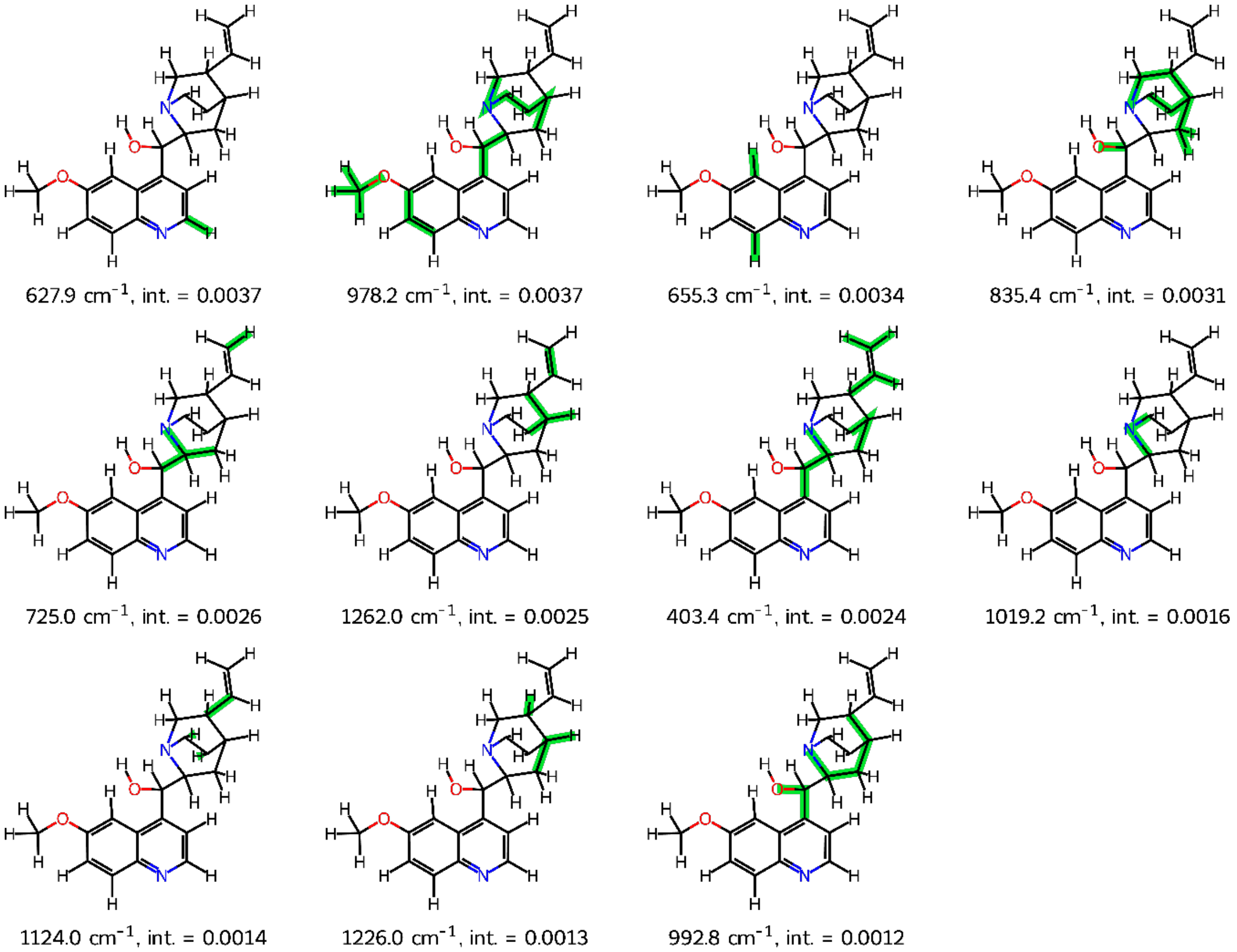

**Fig. S15**. Example of all vibrational bands of quinine in the fingerprint region demonstrating its collective-motion nature. Vibrational modes are ordered by intensity as calculated by xTB. (part 4)

## 2.4 Infrared Spectroscopy – DFT simulations

To validate potential inaccuracies in the predicted frequency spectra using the semiempirical method, a detailed rigorous analysis over a limited set of 101 compounds was performed. The theoretical approach has been used as described in the work by Swart and colleagues.(12) Detailed quantum chemical simulations were performed using Amsterdam Density Functional (ADF 2017)(13) software. In this study, QUILD(14) (Quantum-regions interconnected by Local descriptions) program was used with delocalized coordinates for the optimization of equilibrium structures until the maximum gradient component was less than $10^{-4}$ a.u. Energies, gradients and Hessians for vibrational frequencies including Raman intensities were calculated using BP86-D3(15–17) with a triple/double-zeta valence plus polarization basis set (TZP for metals, DZP for other elements). In all cases, these calculations included solvation effects through the COSMO(18) dielectric continuum model with appropriate parameters for solvent, and scalar relativistic corrections through Zeroth Order Regular Approximation (ZORA).[11]

The number of peaks in the fingerprint region (400–1500 cm$^{-1}$) above an intensity threshold (25 km·mol$^{-1}$) was found to be 0.76 (**Fig. S16**). The histogram of calculated MA *vs*. the expected is depicted in **Fig. S17**.

```
                    Results: Ordinary least squares
==================================================================
Model:              OLS              Adj. R-squared:     0.570
Dependent Variable: y                AIC:                627.4633
Date:               2023-02-08 15:07 BIC:                632.9003
No. Observations:   112              Log-Likelihood:     -311.73
Df Model:           1                F-statistic:        148.3
Df Residuals:       110              Prob (F-statistic): 4.06e-22
R-squared:          0.574            Scale:              15.592
---------------------------------------------------------------------
         Coef.     Std.Err.       t       P>|t|     [0.025    0.975]
---------------------------------------------------------------------
x1       0.4859      0.0399    12.1773    0.0000    0.4068    0.5650
const    5.6173      0.8135     6.9053    0.0000    4.0051    7.2294
------------------------------------------------------------------
Omnibus:              4.323        Durbin-Watson:          1.179
Prob(Omnibus):        0.115        Jarque-Bera (JB):       4.310
Skew:                 0.474        Prob(JB):               0.116
Kurtosis:             2.842        Condition No.:          45
==================================================================
```

**Fig. S16**. Print output from the fit of MA = $x_1 \times n_{\mathrm{IR_peaks}}$ + *const.*; using *statsmodels.api.OLS* in Python.(8)

The list of chemical structures of all compounds used for the DFT study is in **Fig. S18–Fig. S20**.

The best model for inferring the MA based on the DFT-predicted IR spectra is thus:

$$\mathrm{MA} = 0.49 \times n_{\mathrm{IR_peaks}} + 5.6 \quad \textbf{(3)}$$

where $n_{\mathrm{IR_peaks}}$ is the number of IR peaks in the region of 400–1500 cm$^{-1}$ with intensity above 25 km·mol$^{-1}$.

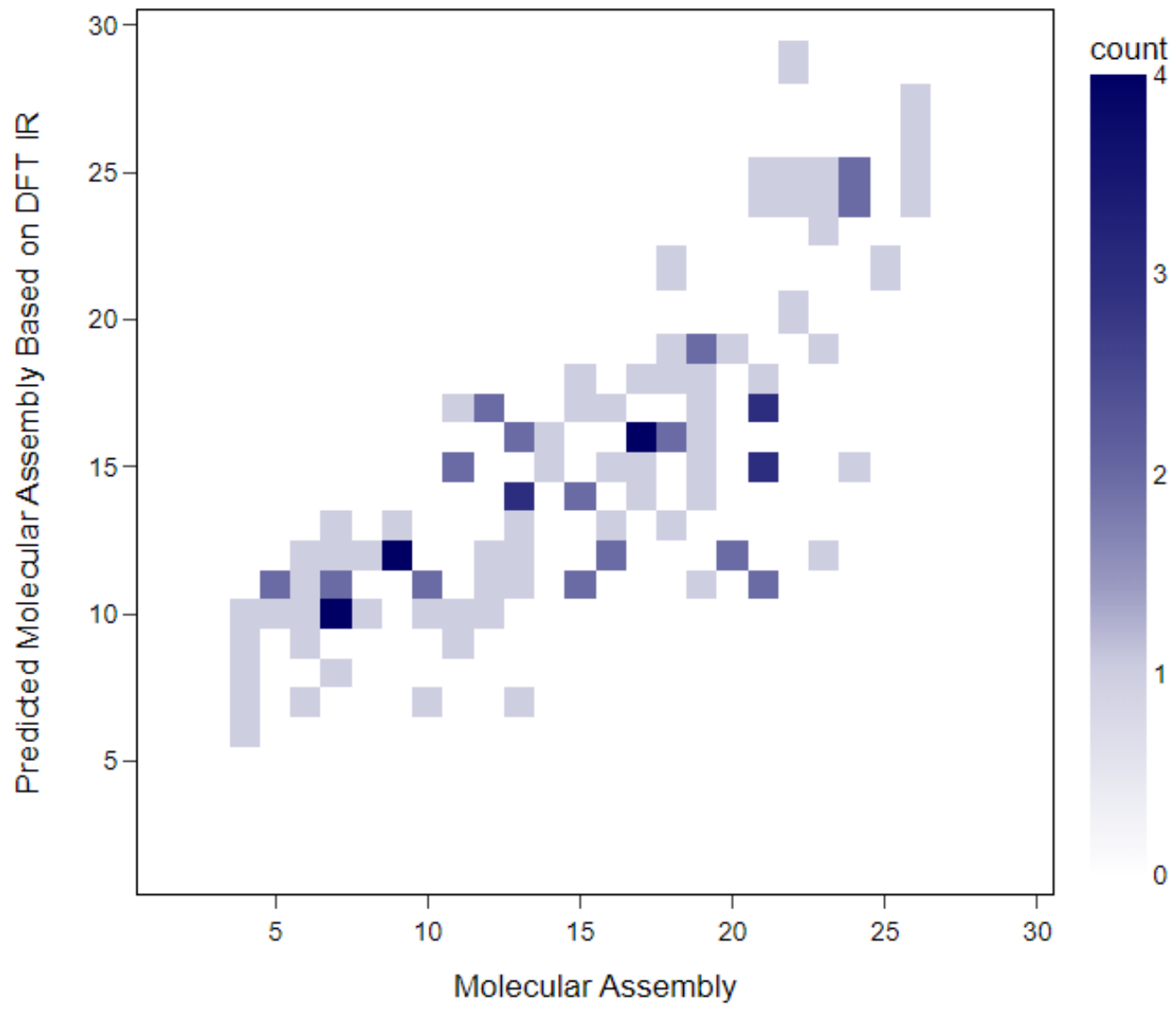

**Fig. S17**. Histogram of predicted MA (based on the **Eq. 3**) *vs.* MA on 112 compounds dataset of DFT calculated IR peaks in the range of 400–1500 $cm^{-1}$ above intensity 25 $km \cdot mol^{-1}$.

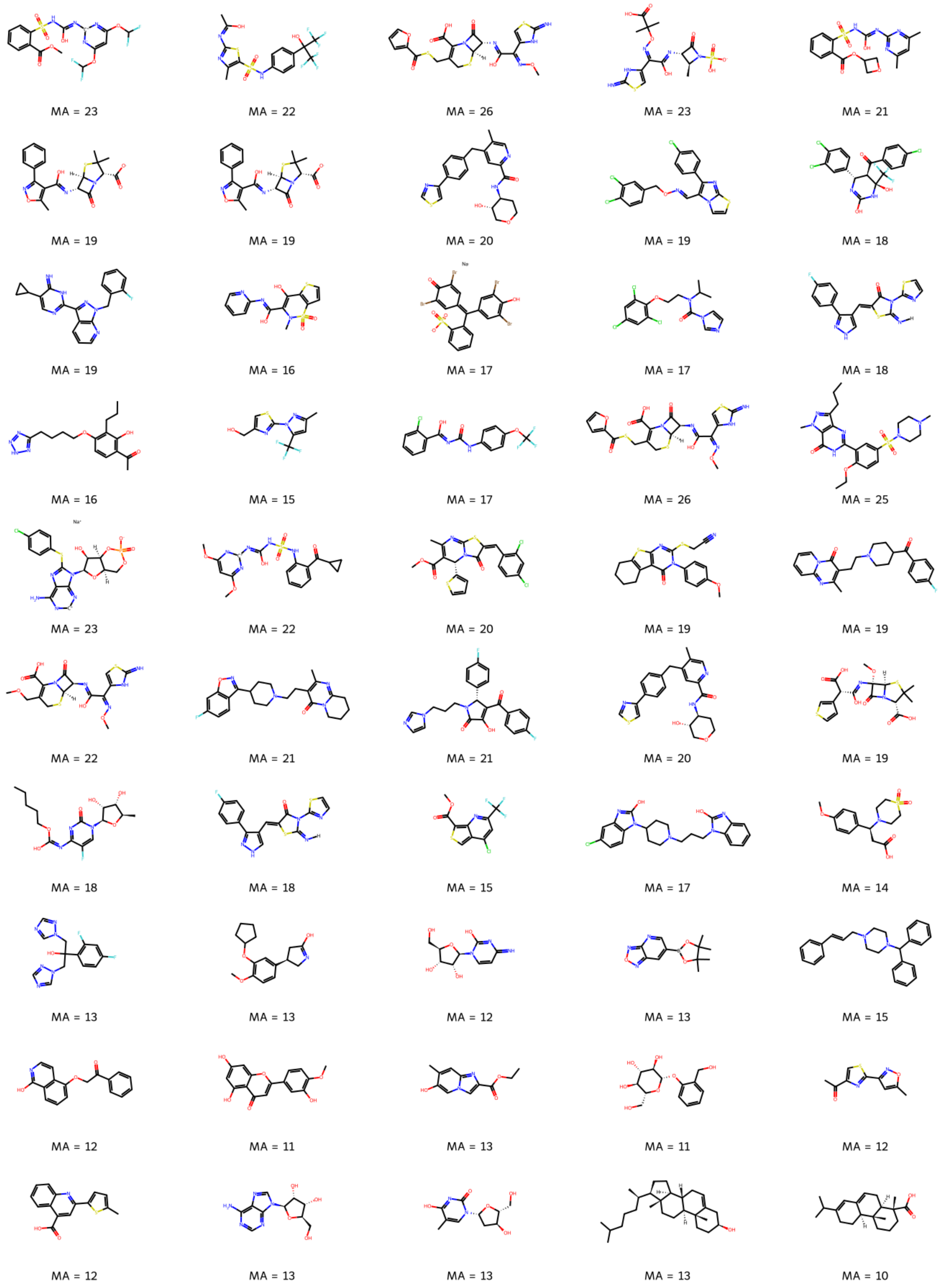

**Fig. S18**. Molecular structures with calculated molecular assembly (MA) were used in the DFT-calculated IR study (**Part 1**).

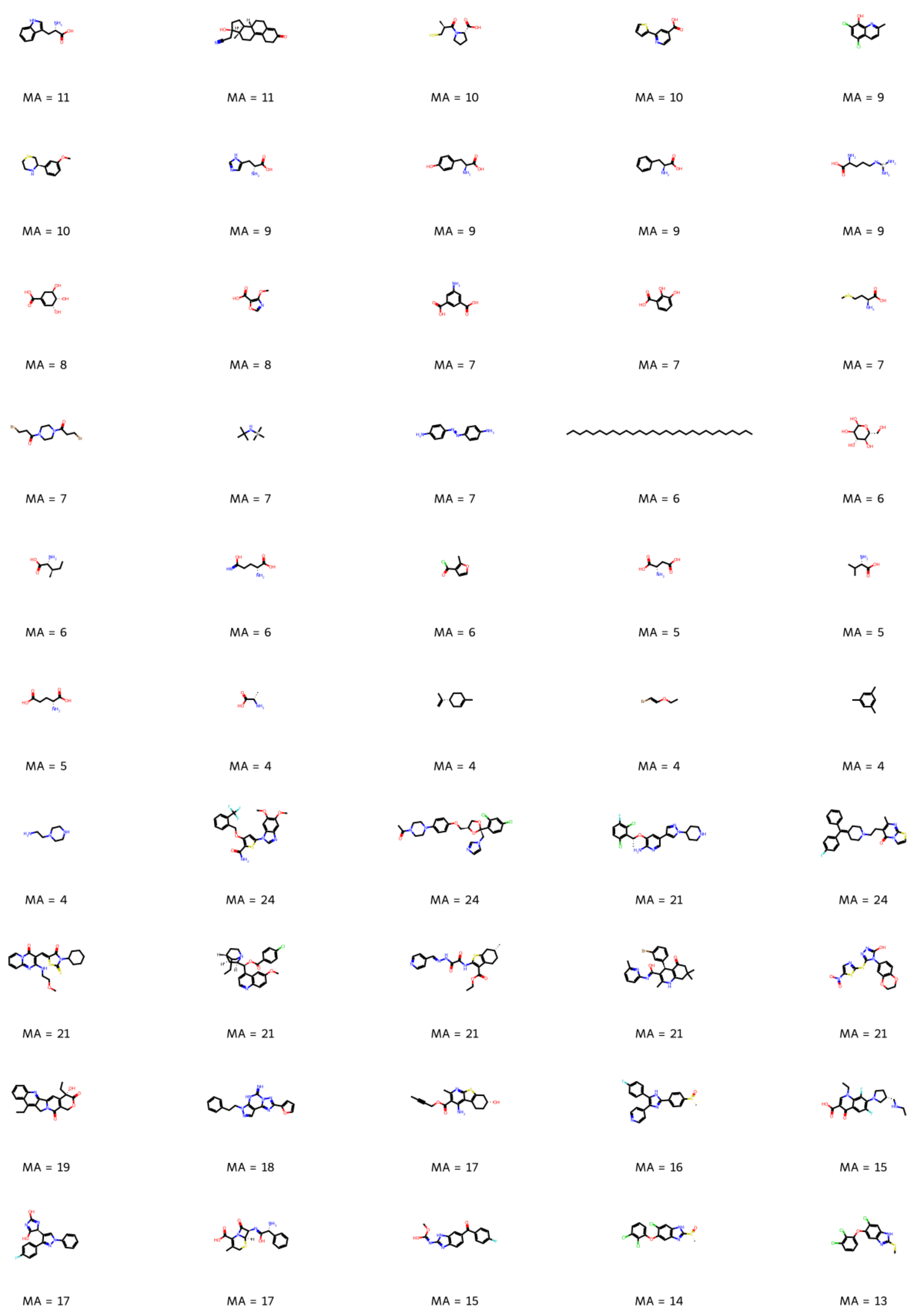

**Fig. S19**. Structures with calculated molecular assembly (MA) used in the DFT calculated IR study (**Part 2**)

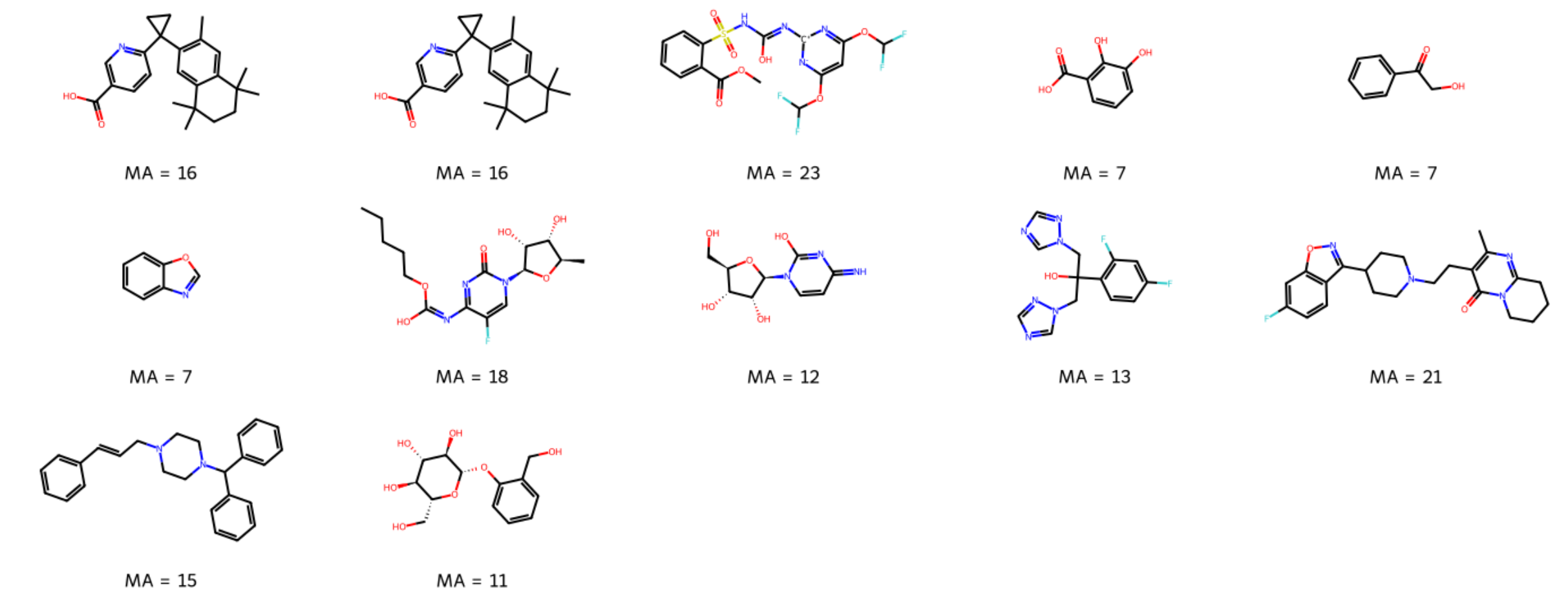

**Fig. S20**. Structures with calculated molecular assembly (MA) used in the DFT calculated IR study (**Part 3**)

# 3 Experimental Infrared Spectroscopy

All experimental IR spectra were acquired on a Thermo Scientific Nicolet iS5 with Specac Golden Gate Reflection Diamond ATR System. All samples were measured in their native state at room temperature (solid state unless liquid at room temperature). The acquired data were processed with Thermo Scientific OMNIC 8.3.103 software; using diamond attenuated total reflectance IR (64 scans, resolution 2 $cm^{-1}$). The spectra were processed at 50% sensitivity and 80% threshold for selecting peaks using OMNIC software (see an example of an acquired spectrum in **Fig. S21**). IR peaks in the fingerprint region (400-1500 $cm^{-1}$) were counted and correlated against the MA of the molecule. To reduce the error between sample screenings, the background IR spectra were recorded after every 3$^{rd}$ sample measurement. Linear regression fit between the experimental IR peaks number in the fingerprint region *vs.* MA agreed provided simple model (**Eq. 4**) with a Pearson's correlation coefficient 0.75:

$$\mathrm{MA} = 0.45 \times n_{\mathrm{IR_peaks}} + 2.26 \quad \textbf{(4)}$$

where $n_{\mathrm{IR_peaks}}$ is the number of IR peaks in the region of 400–1500 $cm^{-1}$. The distribution of MA *vs.* IR-predicted MA is visualised as a histogram in **Fig. S22**. Structures of all compounds used in the study are shown in **Fig. S23** and **Fig. S24**.

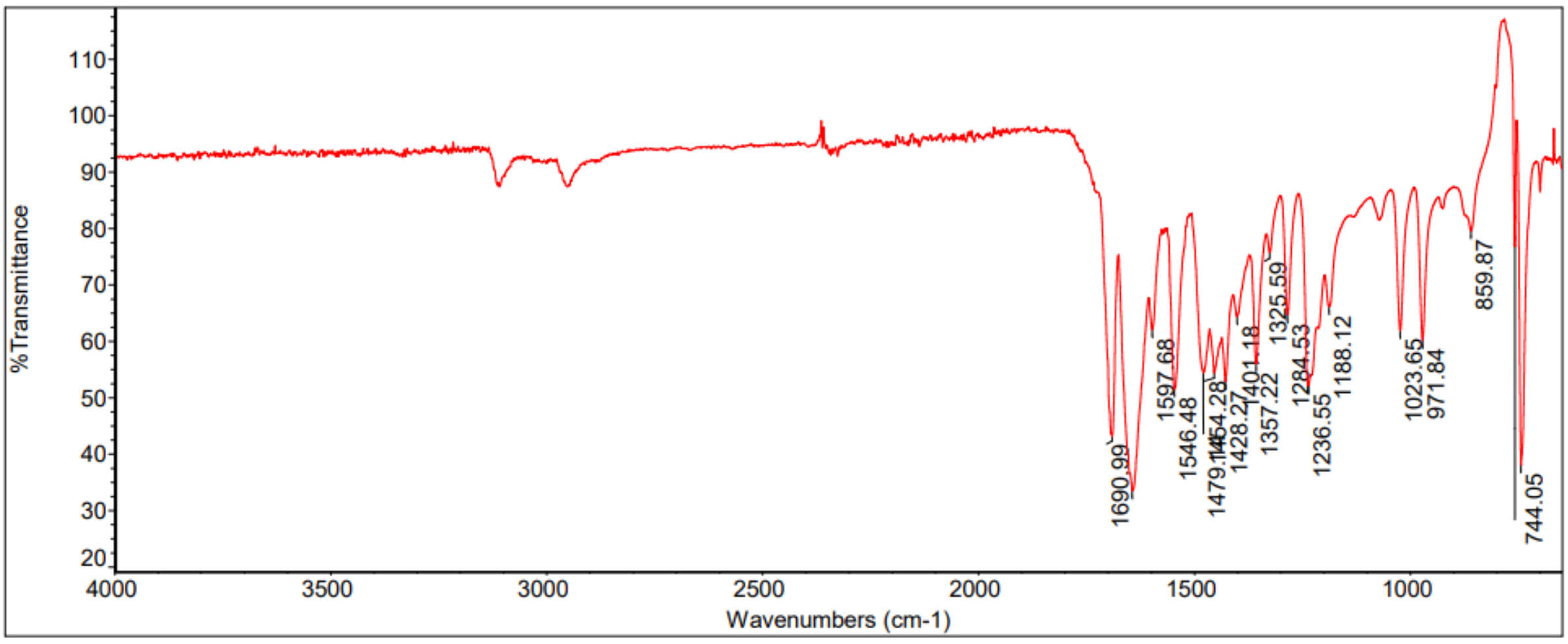

**Fig. S21.** Example of an experimental IR spectrum of Caffeine. Identified peaks in the fingerprint region 400–1500 $cm^{-1}$ considered in the peak count.

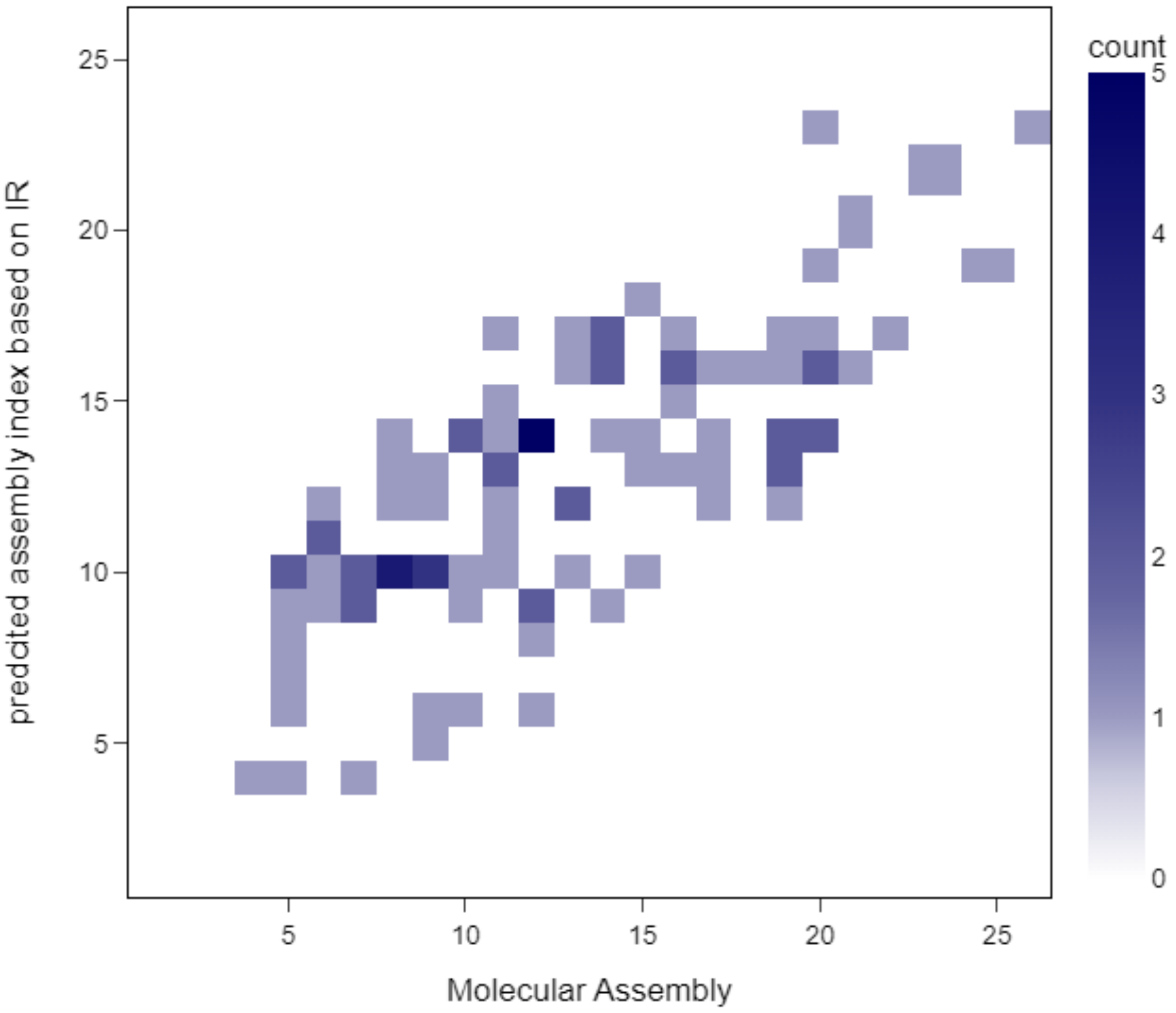

**Fig. S22**. Histogram of predicted MA based on the 99 experimental IR data using **Eq. 4** *vs.* MA.

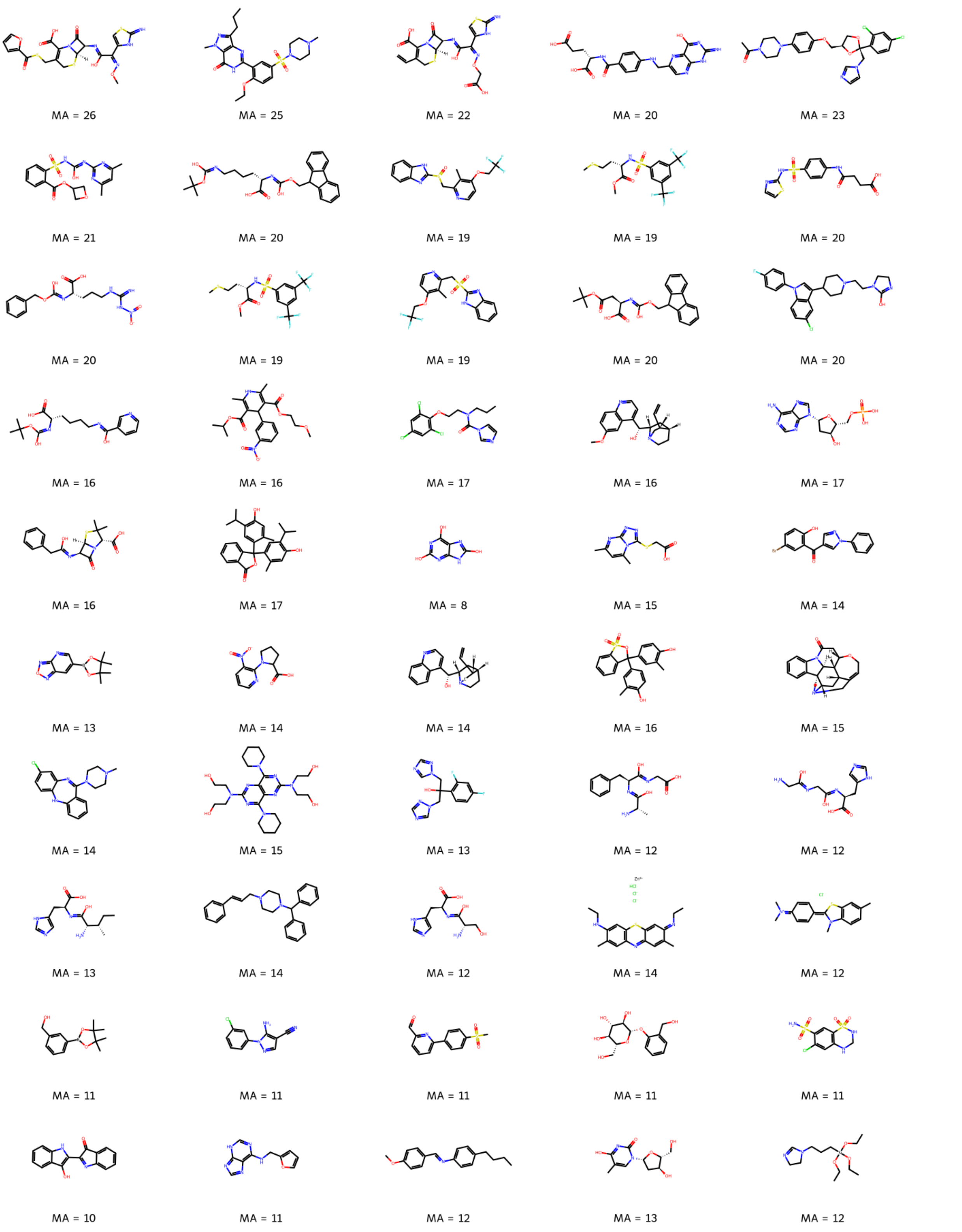

**Fig. S23.** Structures with calculated molecular assembly (MA) used in the experimental IR study (Part 1).

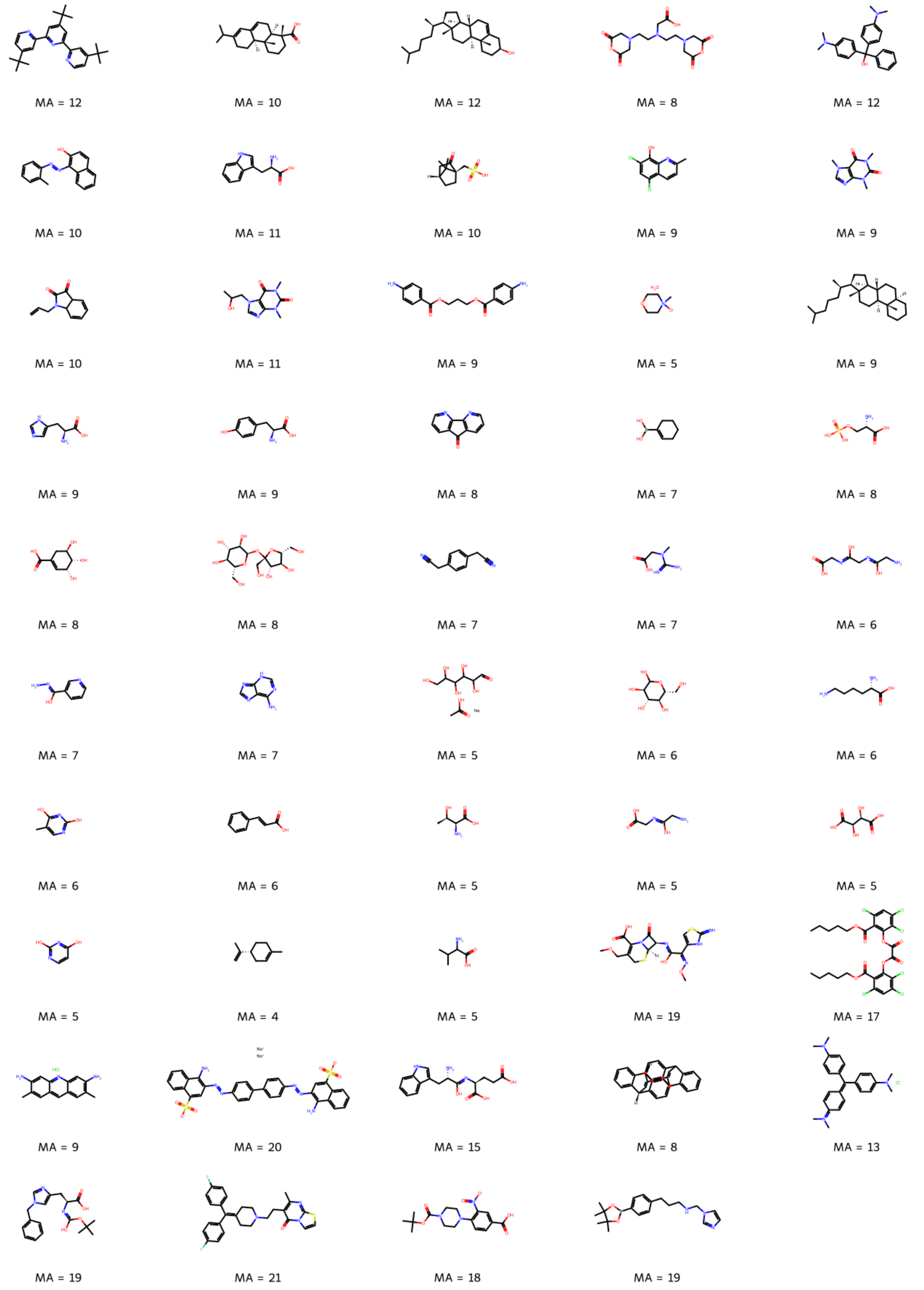

**Fig. S24.** Structures with calculated molecular assembly (MA) used in the experimental IR study (Part 2).

The coefficients for the simple linear function of the number of IR peaks were fitted using the *statsmodels.api.OLS* module in Python was used (**Fig. S25**).(8)

```
                  Results: Ordinary least squares
=================================================================
Model:              OLS              Adj. R-squared:     0.563
Dependent Variable: y                AIC:                528.8456
Date:               2023-02-07 17:18 BIC:                534.0358
No. Observations:   99               Log-Likelihood:     -262.42
Df Model:           1                F-statistic:        127.3
Df Residuals:       97               Prob (F-statistic): 2.33e-19
R-squared:          0.568            Scale:              11.989
--------------------------------------------------------------------
         Coef.     Std.Err.      t       P>|t|     [0.025    0.975]
--------------------------------------------------------------------
x1       0.4531     0.0402    11.2845    0.0000    0.3734    0.5328
const    2.2642     0.9862     2.2960    0.0238    0.3070    4.2215
-----------------------------------------------------------------
Omnibus:              21.605       Durbin-Watson:          1.322
Prob(Omnibus):        0.000        Jarque-Bera (JB):       5.620
Skew:                 0.200        Prob(JB):               0.060
Kurtosis:             1.903        Condition No.:          70
=================================================================
```

**Fig. S25**. Print output from the fit of MA = $x_1 \times n_{\mathrm{peaks}}$ + *const.* for experimental IR spectra; using *statsmodels.api.OLS* in Python.(8)

# 4 Experimental NMR

The investigation of NMR spectroscopy as a prediction tool for MA was experimentally examined on 101 molecules with a range of MA 3–26. Here, the same model that was used in the prediction of MA from the number of different $^{13}C$ resonances in the theoretical dataset was also used to predict MA from was used as developed in the theoretical NMR set (**Eq. 1**).

## 4.1 Sample Preparation details

All samples were prepared with 600 mL $d_6$-DMSO in a 5 mm Bruker 600 MHz rated NMR tube. Concentrations varied from 0.19 mM to 1.17 M due to solubility factors. Several samples (5-aminioisophthalic acid, Oxacillin Sodium Salt, Sildenafil, Triclabendazole) were analyzed *via* NMR at 5, 30 and 300 mM and have shown a minimal effect of the concentration on the extracted number of chemical environment values from spectra. Samples that did not dissolve in $d_6$-DMSO were dissolved in a $D_2O$:$d_3$-MeCN mixture (ratio 75:25).

## 4.2 NMR Experimental Parameters

The NMR used was a Bruker Ascend Aeon 600 MHz NMR spectrometer with a CP DCH 600S3 C/H-D-05 Z cryoprobe installed. All data was processed using Bruker Topspin 3.6.2 and Mestrenova 14.1.1-2451. The NMR experimental parameters are as follows: $^{1}H$ NMR(16 scans, 20 ppm spectral width, 3.46 second acquisition time, 2.00 second relaxation delay), $^{13}C$ NMR(128 scans, 250 ppm spectral width, 1.73 second acquisition time, 0.80 second relaxation delay), $^{13}C$ DEPTQ 90 and DEPTQ 135 (64 scans,250 ppm spectral width,1.73 seconds acquisition time, 1.00 second relaxation delay), $^{1}H$ PSYCHE(16 scans, 12.49 ppm spectral width, 0.89 seconds acquisition time, 1.00 second relaxation delay), HSQC (8 scans, 10 ppm spectral width F2, 250 spectral width F1, 0.09 second acquisition time, 1.49 seconds relaxation delay), $^{13}C$ DOSY (pseudo-2D experiment scans, 200 ppm spectral width F2, 8 TD points, 1.10 second acquisition time, 8.00 second relax delay, 0.80 second diffusion time d20, 1450 μsec gradient pulse P30).

The experiments were tested on several examples to cover a range of concentrations from 5 mM to 300 mM to prove the same number of individual carbon peak types can be achieved, regardless of the concentration.

## 4.3 Classification of the Carbon Types

Accurate counts for all individual $^{13}C$ type environments were obtained using a combination of $^{13}C$, DEPTQ135, DEPTQ90 and HSQC analysis to ensure maximum accuracy before unblinding samples.

To determine the degree of substitution for $^{13}C$ chemical environments, this was first approached using two types of $^{13}C$ DEPT experiments, DEPTQ 135 and DEPTQ 90, that phase carbon signals positive or negative depending on their degree of substitution. The DEPTQ experiment was chosen (not the DEPT) to detect quaternary carbons (otherwise not detected by DEPT). The 135 and 90 portion stands for the final $^{1}H$ tip angle of the pulse in the pulse program before acquisition.(19, 20)

First, $^{13}C$ DEPTQ 135 observes quaternary and $CH_2$ peaks as one phase and the CH and $CH_3$ peaks as the other. The $^{13}C$ DEPTQ 90 then is measured to complement the $^{13}C$ DEPTQ 135 with only the detection of quaternary peaks in one phase and the CH peaks in the other. Using DEPTQ135 and DEPTQ 90 together, all degrees of substitution of the carbons can be identified via NMR. The solvent (expected as quaternary if deuterated) peak must be disregarded in the two counts of carbon peaks. To verify the assignment of the degree of substitution for $^{13}C$ chemical environments in blind samples, $^{1}H$-$^{13}C$ HSQC was used. In the HSQC experiment, the peaks are phased as they are in DEPT with $CH_2$ cross peaks in one phase and CH and $CH_3$ in the other. As $CH_2$ cross peaks are detected in their phase alone, this provides an easy method of counting the number of $CH_2$ chemical environments from $CH_2$ cross peaks. Quaternary $^{13}C$ cross peaks are not detected in HSQC.

The herein described workflow is illustrated on quinine, and its $^{13}C$ NMR (**Fig. S26**), DEPTQ-90 (**Fig. S27**), DEPTQ-135 (**Fig. S28**) and $^{1}H$-$^{13}C$ HSQC (**Fig. S29**) spectra. Structures of all compounds used in the NMR study are shown in **Fig. S30** and **Fig. S31**.

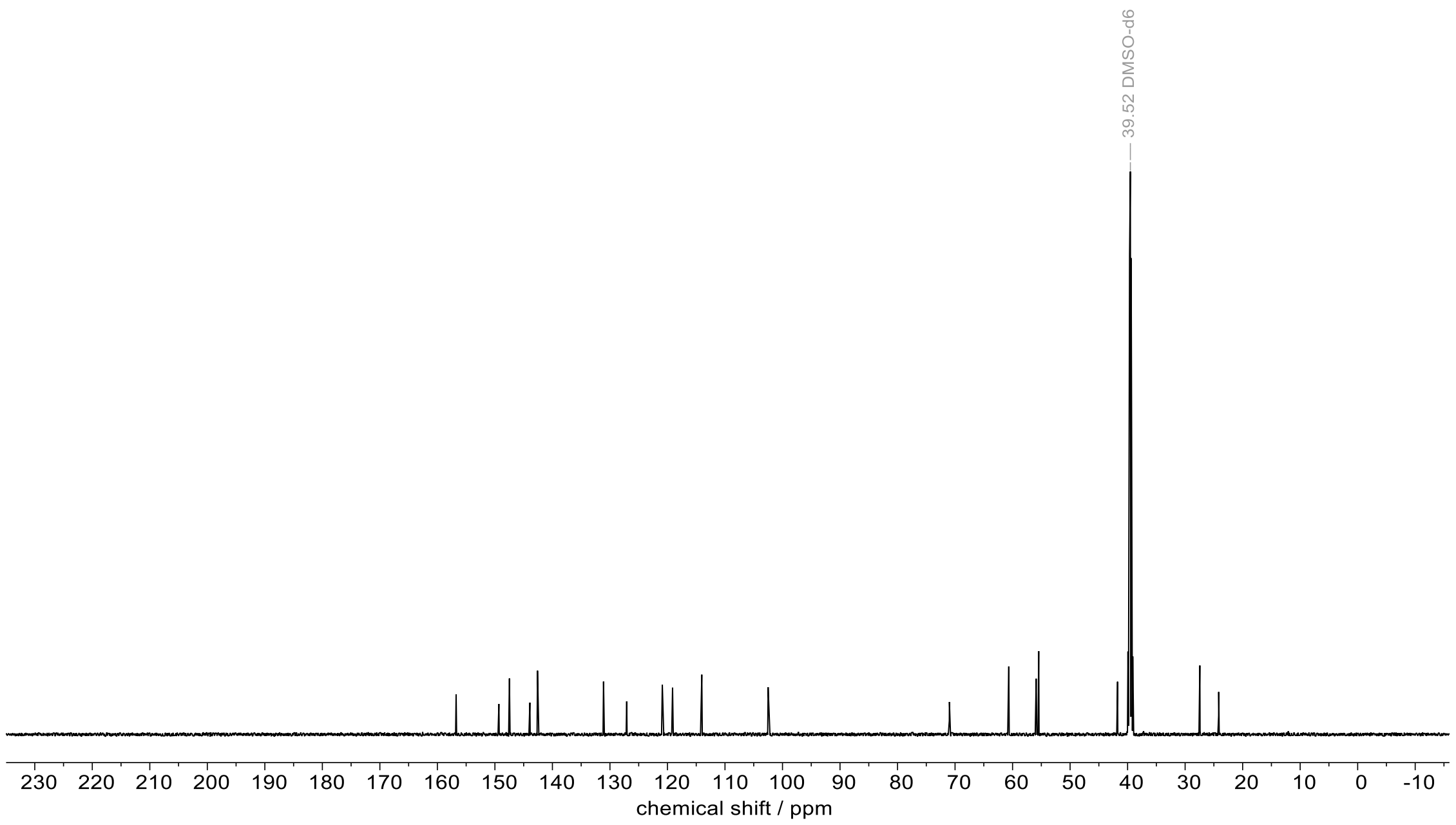

**Fig. S26**. $^{13}C$ NMR (150 MHz) of quinine.

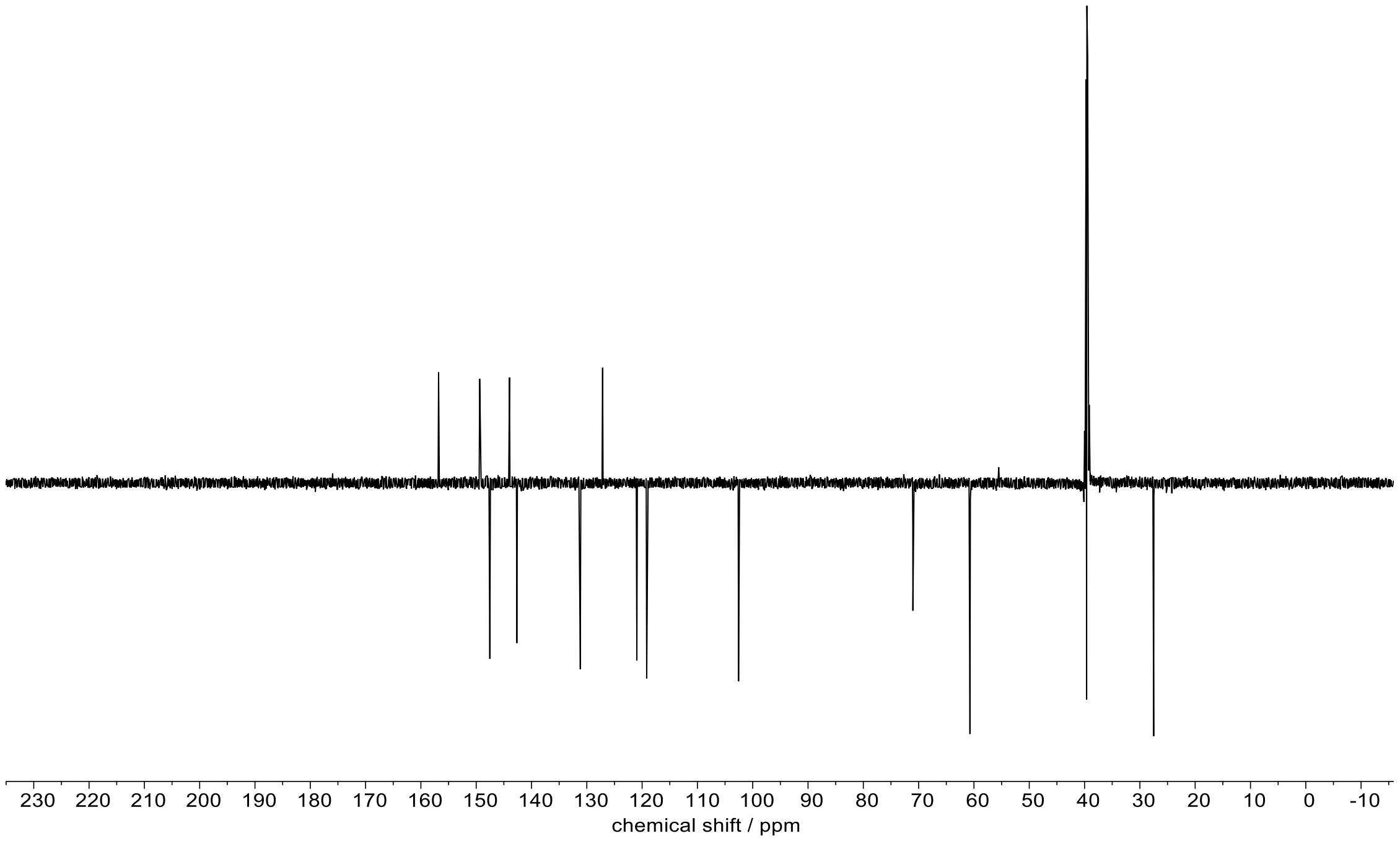

**Fig. S27**. DEPTQ-90 $^{13}C$ NMR (150 MHz) of quinine. Peaks at positive phase are C, peaks at negative are CH. The solvent peak needs to be subtracted from the C count (in this case DMSO-d6 in the positive phase).

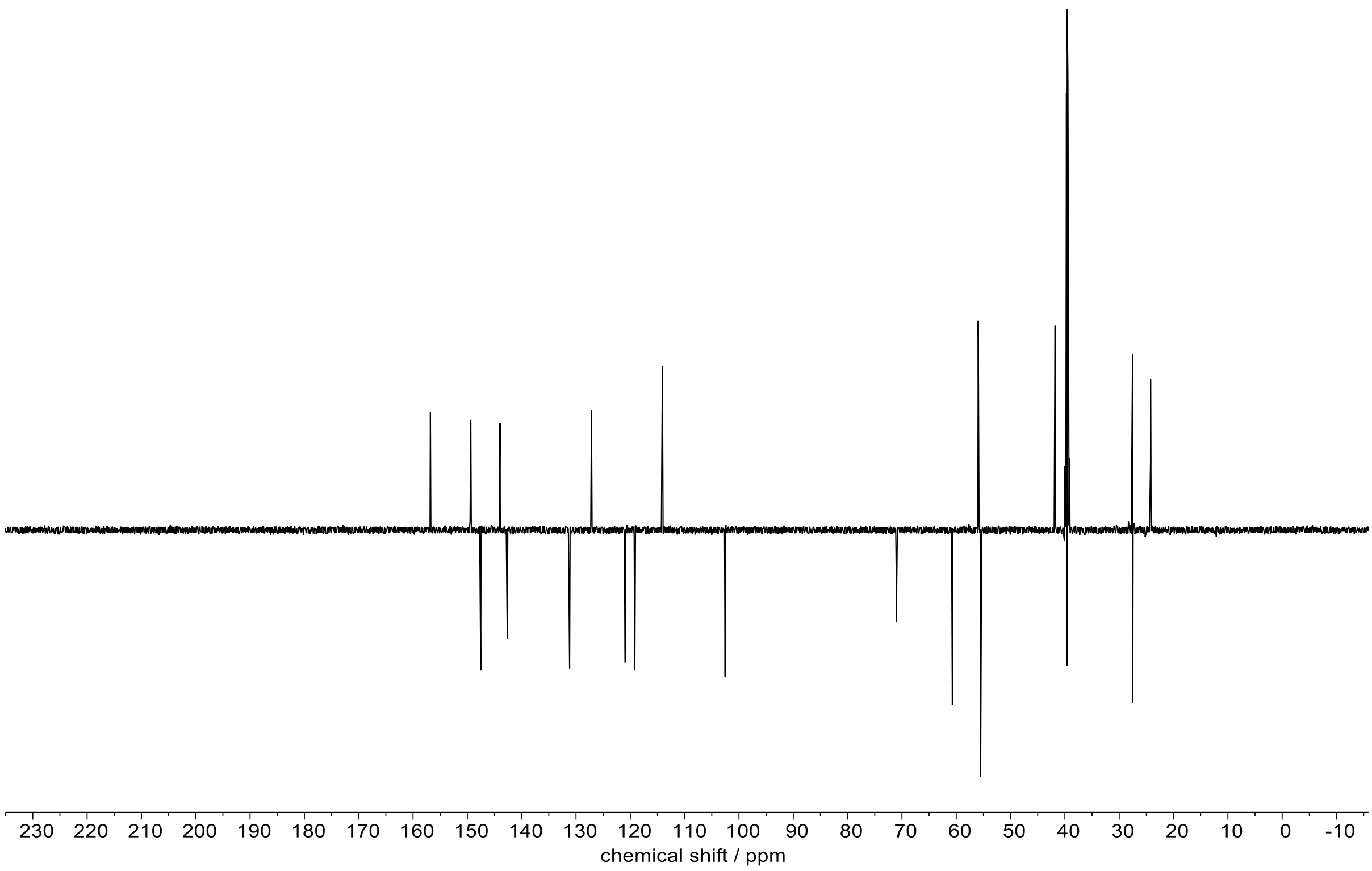

**Fig. S28**. DEPTQ-135 $^{13}C$ NMR (150 MHz) of quinine. Peaks at the positive phase are C and $CH_2$, and peaks at the negative as CH and $CH_3$.

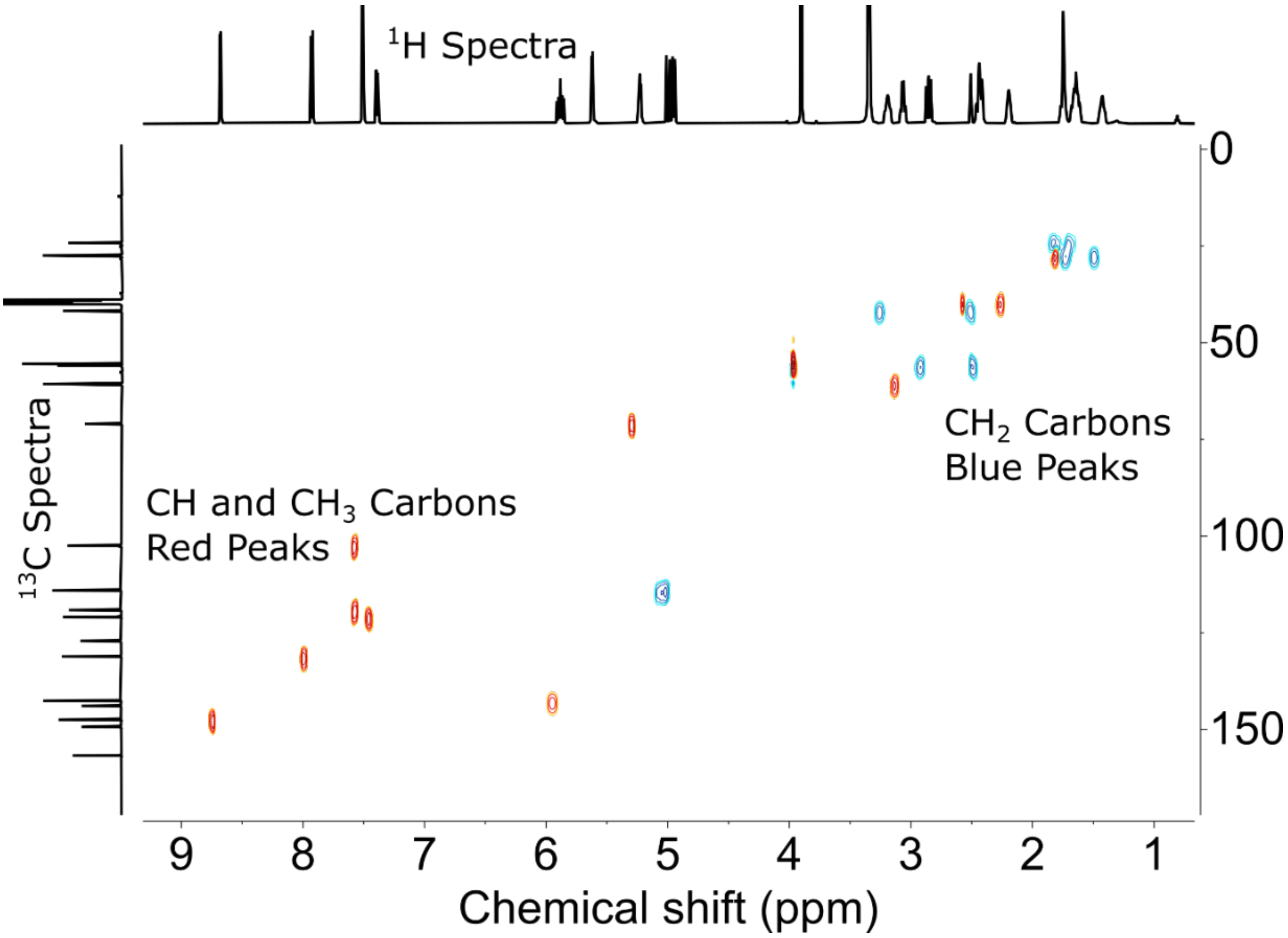

**Fig. S29. $^{1}H$-$^{13}C$ HSQC** Spectrum of quinine highlighting cross peaks of different phases ($CH/CH_3$ in red and $CH_2$ in blue).

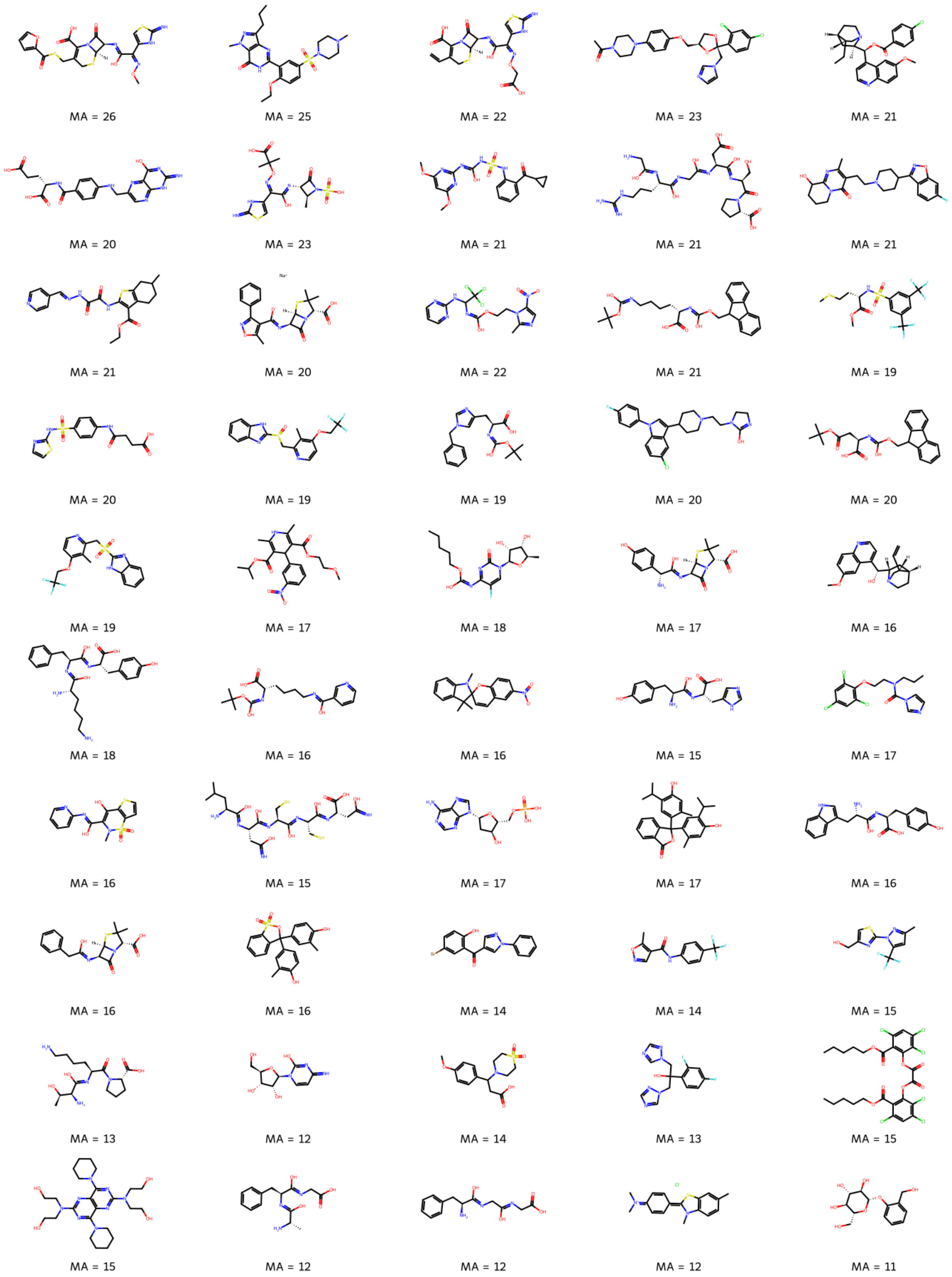

**Fig. S30.** Structures with calculated molecular assembly (MA) used in the experimental NMR study (part 1).

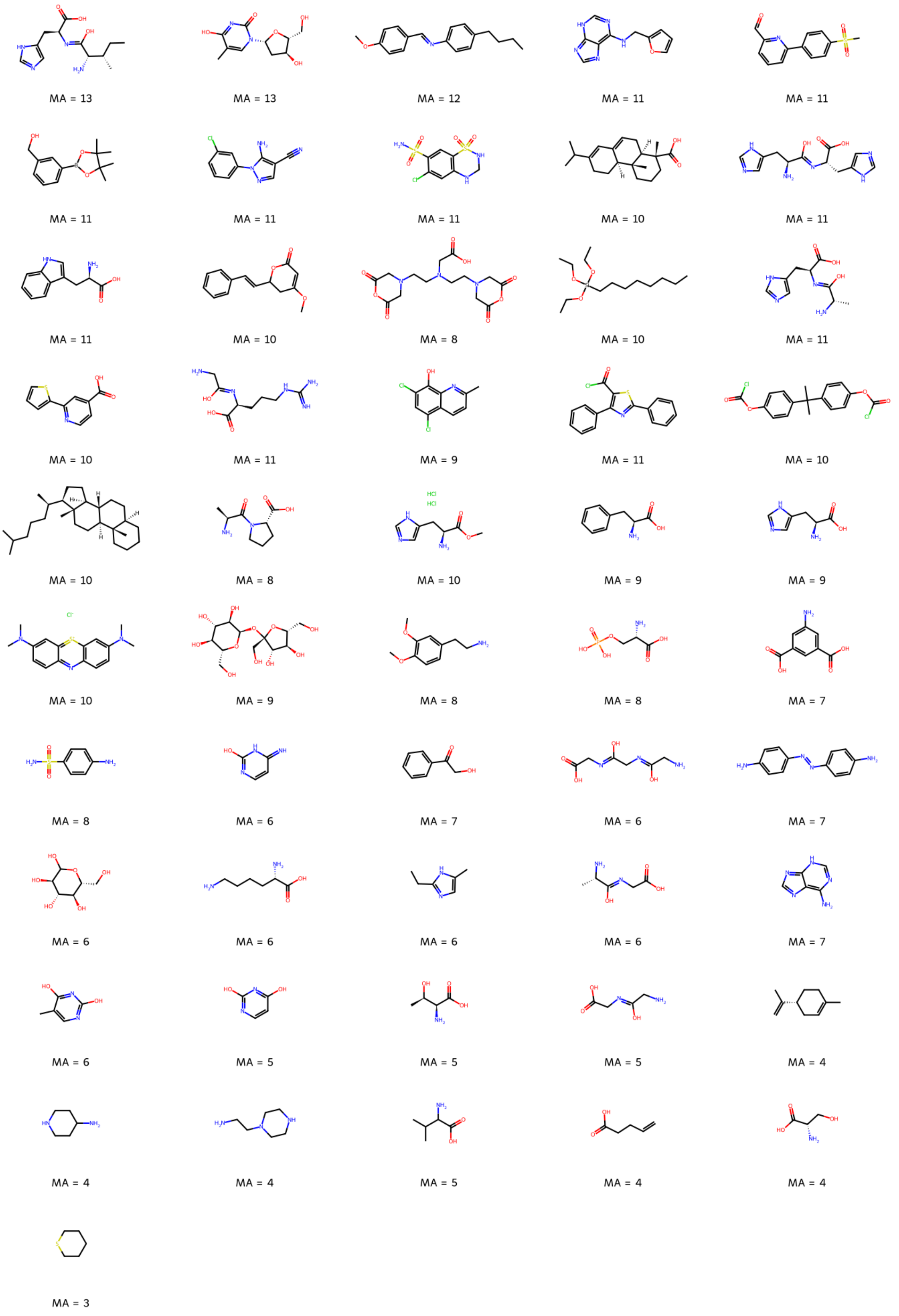

**Fig. S31.** Structures with calculated molecular assembly (MA) used in the experimental NMR study (part 2).

# 5 Combing IR and NMR data

On the set of 10,000 calculated NMR and IR spectra, we have examined our hypothesis that combined information can provide a more reliable MA prediction. We have used the models for the individual spectroscopic techniques (**Eq. 1** and **Eq. 2**) and allowed them to optimise for their relative weighting. The combined model provided a higher correlation of 0.90 using the weighted average of 0.55×NMR and 0.45×IR inferred MA (**Fig. S32**).

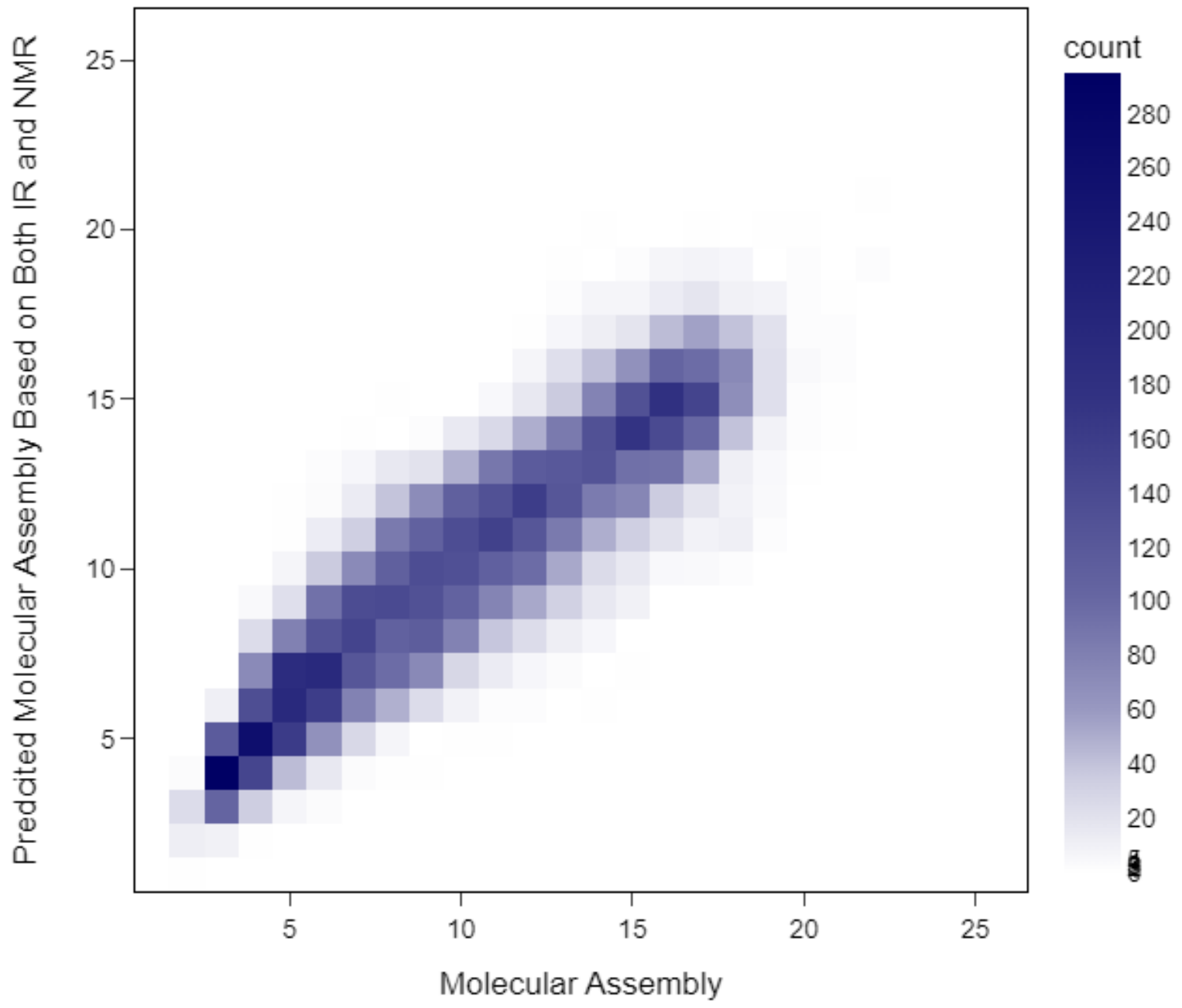

**Fig. S32**. Histogram of predicted MA *vs* expected MA on 10,000 compounds sample based on the fit of NMR and IR prediction using **Eq. 1** and **Eq. 2**, respectively, using a weighted average of NMR and IR of 0.55 and 0.45, respectively.

Analogously to the simulated data, the ratio for a weighted average of the models based on **Eq. 1** for NMR and the experimental model fit for IR **Eq. 4** were optimised for the experimental test sample on the available intersection of the experimental NMR and IR data, comprising 55 molecules. The weighting was 0.7 and 0.3 for the ratio of NMR and IR MA predictions, respectively, yielding a correlation of the predicted and experimental MA of 0.89 (**Fig. S33**).

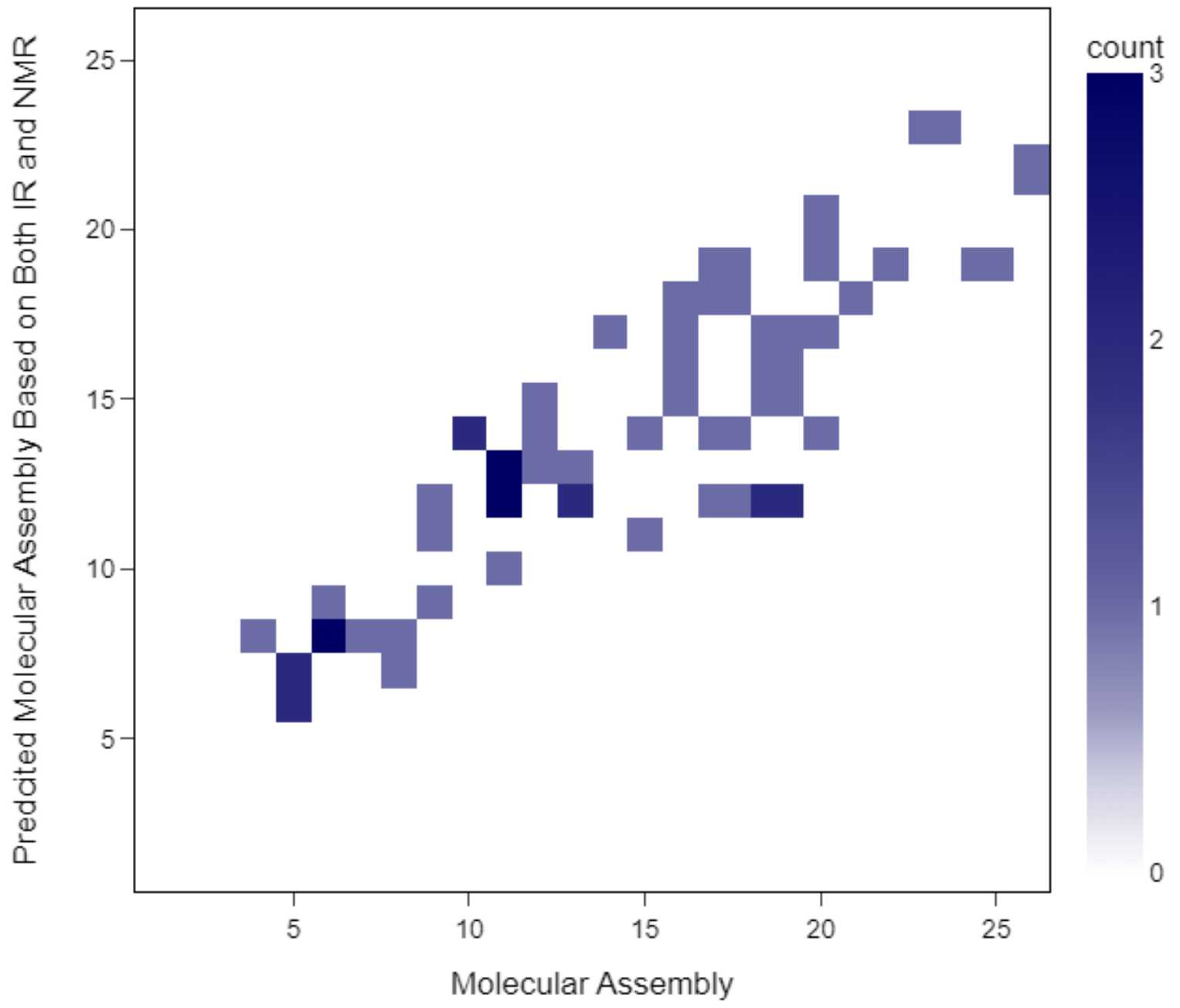

**Fig. S33**. Histogram of predicted MA *vs.* expected MA on 55 compounds sample based on the fit of NMR and IR prediction using **Eq. 1** and **Eq. 4**, respectively, using weighted average of NMR and IR of 0.7 and 0.3, respectively.

In **Table S1** and **S2** are summarised inferred MA from the experimental data. Note that due to experimental limitations, only for 10 compounds the MS$^n$ data suitable for the new recursive algorithm, together with NMR and IR spectroscopies were available. In the rest, MA inference from MS was approximated by corelation of the exact mass of the compounds.

| name | MA inferred from | | | | |
|---|---|---|---|---|---|
| | MS[a)] | NMR | IR | average | MA |
| Ceftiofur | 24.2[a)] | 21.7 | 22.7 | 22.8 | 26 |
| Sildenafil | 21.9[a)] | 19.4 | 19.5 | 20.2 | 25 |
| Cefixime | 20.9[a)] | 19.1 | 17.2 | 19.1 | 22 |
| Ketoconazole | 24.5[a)] | 23.3 | 22.2 | 23.4 | 23 |
| Folic Acid | 20.3[a)] | 18.5 | 22.7 | 20.5 | 20 |
| FMOC-Lys(Boc)-OH | 21.6[a)] | 18.3 | 15.9 | 18.6 | 21 |
| N-[3,5-Bis(trifluromethyl)benzenesulfonyl]-1-methionine | 20.2[a)] | 11.6 | 14.0 | 15.3 | 19 |
| Succinylsulfathiazole | 16.3[a)] | 13.2 | 15.9 | 15.1 | 20 |
| Lansoprazole | 16.9[a)] | 17.7 | 16.8 | 17.1 | 19 |
| Boc-His(Bzl)-OH | 15.8[a)] | 15.1 | 13.1 | 14.7 | 19 |
| Sertindole | 20.3[a)] | 20.8 | 14.5 | 18.5 | 20 |
| Fmoc-D-Asp(OtBu)-OH | 18.9[a)] | 16.4 | 18.6 | 18.0 | 20 |
| 2-(((3-Methyl-4-(2,2,2 trifluroethoxy)pyridin-2-yl)methyl)thio)-1H-benzo[d]imidazole | 17.7[a)] | 16.1 | 15.9 | 16.6 | 19 |
| Nimodipine | 26.4[a)] | 19.8 | 17.2 | 21.1 | 16 |
| Quinine | 14.8[a)] | 18.9 | 15.4 | 16.4 | 16 |
| (2S)-2[(tert-Butoxycarbonyl)amino]-6-[(3-pyridinylcarbonyl)amino] hexanoic acid | 16.1[a)] | 15.6 | 15.9 | 15.8 | 16 |
| Prochloraz | 17.2[a)] | 13.5 | 16.3 | 15.7 | 17 |
| 2'-Deoxyadenosine 5'-monophosphate | 20.3 | 11.4 | 12.2 | 14.7 | 17 |
| Thymolphthalein | 19.8[a)] | 19.1 | 14.5 | 17.8 | 17 |
| Penicillin G | 15.3[a)] | 14.7 | 15.9 | 15.3 | 16 |
| Cresol Red | 17.6[a)] | 19.4 | 13.1 | 16.7 | 16 |
| (5-Bromo-2-hydroxy-phenyl)-(1-phenyl-1H-pyrazol-4-yl) Ketone | 21.5 | 16.5 | 16.8 | 18.2 | 14 |
| Fluconazole | 14.0[a)] | 12.1 | 16.8 | 14.3 | 13 |
| Bis(2-carbopentyloxy-3,5,6-trichlorophenyl) oxalate | 44.3 | 15.0 | 12.7 | 24.0 | 15 |
| Dipyridamole | 32.5 | 9.3 | 14.0 | 18.6 | 13 |

**Table S1**. Inferred MA from the experimental data, average prediction and expected value. a) Values inferred from the exact mass of the compounds.

| name | MA inferred from | | | | MA |
|---|---|---|---|---|---|
| | MS[a)] | NMR | IR | average | |
| H-Ala-Phe-Gly-OH | 13.4[a)] | 12.7 | 13.6 | 13.2 | 12 |
| Thioflavin T | 14.6[a)] | 14.8 | 14.5 | 14.6 | 12 |
| D-(-)-Salicin | 13.0[a)] | 13.3 | 10.4 | 12.2 | 11 |
| H-Ile-His-OH | 12.2[a)] | 12.4 | 11.8 | 12.1 | 13 |
| Thymidine | 11.0[a)] | 10.9 | 15.9 | 12.6 | 13 |
| N-(-4-methoxybenzylidene)-4-butylaniline | 16.7 | 13.9 | 14.0 | 14.9 | 12 |
| Kinetin | 9.7[a)] | 12.1 | 16.8 | 12.8 | 11 |
| 6-[4-(Methylsulfonyl)phenyl]-2-pyridine carboxaldehyde | 15.7 | 12.5 | 12.7 | 13.6 | 11 |
| 3-(Hydroxymethyl)Phenylboronic acid Pinacol ester | 10.6[a)] | 10.2 | 15.4 | 12.1 | 11 |
| 5-Amino-1-(3-chlorophenyl)-1H-pyrazole-4-carbonitrile | 12.1 | 12.7 | 11.3 | 12.1 | 11 |
| Hydrochlorothiazide | 17.4 | 9.7 | 12.2 | 13.1 | 11 |
| Abietic Acid | 13.8[a)] | 17.7 | 6.3 | 12.6 | 10 |
| D-Tryptophan | 9.2[a)] | 12.9 | 13.6 | 11.9 | 11 |
| Diethylenetriaminepentaacetic dianhydride | 16.4[a)] | 7.4 | 10.0 | 11.2 | 8 |
| 5,7-Dichloro-8-hydroxy-2-methyl-quinoline | 13.1 | 12.7 | 10.4 | 12.1 | 9 |
| 5-alpha-Cholestane | 17.1[a)] | 18.0 | 5.4 | 13.5 | 10 |
| L-Histidine | 6.9[a)] | 7.8 | 12.2 | 9.0 | 9 |
| Sucrose | 15.7[a)] | 11.8 | 9.5 | 12.3 | 9 |
| O-Phospho-L-Serine | 8.3[a)] | 4.9 | 12.2 | 8.5 | 8 |
| H-Gly-Gly-Gly-OH | 8.5[a)] | 8.0 | 10.9 | 9.1 | 6 |
| Glucose | 8.1[a)] | 6.8 | 11.8 | 8.9 | 6 |
| L-Lysine | 6.5[a)] | 6.9 | 11.3 | 8.2 | 6 |
| Adenine | 5.9[a)] | 7.7 | 9.5 | 7.7 | 7 |
| Thymine | 5.5[a)] | 7.2 | 10.0 | 7.6 | 6 |
| Uracil | 4.9[a)] | 6.4 | 7.2 | 6.2 | 5 |
| L-Threonine | 5.2[a)] | 5.3 | 9.1 | 6.5 | 5 |
| GlyGly | 5.8[a)] | 6.1 | 10.4 | 7.4 | 5 |
| (R)-(+)-Limonene | 6.0[a)] | 9.5 | 3.6 | 6.4 | 4 |
| DL-Valine | 5.1[a)] | 5.6 | 5.9 | 5.5 | 5 |

**Table S2**. Inferred MA from the experimental data, average prediction and expected value. a) Values inferred from the exact mass of the compounds.

# 6 Mixture Analysis

## 6.1 NMR

To deconvolute the mixture *via* NMR as a proof of concept, the mixture of two compounds was examined using $^{13}C$ DOSY. The experimental setup of the $^{13}C$ DOSY was 200 ppm spectral width, 8 TD points, 1.10 second acquisition time, 8.00 second relax delay, 0.80 second diffusion time d20, 1450 µsec gradient pulse P30 using the $^{13}C$ DOSY-stebpgppg1s routine with 256 scans. An example of quinine and 5-aminoisopthalic acid is presented in **Fig. S34**.

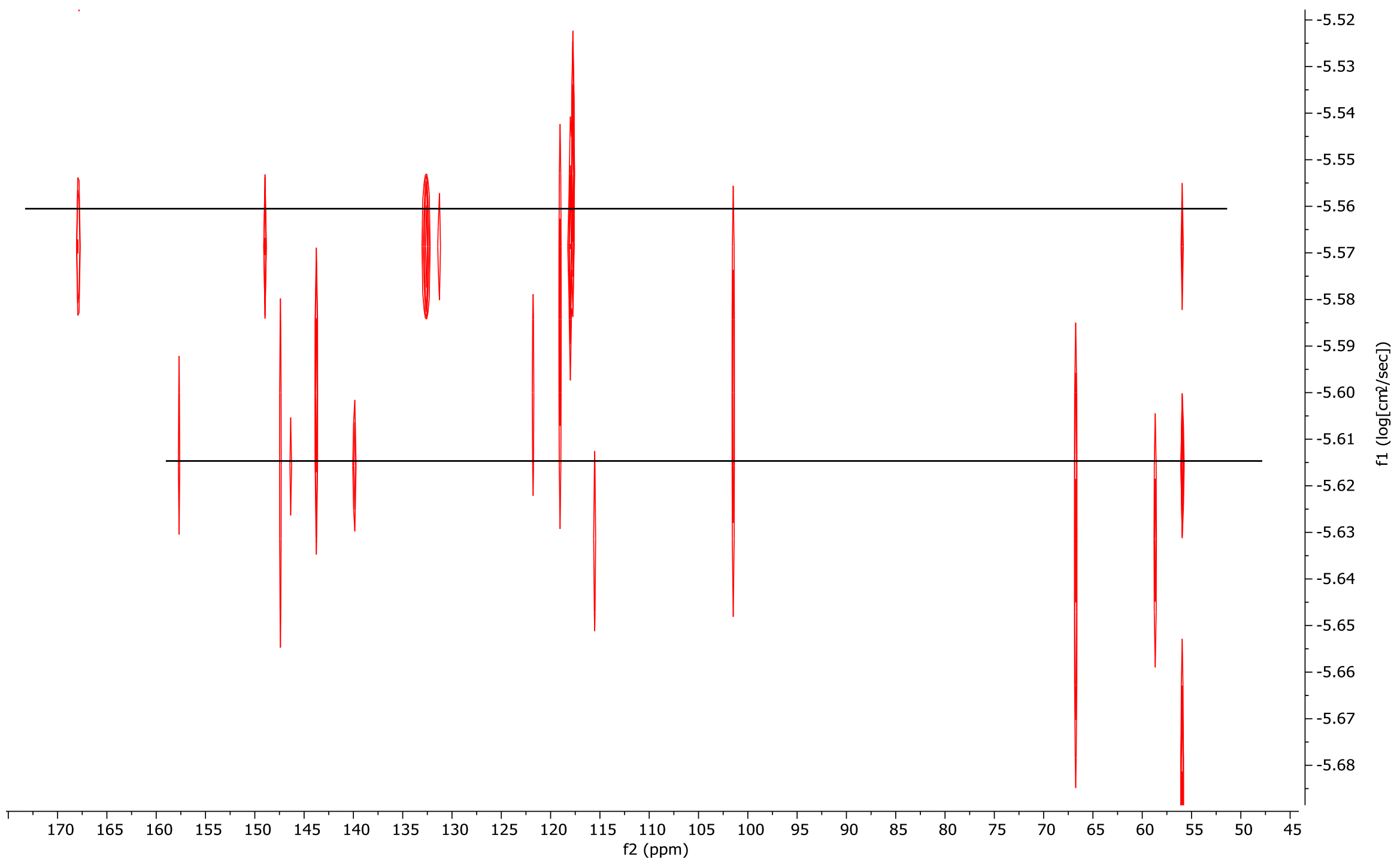

**Fig. S34**. $^{13}C$ DOSY of quinine and 5-aminoisophthalic acid mixture. Two horizontal lines guide the separation of the $^{13}C$ signals.

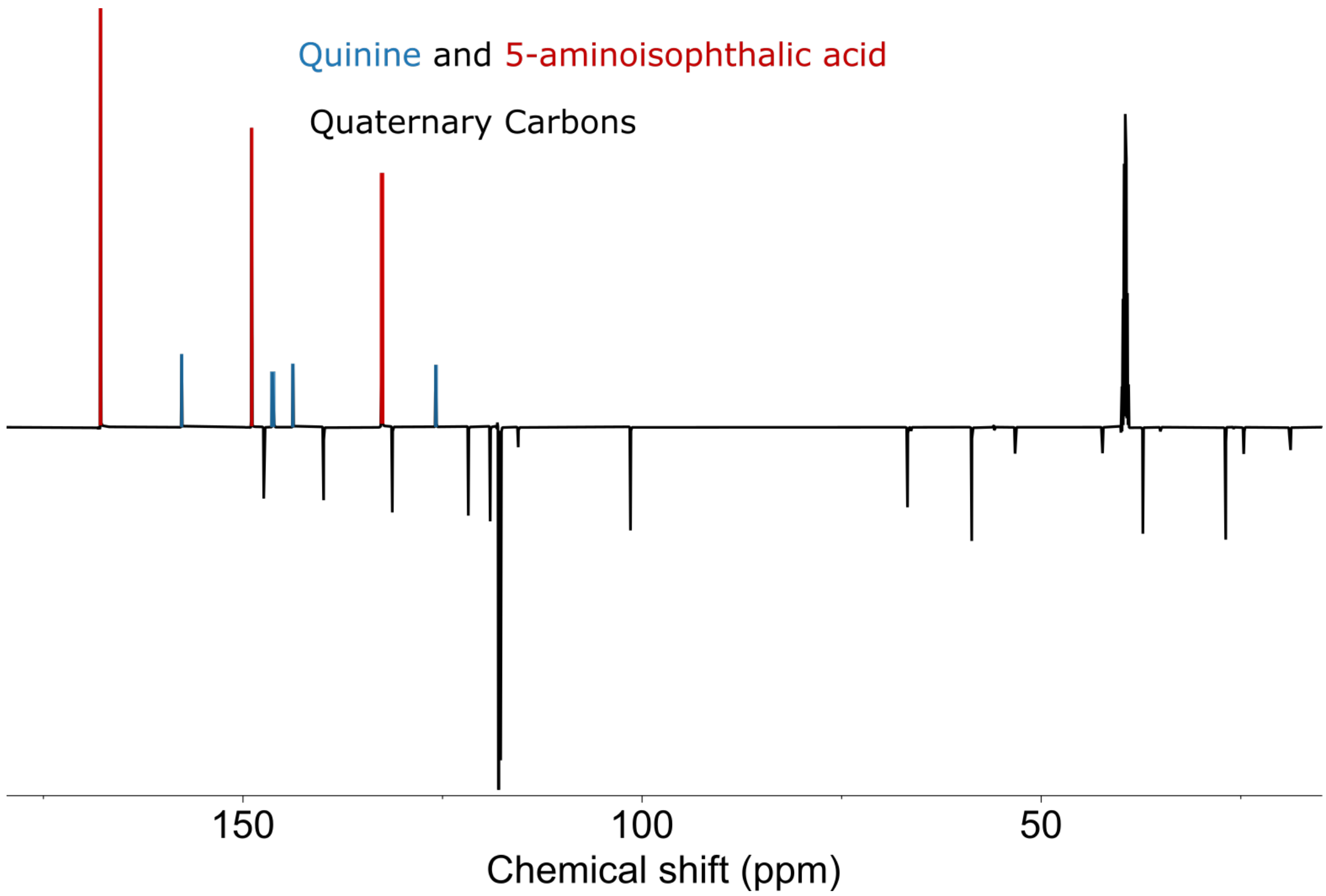

**Fig. S35**. DEPTQ 90 of Quinine (Cq in blue) and 5-aminoisophthalic acid (Cq in red)

# 7 Mass Spectrometry

## 7.1 Theoretical Calculations

To study the relation between MA and molecular weight (MW), a large database (PubChem) was sampled and for nearly 19 million molecules the MA was calculated using assemblyGo algorithm (as described previously) with a short timeout of 10 seconds. As the algorithm incrementally decreases MA over the duration of the calculation, and the initial values are close to the naïve MA, it could be expected that with a shorter timeout, a larger error will be observed for larger molecules. Therefore, we eventually considered only molecules with MW below 500 g·mol$^{-1}$. The number of molecules in that dataset was ~16.7 million. General trends of the MA and MW relation could be characterised with a linear function, as a first-order approximation, as MA = 0.0468×MW – 0.4070. The upper limit was approximated as a linear function fitted on the 99 percentile of the MA values per MW bin as $MA_{max}$ = 0.0547×MW + 0.9458 (**Fig. S36**).

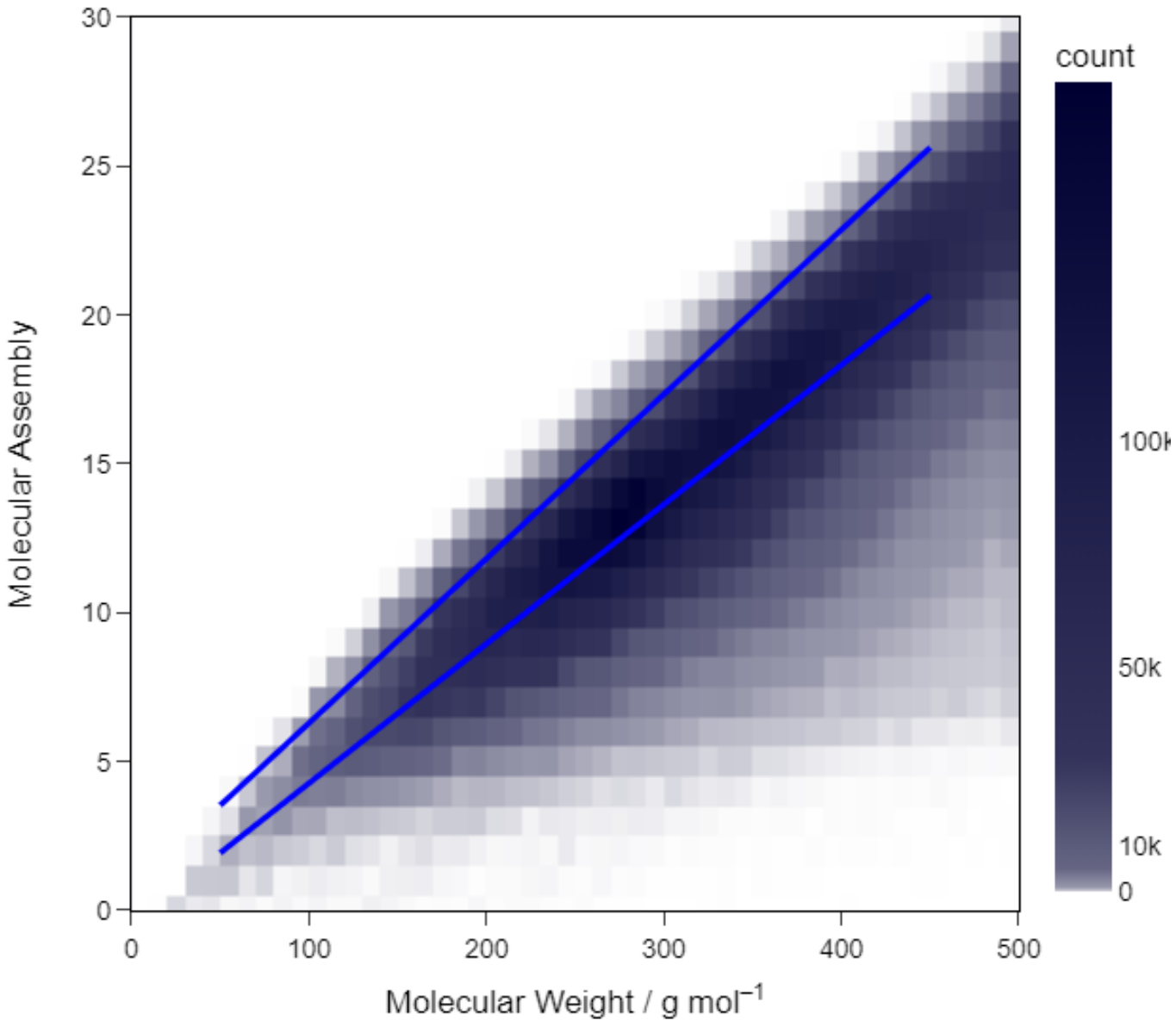

**Fig. S36** Histogram of MA *vs*. MW of 16.7 million compounds sampled from PubChem database calculated using 10 s timeout. The upper limit is characterised by $A_{max}$ = 0.0547×MW + 0.9458; the average MW-to-MA relation as MA = 0.0468×MW – 0.4070

```
                     Results: Ordinary least squares
======================================================================
Model:              OLS              Adj. R-squared:     0.789
Dependent Variable: y                AIC:                71982652.2227
Date:               2023-11-01 11:19 BIC:                71982681.4870
No. Observations:   16720274         Log-Likelihood:     -3.5991e+07
Df Model:           1                F-statistic:        6.255e+07
Df Residuals:       16720272         Prob (F-statistic): 0.00
R-squared:          0.789            Scale:              4.3372
-------------------------------------------------------------------------
          Coef.     Std.Err.        t        P>|t|      [0.025     0.975]
-------------------------------------------------------------------------
x1        0.0468      0.0000    7908.7996    0.0000     0.0468     0.0468
const    -0.4070      0.0020    -200.3950    0.0000    -0.4110    -0.4031
----------------------------------------------------------------------
Omnibus:              2930137.758     Durbin-Watson:        1.870
Prob(Omnibus):        0.000           Jarque-Bera (JB):     7013689.075
Skew:                 -0.994          Prob(JB):             0.000
Kurtosis:             5.473           Condition No.:        1368
======================================================================
```

**Fig. S37**. Print output from the fit of MA = $x_1$×MW + *const*. on the whole dataset; using *statsmodels.api.OLS* in Python.(8)

```
                  Results: Ordinary least squares
=================================================================
Model:              OLS              Adj. R-squared:     0.994
Dependent Variable: y                AIC:                95.8801
Date:               2023-11-01 11:19 BIC:                99.6637
No. Observations:   49               Log-Likelihood:     -45.940
Df Model:           1                F-statistic:        7370.
Df Residuals:       47               Prob (F-statistic): 2.57e-53
R-squared:          0.994            Scale:              0.39808
-------------------------------------------------------------------
         Coef.     Std.Err.       t       P>|t|     [0.025    0.975]
-------------------------------------------------------------------
x1       0.0547      0.0006    85.8476    0.0000    0.0534    0.0560
const    0.9458      0.1858     5.0893    0.0000    0.5719    1.3197
-----------------------------------------------------------------
Omnibus:              20.644       Durbin-Watson:          0.734
Prob(Omnibus):        0.000        Jarque-Bera (JB):       33.993
Skew:                 -1.290       Prob(JB):               0.000
Kurtosis:             6.162        Condition No.:          601
=================================================================
```

**Fig. S38** Print output from the fit of MA = $x_1$×MW + *const*. on the 99 percentile per MW bin; using *statsmodels.api.OLS* in Python.(8)

To characterise the typical MA distribution of specific MW, skew-normal distribution provided a good fit. The fit was performed for every MW bin and the extracted parameters: skewness *a*, location *loc* and *scale* were used to fit the general trend. Simplified prediction of the *a*, *loc* and *scale* parameters for specific MW were fitted as: $a = -0.0083\times\mathrm{MW} + 0.1117$ (**Fig. S39**); $loc = 0.0539\times\mathrm{MW} - 0.4061$ (**Fig. S40**); $scale = 0.0074\times\mathrm{MW} + 0.5108$ (**Fig. S41**).

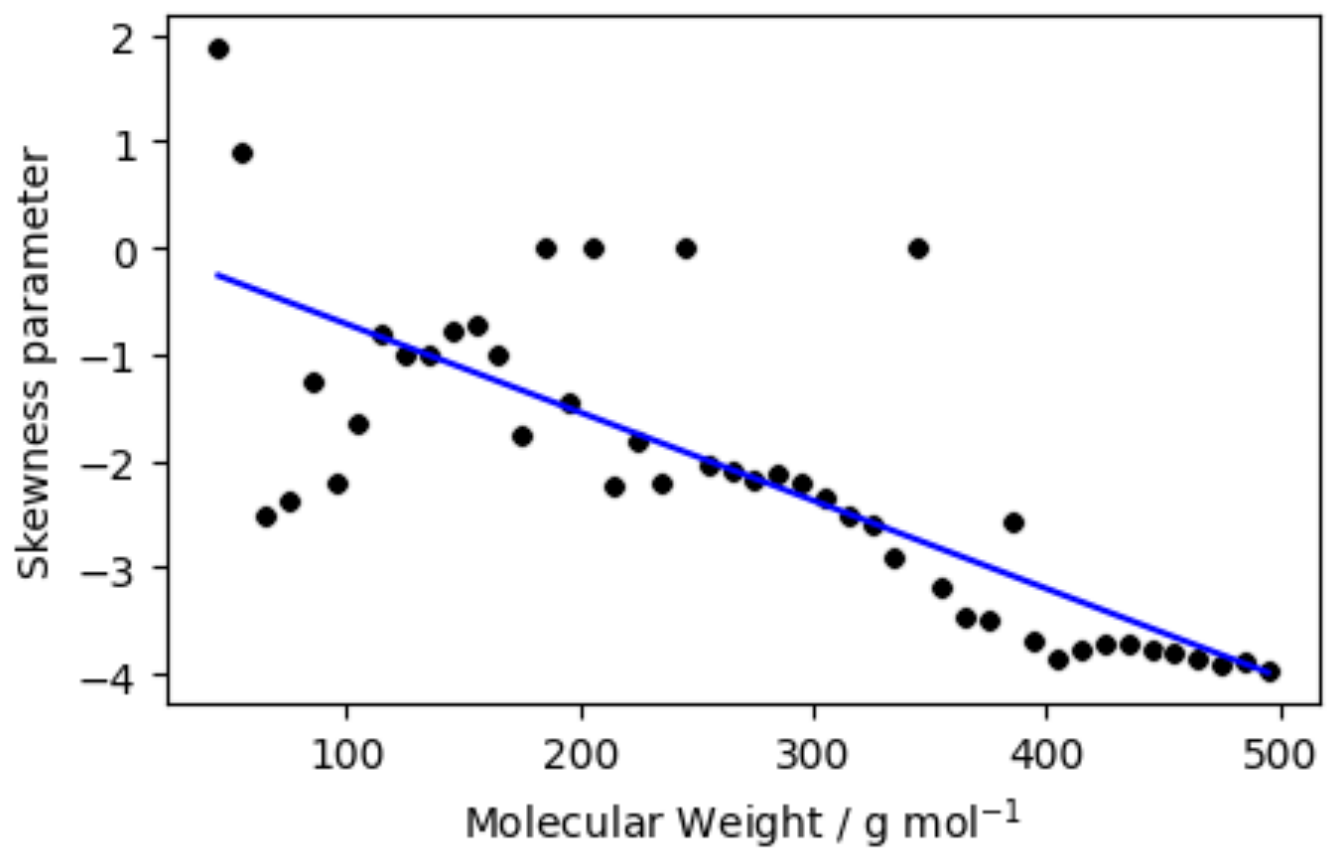

**Fig. S39** General trend of the skewness parameter *vs*. MW, fitted as: $a = -0.0083\times\mathrm{MW} + 0.111$

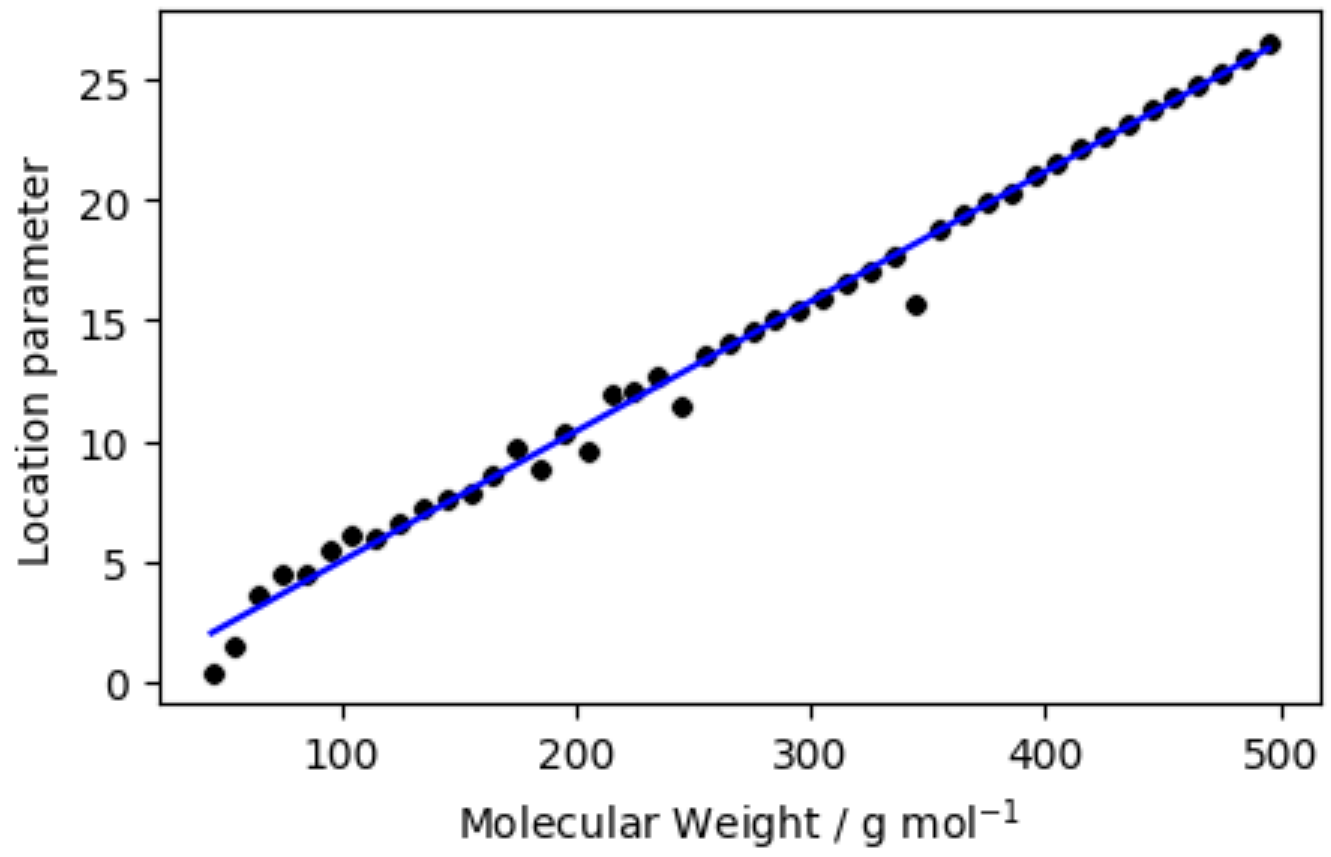

**Fig. S40** General trend of the location parameter *vs*. MW, fitted as: $loc = 0.0539\times\mathrm{MW} - 0.4061$.

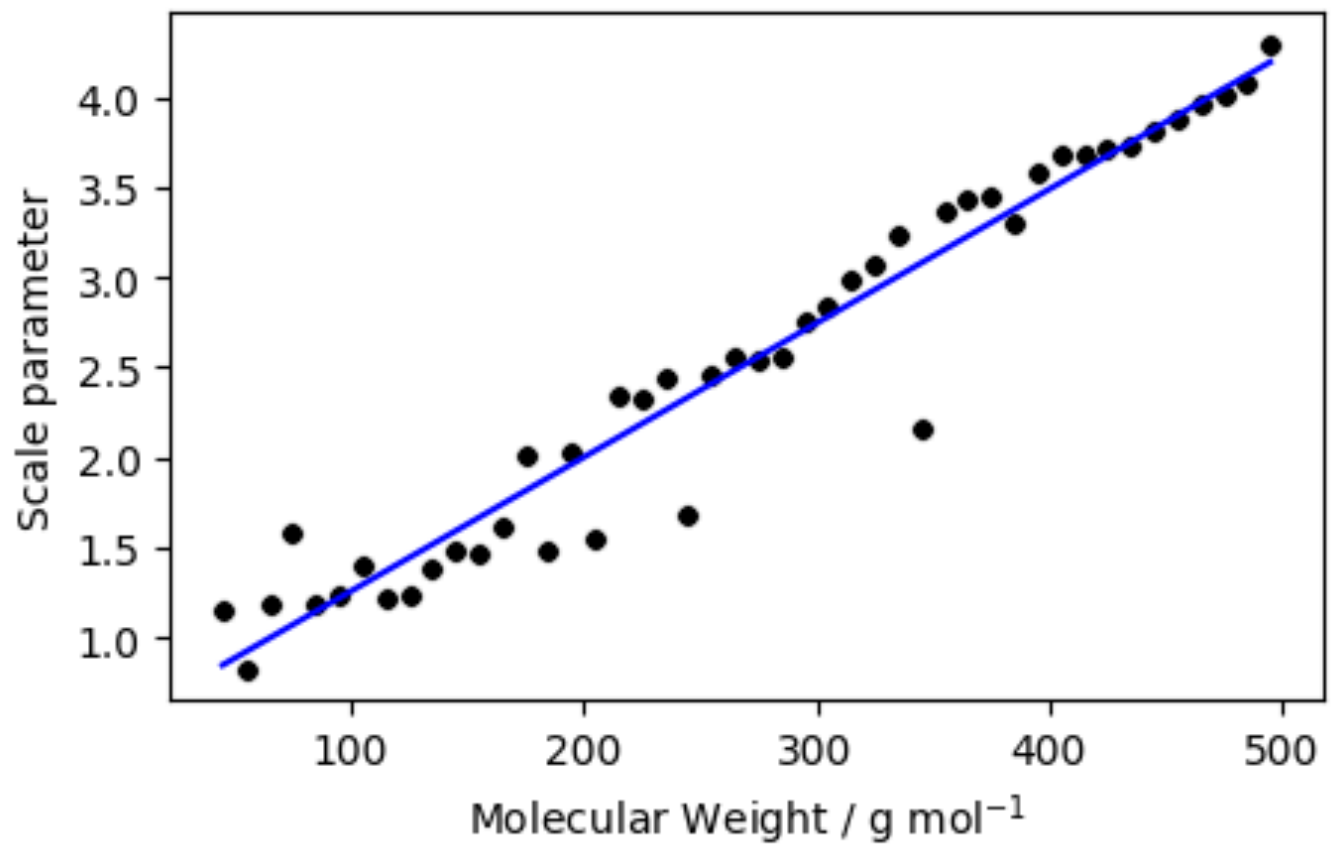

**Fig. S41** General trend of the scale parameter *vs*. MW, fitted as $scale = 0.0074\times\mathrm{MW} + 0.5108$.

## 7.2 Recursive MA algorithm

Examining the distribution of MA versus molecular weight (MW) reveals that the uncertainty in MA is reduced at low MW. When multi-level fragmentation data is available *via* tandem mass spectrometry, a recursive logic can take advantage of the tighter uncertainty bounds at low molecular weight to yield an improved MA estimate. Our recursive MA algorithm, represented as the mathematical function *MA* operating on a molecule or fragment *f*, hinges on the following operating principles, summarized in **Fig. S42**.

1. Tabulated monoisotopic masses for period table elements. Fragments exactly matching an isotope trigger a base case: $MA(f) = 0$.
2. A second base case is used when *f* has no child ions and cannot be identified as an element. An upper bound that can be calculated solely based on the molecular weight of a molecule or fragment (discussed below).
3. The recursive case is used when at least one child ion is present. For each child ion *c*, $MA(c)$ along with the MA of its associated complement mass $MA(c') = MA(f - c)$ is calculated. Note that the complement ion may not be observed.
4. If requested by the user, the algorithm can extend the search for child ions to the current MS level by examining pairs of ions $i_1$ and $i_2$ among *M*'s siblings that are complementary with respect to *M*, *i.e.* $i_1 + i_2 = M$. This functionality is essential when higher level MS data is not available.

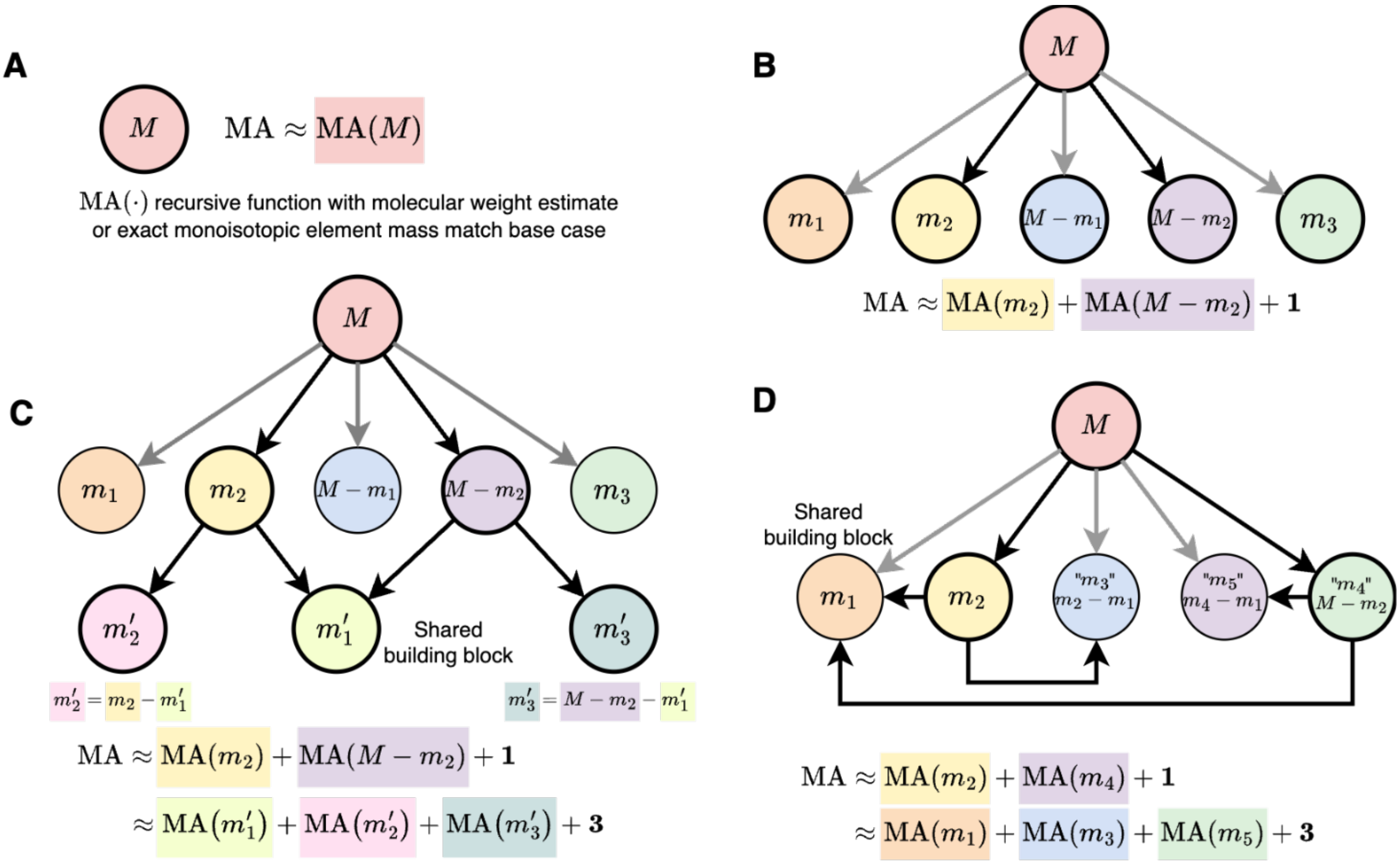

**Fig. S42** . Scenarios encountered during execution of the recursive MA algorithm. (A) Base case: for a molecule or fragment with no associated child ions, MA is estimated using molecular weight, either via an exact match to the monoisotopic mass of an element (MA = 0) or the statistical MA ~ MW relationship. (B) When child ions are present, the algorithm recursively evaluates the possibility of forming *M* from every pair of child ions .

The molecular weight estimate base case is informed by the empirical relationship between MW and maximum (worst-case) MA, depicted in **Fig. S43**, where each point represents the 99$^{th}$ MA percentile within a given mass bucket. Bucket boundaries follow logarithmic spacing to give equal priority to different mass scales. This empirical distribution of points can be fit to a piecewise linear function of the following approximate form.

$$\mathrm{MA(MW)} = \begin{cases} 0, & x < 19 \\ \dfrac{(\mathrm{MW} - 19)}{13.5}, & x \geq 0 \end{cases}$$

It can be shown that this empirical fit closely follows a form mandated by assembly theory. This derivation is based on two fundamental assumptions.

1. The MA of fundamental building blocks is 0.
2. In the absence of structural reuse (worst-case or maximum MA), fragmenting a molecule into roughly equal parts gives the following relationship (constant 1 representing one composition step).

$$\mathrm{MA(MW)} = \mathrm{MA}\left(\frac{\mathrm{MW}}{2}\right) + \mathrm{MA}\left(\frac{\mathrm{MW}}{2}\right) + 1$$
$$= 2\mathrm{MA}\left(\frac{\mathrm{MW}}{2}\right) + 1$$

Recursive application of this rule *N* times gives

$$\mathrm{MA(MW)} = 2^{\mathrm{MA}\left(\frac{\mathrm{MW}}{2^N}\right)} + \sum_{n=0}^{N-1} 2^n$$
$$= 2^N \mathrm{MA}\left(\frac{\mathrm{MW}}{2^N}\right) + 2^N - 1.$$

When subdivision reaches the level of any building block with molecular weight $\mathrm{MW}_0$, i.e. $\frac{\mathrm{MW}}{2^N} = \mathrm{MW}_0$, given $\mathrm{MA}(\mathrm{MW}_0) = 0$ we obtain

$$\mathrm{MA(MW)} = 2^N \mathrm{MA}(\mathrm{MW}_0) + 2^N - 1$$
$$= 2^N - 1$$
$$= \frac{\mathrm{MW} - \mathrm{MW}_0}{\mathrm{MW}_0}.$$

In light of this derivation, the empirical formula seems to indicate that typical values of $\mathrm{MW}_0$ lie between 13 and 19, in line with the composition of typical organic compounds from the elements boron to fluorine.

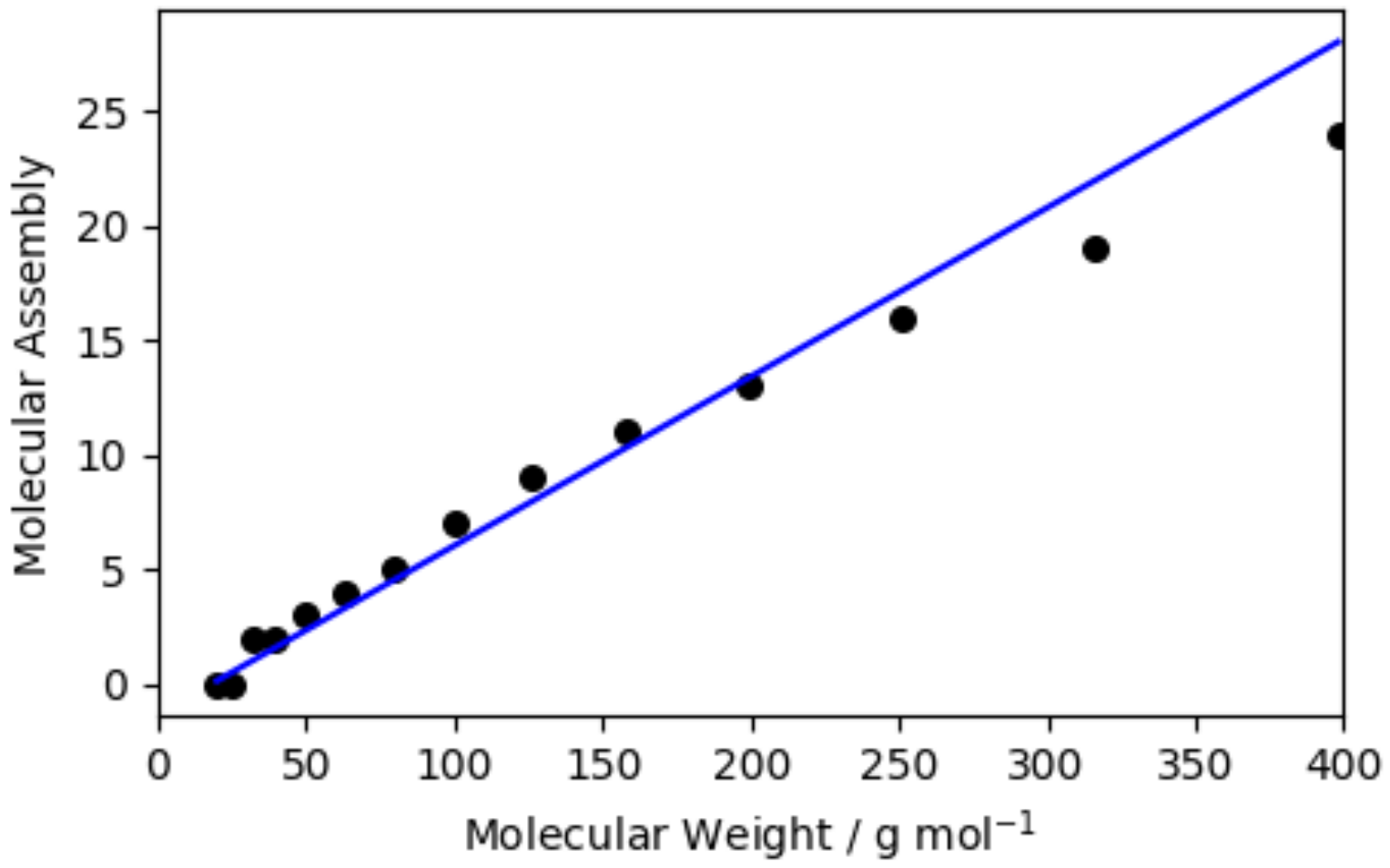

**Fig. S43** Fit of the piecewise linear function of molecular weight to logarithmically-binned 99 percentile MA values, as an upper limit to MA(MW).

## 7.3 Experimental Mass Spectrometry

For the experimental multi-level fragmentation mass spectrometry, stock solutions were prepared using 10 mg of the sample per 10 mL of acetonitrile and further diluted using 250 µL of stock solution and 9.75 mL of acetonitrile yielding a final sample concentration of approx. $8\times10^{-5}$ M. These samples were filtered through a 0.45 µm nylon syringe filter prior to MS analysis. Samples were analysed for 6 minutes by direct injection using a Thermo Scientific Orbitrap Fusion Lumos Tribrid mass spectrometer with a HESI ion source applying a +5.0 kV voltage. The ion transfer tube temperature was set to 290 °C and the vaporiser temperature was set to 250 °C. Samples were analysed *via* two different Single Ion Monitoring (SIM) methods investigating firstly a stepped $MS^3$ procedure and secondly an $MS^5$ procedure. The stepped $MS^3$ procedure involved fragmenting the target $MS^1$ analyte *via* HCD at 6 discrete energy levels (15, 25, 35, 45, 55, 65%, normalised values as implemented in the machine) before collating them into 2 $MS^2$ spectra (15, 25, 35%) and (45, 55, 65%). The $MS^2$ spectra were each analysed via a data-dependent acquisition (DDA) method where the 40 most abundant ions from each were selected for $MS^3$ fragmentation. Dynamic exclusion of ions was set for 120 seconds after the ion had been selected six times in 60 seconds. The resolution of the SIM scan was 120000 and the $MS^2$ and $MS^3$ fragmentation spectra were set at 30000 with an isolation window of 5 ppm. The $MS^5$ procedure achieved ion fragmentation *via* HCD set to 35% with an isolation window of 10 ppm. The spectra were analysed via a DDA method where the 40 most abundant $MS^2$ ions were selected for $MS^3$ fragmentation, the 30 most abundant $MS^3$ ions were selected for $MS^4$ fragmentation, and the 20 most abundant $MS^4$ ions were selected for $MS^5$ fragmentation. The dynamic exclusion and resolution remained the same as the stepped $MS^3$ method with the $MS^4$ and $MS^5$ also set at 30000.

All molecules with the MW 300 ± 5 $g\cdot mol^{-1}$ used in the multi-level fragmentation mass spectrometry are depicted in **Fig. S44**.

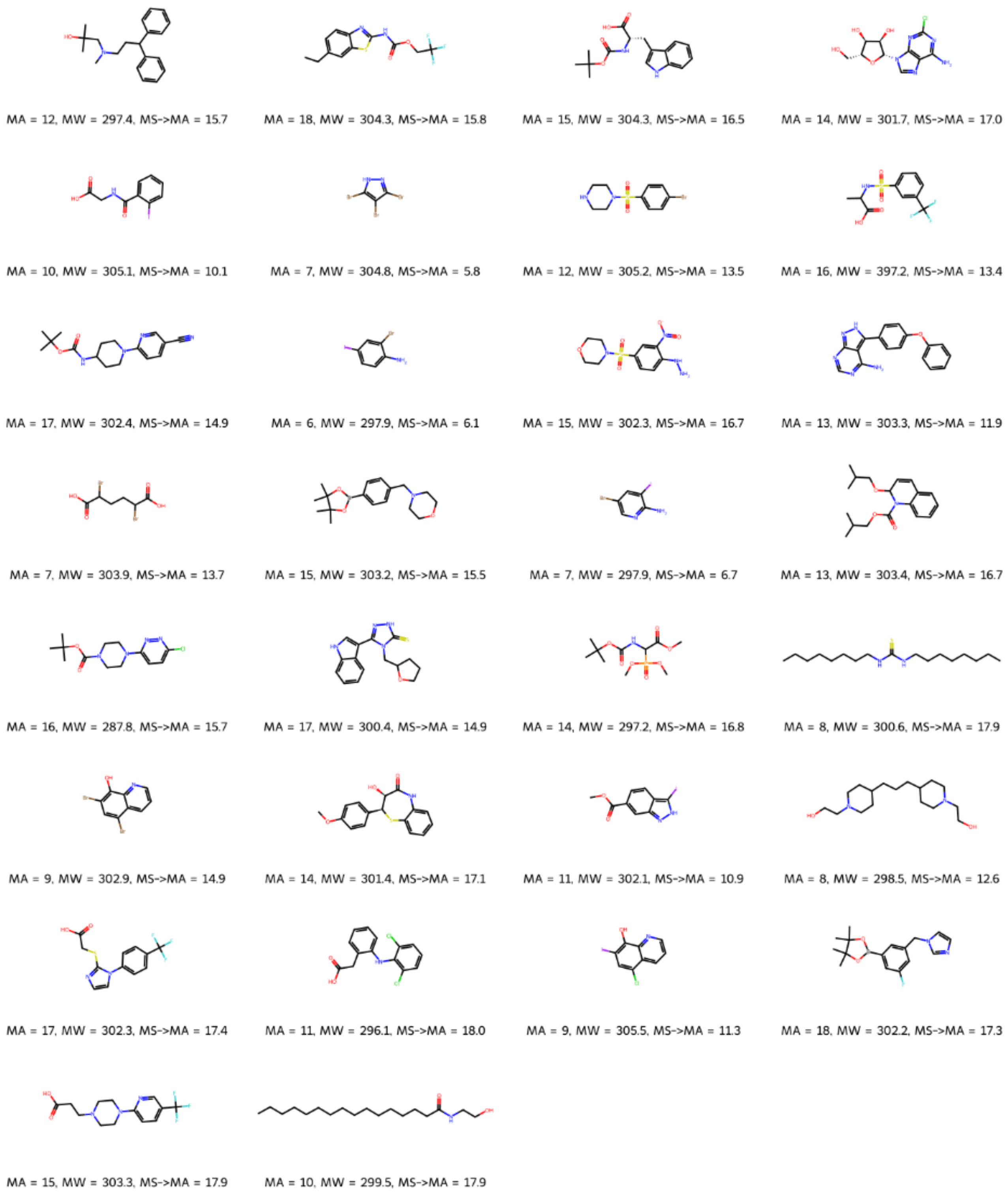

**Fig. S44** 30 Molecules with up to MS5 data with a similar mass of 300 ±5 g·mol$^{-1}$ and various MA from 6 to 18.